\documentclass[journal]{IEEEtran}

\usepackage{graphicx} 
\usepackage{subcaption}

\usepackage{cite}
\usepackage{hyperref}
\hypersetup{
    colorlinks,
    linkcolor={red!50!black},
    citecolor={blue!50!black},
    urlcolor={blue!50!black}
}
\usepackage{amsmath, amssymb, amsfonts, enumerate, bbold, multirow,booktabs}
\usepackage{wasysym}
 \usepackage[normalem]{ulem}

\providecommand{\argmax}{\mathop\mathrm{arg max}} 

\def\reals{\mathbb R}

 \def\sfa{\widetilde T}

\def\sfa{\widetilde T}
\def\reals{\mathbb R}
\newtheorem{theorem}{Theorem}
\newtheorem{lemma}{Lemma}
\newtheorem{proposition}{Proposition}

\newtheorem{remark}{Remark}
\newtheorem{definition}{Definition}

\usepackage[usenames,dvipsnames,table]{xcolor}

\newcommand{\blue}[1]{{\color{black} #1}}
\newcommand{\purple}[1]{{\color{black} #1}}
\newcommand{\mehdi}[1]{{\color{black}#1}}
\newcommand{\brown}[1]{{\color{black}#1}}
\newcommand{\orange}[1]{{\color{black}#1}}

\newcommand{\revise}[1]{{\color{black}#1}}
\newcommand{\rev}[1]{{\color{black}#1}}

\author{Mehdi Davoudi,  Junjie Qin, and Xiaojun Lin
\thanks{M. Davoudi and J. Qin are with the Elmore Family School of Electrical and Computer Engineering, Purdue University. X. Lin is with the Department of Information Engineering, The Chinese University of Hong Kong.
	 Emails: 
        {\tt\small \{mdavoudi , jq, linx\}@purdue.edu}}%
}

\begin{document}
\title{
	\textcolor{black}{Market-Driven Equilibria for Distributed Photovoltaic Panel Investment}
}

\author{Mehdi Davoudi,
	Junjie Qin,
	and Xiaojun Lin
	\thanks{M. Davoudi and J. Qin are with the Elmore Family School of Electrical and Computer Engineering, Purdue University, West Lafayette, IN, USA. X. Lin is with the Department of Information Engineering, the Chinese University of Hong Kong.
		 Emails:
		{\tt\small mdavoudi@purdue.edu, jq@purdue.edu, xjlin@ie.cuhk.edu.hk}}%
}


\maketitle

\addtolength{\textfloatsep}{-2mm}
\begin{abstract}
	This study investigates long-term investment in distributed photovoltaic panels by individual investors. We consider a setting where investment decisions are driven by expected revenue from participating in short-term electricity markets over the panel lifespan. These revenues depend on short-term market equilibria, including prices and allocations, which are endogenously influenced by the aggregate installed panel capacity. We model interactions among investors as a non-atomic game and develop a framework that links short-term market equilibria to the resulting long-term investment equilibrium. Within this framework, we analyze three market mechanisms: (a) a single-product real-time energy market, (b) a product-differentiated real-time energy market that treats solar energy and grid energy as different products, and (c) a contract-based panel market that trades rights to future production from panel capacity ex ante, rather than realized solar production ex post. For each mechanism, we derive short-term equilibrium outcomes and associated expected revenues, and analytically characterize the corresponding long-term Nash equilibrium capacity. \rev{We compare these investment equilibria with a benchmark social optimum and establish that, in the baseline risk-neutral setting, the product-differentiated market attains this benchmark, while the single-product market induces lower investment.} We also show that the contract-based market can lead to over-investment when users' additional valuations for solar energy are small. We further study extensions incorporating time-varying operating conditions and investor risk aversion, showing how these considerations affect expected revenues and investment incentives while broadly preserving key insights from the baseline analysis. Finally, we \rev{evaluate our theoretical findings through a numerical case study.}
\end{abstract}
\section{Introduction}\label{sec:intro}
The increasing adoption of renewable energy, particularly \textcolor{black}{photovoltaic (PV) panels}, plays an important role in modern power systems. 
While utility-scale solar projects contribute significantly to overall capacity, distributed rooftop installations are playing an increasingly prominent role in reshaping the electricity landscape. \textcolor{black}{Notably, by the end of 2024, the U.S. had installed about 175~GW of solar capacity, roughly 31\% of which came from residential and commercial installations~\cite{9e3161cca1ff4906b8149e1c4c148787}.}

Unlike centralized projects planned and operated by utilities, distributed solar investments at the distribution level reflect individual adoption and investment decisions by households and businesses.
Several factors influence these decisions. \revise{A key factor is the financial return from the investment. We acknowledge that, in many existing residential and commercial PV programs, this return is primarily shaped by regulated or retail compensation schemes, such as feed-in tariffs, net-metering programs, or other administratively determined rates for PV generation injected into the grid. These compensation mechanisms are relatively static in nature \cite{flores2015compensation, zinaman2017grid}.
	However, emerging research on transactive energy and local or distribution-level market designs has also examined more dynamic, market-based mechanisms for coordinating distributed energy resources (DERs) and local electricity users. These studies have shown that, under appropriate market and regulatory conditions, such mechanisms can create additional revenue streams and improve financial outcomes for DER participants beyond those available under conventional tariff-based compensation schemes \cite{azim2025tariffs,park2024p2p}.
	
	Motivated by this latter line of research, this paper focuses on settings in which the financial return on PV investment is driven by the expected revenue earned through participation in short-term distribution-level markets over the operational lifetime of the PV panels. In these settings, the resulting revenue is determined by the corresponding market equilibria. These short-term equilibria, in turn, depend on the aggregate installed PV capacity, which determines the amount of PV generation available as market supply. This aggregate installed capacity is itself determined by the long-term investment equilibrium arising from the interactions among individual investors.}
Therefore, there is a \emph{feedback loop} between the equilibria of short-term markets and the long-term investment equilibrium, as each influences the other.
A precise characterization of this relationship is crucial for understanding the emergence of long-term investment equilibria under different market designs.
 Such a characterization is not only important for utilities and system operators, who may need to account for distributed capacity growth in their planning studies, but also for policymakers and market designers seeking to understand the long-term impact of alternative market designs on PV panel adoption.
This motivates us to study the following critical open\footnote{\textcolor{black}{We note that a few studies in related settings, such as conventional generation investment, also link long-run investment equilibria to short-run operational equilibria. However, they largely focus on spot markets~\cite{bushnell2007equilibrium,mousavian2020equilibria} and either do not yield closed-form characterizations~\cite{bushnell2007equilibrium} or do not incorporate detailed consumer modeling, e.g., heterogeneous preferences or valuations~\cite{bushnell2007equilibrium,mousavian2020equilibria}.}} question:
\emph{How do short-term market equilibria, arising from diverse solar market mechanisms, shape the corresponding long-term distributed investment equilibrium?}

\subsection{Organization and Contributions}\label{contributions}
To address this \textcolor{black}{open} problem, we consider a large population of individual investors and aim to characterize the long-term investment equilibrium that emerges from the previously described feedback loop. We first develop an investment model, formulated as a non-atomic game, \textcolor{black}{where each player is infinitesimal with negligible market power, reflecting the large number of small investors in distribution-level settings. This model links long-term investment decisions to expected revenues under a given market design (Section~\ref{sec:model}).} To characterize these revenues, we introduce a \emph{unified} modeling framework for \textcolor{black}{different types of} short-term markets (Section~\ref{sec:mechanisms}) that allows us to formally define and compare different market designs. \textcolor{black}{Our unified market model captures} key features such as product differentiation and ex-ante/ex-post trading. We then apply it to three representative mechanisms that differ in these features (Section~\ref{section3})\textcolor{black}{.} First, we study (a) a single-product real-time market similar to current
	real-time electricity markets; then (b) a new product-differentiated real-time market to assess the impact of distinguishing clean solar energy from conventional grid energy; and finally (c) a new contract-based panel market that operates on a much slower timescale than real-time markets. For each, we first derive analytical characterizations of the short-term market equilibria and then use the investment model to \textcolor{black}{obtain and compare} the associated long-term equilibrium aggregate capacities (Section~\ref{section4}). Our analysis, initially developed under homogeneous operation periods and
	risk-neutral investors, is then extended to incorporate heterogeneous operation
	periods and risk-sensitive investment behavior (Section~\ref{extensions}),
	followed by numerical experiments (Section~\ref{Numerical}).

We contribute to the literature in the following key ways:
\begin{enumerate}[(a)]
\item
We offer an analytically tractable way to link short-term market equilibria to long-term aggregate investment equilibrium through a non-atomic game-based investment model that captures how individual investors affect one another through aggregate investment decisions. The framework can also accommodate common regulated pricing schemes, such as feed-in tariffs and time-of-use rates, as special cases.

\item We develop a unified market-modeling framework that delivers closed-form characterizations of short-term market equilibria across diverse market mechanisms, accommodating both ex-ante and ex-post trading as well as heterogeneous consumer preferences.

\item These analytical characterizations yield new insights into how different market mechanisms shape PV investment. Specifically, we theoretically establish
that the single-product real-time market, where solar energy is pooled with conventional grid energy and traded \emph{ex-post}, leads to under-investment. Thus, we believe there is significant value in considering alternative market mechanisms. In particular, in the baseline risk-neutral setting, we find that the product-differentiated market, where solar energy is traded \emph{ex-post} and may be sold at a premium due to heterogeneous user preferences, yields the socially optimal capacity \rev{under the welfare framework considered in this paper}. Finally, the contract-based market, where \emph{ex-ante} panel capacity is traded instead of realized generation, can result in over-investment when users' extra valuations for solar energy are small. Such over-investment can be desirable under economies of scale, where higher aggregate deployment lowers the cost of PV investment.
\end{enumerate}

\subsection{Related Literature}
\mehdi{This work is related to the intersection of two broad lines of literature: distribution-level electricity market mechanisms and long-term investment in solar panels.} \mehdi{Most existing studies on the investment side of the literature adopt a centralized perspective, focusing on the optimal strategy of a single decision-maker such as a utility planner or an individual investor (see, e.g., \cite{bauner2015adoption,el2004optimal, goop2021impact}). However, this line of work neglects the interdependencies of investment decisions in distributed settings, where the actions of one investor can influence the returns of others through shared market mechanisms.}

\purple{On the market side, a variety of distribution-level electricity market designs have been proposed. Understanding how these structures differ is essential for our analysis, as their design features shape the expected revenue of investors.}
 \purple{In this context, }\mehdi{prior research has proposed a range of market designs, with particular emphasis on real-time pricing and service contract-based approaches. In a broad sense, real-time pricing mechanisms involve the determination of time-varying electricity prices \purple{based on} the \emph{ex-post} realized values of supply and demand. In contrast, service contract-based mechanisms typically offer less volatile prices in exchange for a fixed level of electricity service, relying on \emph{ex-ante} statistics such as those related to random solar generation. Extensive research has explored the relative merits of these retail market designs, examining factors such as economic efficiency~\cite{chao2012competitive, LO201939}, price and bill volatility~\cite{6197252, 10.1145/3538637.3538846}, and environmental impacts~\cite{holland2008real}. In addition, some studies have examined another facet of market design: whether electricity from solar and conventional sources should be treated as differentiated products, given their distinct operational characteristics and emissions~\cite{woo2014review,sorin2018consensus}. Despite the contributions of these studies~\cite{chao2012competitive, LO201939,6197252, 10.1145/3538637.3538846,holland2008real,woo2014review, sorin2018consensus}, they all focus primarily on short-term performance criteria and overlook the long-term impacts of their designed markets on investment decisions.}

The most closely related paper to ours is \cite{henriquez2021sharing}, which studies equilibrium PV investment capacities under feed-in-tariff and sharing-economy markets. However, it adopts a finite-player game with large investors, whereas we consider a large population of small investors. Our \orange{non-atomic} game formulation enables analytical comparisons of equilibrium capacities \orange{driven} by different market mechanisms, whereas \cite{henriquez2021sharing} relies on numerical simulations for such comparisons. It is also worth noting that, while non-atomic game formulations have been used in other contexts, such as workplace EV charging pricing~\cite{mou2024nexus} and distributed storage investment~\cite{qin2019distributed}, they have not been used in an investment framework that links short-term market equilibria under diverse market designs to long-term investment equilibrium.

This paper generalizes our previous conference work~\cite{davoudi2024role} by
(a) extending our theoretical results to settings with heterogeneous operation
periods and risk-sensitive investors, (b) presenting rigorous mathematical
proofs, and (c) demonstrating our theoretical results through numerical
experiments.

\section{Investment Model}\label{sec:model}

\begin{figure}[h]
	\centering
	\includegraphics[width=0.49\textwidth]{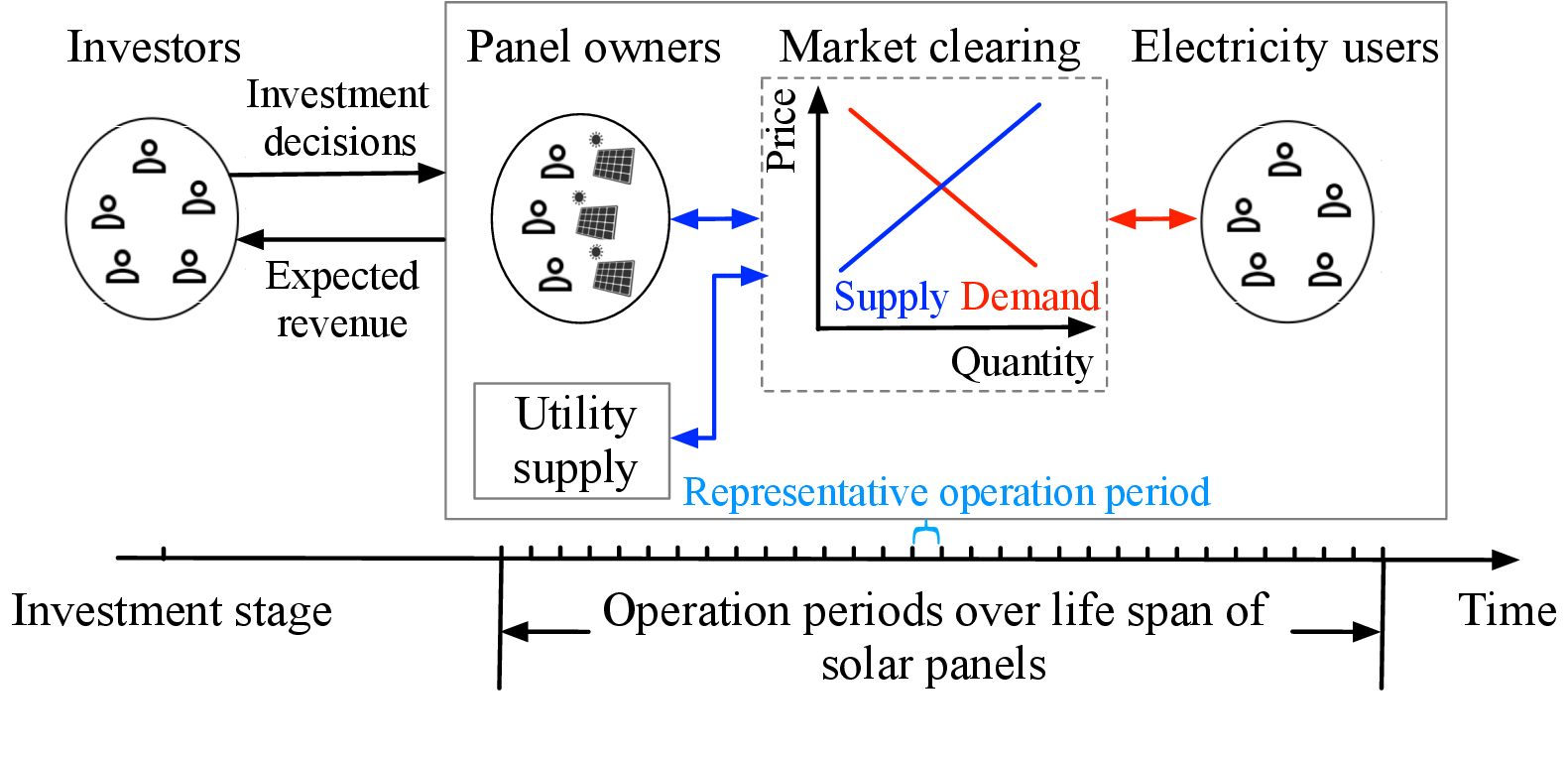}
	\caption{Schematic for distributed panel investment}
	\vspace{-1.6em}
	\label{fig:schematics}
\end{figure}

A large group of potential investors considers installing PV panels and becoming panel owners. \rev{Their investment decisions weigh the expected revenues from participating in short-term electricity markets over the panel lifespan against the upfront capital and installation costs and the operation and maintenance expenses incurred over that lifespan. Since the primary objective of this paper is to understand how different short-term market mechanisms affect these long-term investment decisions, we primarily focus on expected market revenues relative to these costs and assume that investors are risk neutral.} \rev{We later relax the risk-neutrality assumption in Section~\ref{extensions}, where we also examine the effect of downside revenue risk on investment decisions rather than considering expected revenue alone.} An overview of the setup is shown in Fig.~\ref{fig:schematics}.

\subsection{Connecting Planning and Operation Timescales}

PV panels are typically assumed to have a 25-year lifespan~\cite{sodhi2022economic}. Accordingly, at the investment stage, estimating revenue over such a long horizon is challenging because long-term forecasts of solar generation and market conditions are rarely available. A common approach is therefore to use a shorter planning period (e.g., one year), estimate expected revenue from historical data, and scale the estimate by the number of such intervals over the panel lifespan. \rev{Following this approach, we next describe how expected revenue is evaluated within such a planning period and how this estimate connects to lifetime expected revenue.}

\rev{Throughout the paper, we take the first planning period over the panel lifespan as the base planning period for our calculations (e.g., the first year of a 25-year lifespan).} Since such a planning period is long, it contains multiple \emph{operation periods}, each on a shorter timescale (e.g., an hour or a day). We consider a finite-horizon discrete-time planning window consisting of operation periods $\mathcal{T}=\{1,\dots,T\}$, indexed by $t$, where $T$ is the total number of such periods. We assume all short-term operating conditions other than solar generation, such as load and utility prices, are fixed and common across operation periods \rev{within the base planning period}. Solar generation is instead uncertain and may vary from one operation period to another, \rev{with its realization in each operation period drawn from the same probability distribution}. Therefore, although realized revenue may differ across operation periods due to different solar-generation realizations, the \emph{expected revenue} is the same for every operation period. This allows us to first estimate the expected revenue over a single operation period, e.g., the \emph{representative operation period} shown in Fig.~\ref{fig:schematics}, and then multiply this value by $T$ to obtain the expected revenue over the base planning period. \rev{The revenues associated with the remaining planning periods are then related to this base-period revenue by accounting for inflation-driven changes in monetary quantities, such as market-clearing prices, and discounting the resulting future revenues to their present values. Equivalently, these steps can be combined into a single scaling factor, denoted by $\sfa$, such that multiplying the expected revenue from the representative operation period by $\sfa$ directly yields the present value of the expected revenue over the full panel lifespan. The calculation of $\sfa$ is provided in the next subsection.}

\begin{remark}[More granular operation model]
	\mehdi{A more detailed operational model for the base planning period can be implemented by incorporating multiple operation periods, each characterized by distinct supply and demand values/distributions. For instance, separate solar generation profiles can be used for daytime and nighttime hours, along with different load patterns for weekdays and weekends. The expected revenues from these operation periods can then be aggregated to obtain the expected revenue over the base planning period and subsequently scaled to obtain the present value of expected revenue over the full panel lifespan. These settings can be accommodated within our framework; see Section~\ref{heterg:time} for further discussion.}
\end{remark}

\rev{\subsection{Present-Value Adjustment and Investment Cost}\label{lifecycle:financial}}

\rev{Let $N_{\mathrm{life}}$ denote the number of equal-length planning periods over the panel lifespan; for example, $N_{\mathrm{life}}=25$ when each planning period is one year. Let $\rho\geq0$ denote the nominal discount rate per planning period, and let $\kappa_y>0$ denote the cumulative inflation adjustment factor from the base planning period to planning period $y\in\{1,\ldots,N_{\mathrm{life}}\}$, with $\kappa_1=1$. We assume that inflation is the only source of long-term changes in monetary quantities such as electricity prices, so their values in planning period $y$ are obtained by multiplying their base-period values by $\kappa_y$. Consequently, the expected revenue in planning period $y$ is $\kappa_y$ times the expected revenue in the base planning period.}

\rev{These planning-period expected revenues are then discounted to obtain their present values. Accounting for discounting at rate $\rho$ and noting that each planning period contains $T$ operation periods, the scaling factor introduced in the previous subsection is}
\rev{
	\begin{equation}\label{eq:lifecycle_scaling}
		\sfa
		:=
		T\sum_{y=1}^{N_{\mathrm{life}}}
		\frac{\kappa_y}{(1+\rho)^y}.
	\end{equation}
}

\rev{Thus, multiplying the expected revenue of the representative operation period by $\sfa$ gives the present value of the expected revenue over the panel lifespan.}

\rev{
	\begin{remark}[Long-term changes beyond inflation]
		The above treatment captures long-term changes in monetary quantities associated with general inflation. Other changes in monetary or physical conditions over the panel lifespan, including module degradation, would require more detailed analysis to account for their long-term evolution and are left for future work.
	\end{remark}
}

\rev{We next account for the corresponding investment costs. These costs consist of the upfront capital and installation cost, as well as the operation and maintenance (O\&M) expenses incurred throughout the panel lifespan. Let $\pi_{\mathrm{cap}}$ denote the upfront capital and installation cost per unit panel capacity. Moreover, let $\pi_{\mathrm{om}}$ denote the recurring O\&M cost per unit installed capacity over one planning period, expressed at the base-period price level; consistent with relevant PV techno-economic studies~\cite{nrel_panel}, we consider this cost on a per-capacity basis. Assuming that inflation is the only source of long-term changes in O\&M expenses, their values in planning period $y$ are obtained by multiplying $\pi_{\mathrm{om}}$ by $\kappa_y$. Their present value can therefore be incorporated into the total investment cost per unit panel capacity as}
\rev{
	\begin{equation}\label{eq:lifecycle_cost}
		\pi_0
		:=
		\pi_{\mathrm{cap}}
		+
		\pi_{\mathrm{om}}
		\sum_{y=1}^{N_{\mathrm{life}}}
		\frac{\kappa_y}{(1+\rho)^y}
		=
		\pi_{\mathrm{cap}}
		+
		\frac{\sfa}{T}\pi_{\mathrm{om}}.
	\end{equation}
}

\rev{Accordingly, $\pi_0$ used throughout the remainder of the paper denotes the total investment cost per unit panel capacity, defined as the sum of the upfront capital and installation cost and the present value of O\&M expenses over the panel lifespan.}

\subsection{Panel Investment Game} \label{investment:game}

We consider a large number of small investors, each of whom is infinitesimal relative to the market and therefore has no individual market power, while their aggregate investment decisions affect market outcomes. This motivates modeling their interaction as a non-atomic game \cite{aumann2015values}. Consequently, we represent the population of investors as the continuum $\mathcal{I}_\mathrm{inv} := [0,1]$. We consider the case where each investor $i\in\mathcal I_\mathrm{inv}$ decides whether to install a PV panel with a fixed capacity $\overline c$. \rev{Panel capacity can be represented using different consistent measures; throughout this paper, we represent it by panel area, measured in $\mathrm{m}^2$.} Let the decision of investor $i$ be $x_i\in\{0,1\}$, where $x_i=1$ indicates investment and $x_i=0$ indicates no investment. The total installed panel capacity is then 
\begin{equation}\label{eq:agg:c}
	c = \overline c \int_{\mathcal I_\mathrm{inv}} x_i \,\, \mathrm{d}i.
	\vspace{-0.25em}
\end{equation}

\rev{We assume that the investment cost is attractive, at least when the solar production can be sold at the utility rate:}
\begin{equation}\label{eq:pi0cond}
	\blue{c}\pi_0 < \sfa \pi_\mathrm{u} \mathbb E [\blue{c}G],
\end{equation}
\rev{where $\pi_{\mathrm{u}} > 0$ is the per-unit electricity price faced by consumers when purchasing electricity supplied by the utility during any operation period of the base planning period}, and $G$ denotes solar energy generated per unit panel capacity over an operation period, measured in $\mathrm{kWh}/\mathrm{m}^2$. Thus, $cG$ represents the total solar energy generated during an operation period. \rev{Here, $G$ is modeled as a random variable with cumulative distribution function $F_G$ and probability density function \(f_G\).}
\revise{For simplicity, we assume that \(G\) has bounded support \([0,\overline g]\), where \(\overline g<\infty\), admits a continuously differentiable density \(f_G\), and has finite first and second moments. We further assume that \(f_G(g)>0\) for all \(g\in(0,\overline g)\).}

\rev{Although each investor is individually infinitesimal, investors interact through the aggregate installed capacity $c$: their collective investment decisions determine $c$, which in turn affects the short-term market outcomes and the revenues earned from installed panels. Therefore, the expected return of investor $i$ depends on both their own investment decision $x_i$ and the aggregate investment decisions of the other investors.} Accordingly, the payoff for investor \(i\) from installing panel capacity \(\overline{c}\) and selling the generated electricity in short-term market \(m\), where \(m\) indexes the different short-term market mechanisms, throughout the panel lifespan is defined as the present value of the expected revenue over the panel lifespan minus the \rev{investment cost}:
\begin{equation}\label{eq:inv:po}
	\Pi^\mathrm{inv}_{m,i} (x_i, c) = \left[
	(\Pi_{m}^\mathrm{s}(c)/c)- \pi_0 \right] \overline c\,x_i. 
\end{equation}
Here, \(\Pi_m^\mathrm{s}(c)\) denotes the \rev{present value of the expected total revenue received by all panel owners over the panel lifespan} under short-term market \(m\), which will be defined separately for each market in Section~\ref{sec:mechanisms}. Thus, \(\Pi_m^\mathrm{s}(c)/c\) is the \rev{present value of expected revenue per unit of installed capacity over the panel lifespan}, and \((\Pi_m^\mathrm{s}(c)/c)\,\overline c x_i\) gives investor \(i\)'s \rev{present value of expected revenue over the panel lifespan}. Subtracting the \rev{investment cost} \(\pi_0 \overline c x_i\) yields the net payoff \(\Pi^\mathrm{inv}_{m,i}(x_i,c)\). Each investor \(i \in \mathcal I_\mathrm{inv}\) chooses their investment decision to maximize \(\Pi^\mathrm{inv}_{m,i}(x_i,c)\).

Given the players $\mathcal I_\mathrm{inv}$, actions $x = \{x_i\}_{i \in \mathcal I_\mathrm{inv}}$, and payoff \eqref{eq:inv:po}, we have defined the \emph{\textcolor{black}{panel} investment game}, which is an \emph{aggregate game} where \textcolor{black}{investor $i$'s payoff depends on the other investors' decisions only through the total panel capacity $c$}. A Nash equilibrium (NE) for this game is defined as: 
\begin{definition}[NE for panel investment game]\label{def1}
	\textcolor{black}{The aggregate panel capacity $c$
		constitutes an NE for the \textcolor{black}{panel} investment game under market mechanism $m$ if there exists a collection of investment decisions $x$ such that $\Pi_{m,i}^\mathrm{inv}(x_i, c)\ge \Pi_{m,i}^\mathrm{inv}(x_i',c)$, 
		for all $x_i' \in \{0,1\}$ and $i\in \mathcal I_\mathrm{inv}$, and
		\eqref{eq:agg:c} holds.}
\end{definition}

Under this condition, no investor will have an incentive to unilaterally deviate from the NE decision given the aggregate capacity that is consistent with everyone's investment decision.

Building on this investment model, we next examine several representative solar market mechanisms to assess how their structural differences influence investors’ payoffs and, in turn, the resulting long-term NE aggregate capacities.

\section{Unified Modeling Framework for Short-term Market Mechanism Analysis}\label{sec:mechanisms}
 \purple{We present generic modeling elements applicable across diverse market settings and then provide a detailed description of the specific market mechanisms analyzed in this paper.}
 \vspace{-0.5em}
\subsection{Common Model Elements}\label{common:models}
\purple{Consider an operation period (e.g., \brown{the} representative operation period in Fig.~\ref{fig:schematics}) where the expected revenue of investors is shaped by short-term conditions in the distribution-level solar energy market.}
 \mehdi{The market consists of buyers (i.e., electricity consumers), sellers\footnote{We acknowledge the possibility that electricity prosumers can be both buyers and sellers at the same time. For the market mechanisms that we will consider in this paper, a prosumer can be modeled as the superposition of a buyer and a seller \textcolor{black}{\textcolor{black}{since} buying and selling prices are \textcolor{black}{uniform} within each market}. This observation may not apply to other markets where buying and selling prices are \textcolor{black}{non-uniform}.} (i.e., \textcolor{black}{the collection of investors who choose to invest}), and a utility company that provides supply when solar generation alone cannot fully satisfy demand. \orange{Similar to the investment game, we model the groups of a large number of buyers and sellers as continuous intervals \textcolor{black}{through a non-atomic game formulation.}\footnote{\textcolor{black}{Non-atomic game formulations \cite{aumann2015values} differ from non-atomistic market models \cite{mongey2021market}, where some players are large enough to exercise market power.}}}}

Define the set of buyers by the interval $\mathcal I_\mathrm{b}:= [0,1]$. 
In our analysis, we consider a situation where environmentally conscious electricity consumers have an extra \emph{heterogeneous} valuation for solar energy over the conventional energy supplied by the utility. For a buyer $i \in \mathcal I_\mathrm{b}$, we denote \purple{their} \textcolor{black}{fixed} extra valuation for each unit of load served by solar generation as $v_i \in \reals_+$, \mehdi{referred to as the \emph{solar premium}. \textcolor{black}{Accordingly, their willingness to pay for a unit of solar-served load is $\pi_{\mathrm{u}}+v_{i}$.
Heterogeneity is captured by the distribution of $\{v_i\}_{i\in\mathcal I_\mathrm{b}}
$ across buyers}}, characterized by a function $F_V$ defined as
 \mehdi{$F_V(v) = \int_{i \in \mathcal{I}_\mathrm{b}}\, \mathbb{1}\{v_i \le v\} \, \, \mathrm{d}i$},
where $\mathbb 1$ is \purple{the} indicator function. \mehdi{For simplicity, we assume that \( F_V \) is differentiable and strictly increasing on \([0, \overline{v}]\), where \( \overline{v} \ge 0 \).} \purple{We denote its derivative by \( f_V \), supported on \([0, \overline{v}]\), and assume \( f_V(v) \) is bounded away from zero for all \( v \in [0, \overline{v}] \).}  
Let the total electric load of all consumers be $L$. We assume $L$ is \emph{inelastic}, deterministic, and equally distributed across all consumers.

\orange{Let the set of sellers be $\mathcal I_\mathrm{s} := [0,1]$. Note that although only a subset of $\mathcal I_\mathrm{inv}$ decides to invest and thereby constitutes $\mathcal I_\mathrm{s}$, we relabel this subset to form a normalized interval $[0,1]$. We specifically examine the scenario where the sellers are homogeneous\footnote{Suppliers with heterogeneous panel characteristics have been studied in, e.g., \cite{10.1145/3538637.3538846}.}. Given the aggregate capacity \( c \) from~\eqref{eq:agg:c}, the homogeneity of sellers and unit length of \( \mathcal{I}_\mathrm{s} \) imply that the panel capacity of each seller is also $c$, as $c=\bar{c} \int_{\mathcal{I}_{\mathrm{inv}}} x_i \, \mathrm{d}i=c \int_{\mathcal{I}_s}  \, \mathrm{d}i$. Thus, the total solar generation from all sellers is \( cG \).} \mehdi{We now proceed with detailed modeling of the selected markets.}

\revise{
	\begin{remark}[Scope of the non-atomic modeling framework]
		The non-atomic formulation in both the short-term market and the investment model is motivated by the distribution-level setting considered in this paper, where a large number of small residential and commercial participants interact through aggregate market outcomes. This setting differs from markets with large strategic participants, such as large prosumers, utility-scale solar farms, or distributed energy resource aggregators. In such cases, strategic formulations such as Cournot competition or Stackelberg games may be more appropriate for the short-term market, and the long-term investment problem may become an atomic or mixed atomic--non-atomic game. The resulting revenue functions and equilibrium existence may therefore change. We leave such strategic-participation models for future work.
	\end{remark}
}

\subsection{Real-time Pricing Mechanisms} 
We elaborate on two alternative types of real-time pricing mechanisms proposed in the existing literature. In both mechanisms, the market price is determined by the realized values of supply and demand. The primary distinction between these two mechanisms lies in whether solar generation is treated as \purple{the same product as utility-supplied electricity}. This aspect becomes crucial when consumers place additional value on renewable generation. We denote the mechanism where solar is traded as a distinct product as the \emph{product-differentiated real-time market} (\texttt{prt}), while the other is referred to as the \emph{single-product real-time market} (\texttt{srt}).

Within the product-differentiated real-time market, the traded product is solar electricity.  Each seller $i\in \mathcal I_\mathrm{s}$ can sell any quantity of solar generation within the set $ \mathcal Q_i^\mathrm{s} = [0, cG]$, which depends on the realized value of $G$.  The seller payoff function \textcolor{black}{for the representative operation period} is 
\begin{equation}\label{seller:payoff:real}
\Pi_i^\mathrm{s}(q_i^\mathrm{s},\pi) = \pi  q^\mathrm{s}_i,
\end{equation}
where market equilibrium price $\pi$ and cleared quantity $q^\mathrm{s}_i$ will be defined momentarily. 
\purple{Given} the inelastic load $L$, \purple{a buyer $i \in \mathcal I_\mathrm{b}$} can purchase any quantity of solar within the range $\mathcal Q^\mathrm{b}_i= [0,L]$. Any remaining demand not met by solar will be fulfilled by the utility supply at the price $\pi_{\mathrm{u}}$. The buyer payoff function \textcolor{black}{for the representative operation period} is
\begin{equation}\label{cons:payoff}
	 \Pi_i^\mathrm{b}(q_i^\mathrm{b},\pi) = (v_i- \pi) q^\mathrm{b}_i - \pi_\mathrm{u} (L-q^\mathrm{b}_i),
\end{equation}
\textcolor{black}{where $q_i^\mathrm{b}$ is the cleared quantity for the buyer. The first term in~\eqref{cons:payoff} is the buyer’s surplus in the market, and the second term accounts for the payment for the residual demand served by the utility.}

Denote $q^\mathrm{s}:= \{q^\mathrm{s}_i\}_{i \in \mathcal I_\mathrm{s}}$ and $q^\mathrm{b}:= \{q^\mathrm{b}_i\}_{i \in \mathcal I_\mathrm{b}}$. 
We use the following equilibrium notion to characterize the market outcome:
\begin{definition}[Competitive equilibrium]\label{def:ce}
A price-allocation tuple $(\pi, q^\mathrm{s}, q^\mathrm{b})$ constitutes a competitive equilibrium (CE) of the market if the following conditions are met\orange{ \cite{mas1995microeconomic}:}
\begin{itemize}
	\item \emph{Individual rationality}: Given price $\pi$, for every seller $i \in \mathcal I_\mathrm{s}$,  $q_i^\mathrm{s} \in \argmax_{q \in \mathcal Q^\mathrm{s}_i}\Pi_i^\mathrm{s}(q,\pi)$, and for every buyer $i \in \mathcal I_\mathrm{b}$,  $q_i^\mathrm{b} \in \argmax_{q \in \mathcal Q^\mathrm{b}_i}\Pi_i^\mathrm{b}(q,\pi)$. 
	\item \emph{Market clearing condition}: 
	\[
	\int_{ \mathcal I_\mathrm{s}} q^\mathrm{s}_{i} \,\,\mathrm{d} i = \int_{\mathcal I_\mathrm{b}} q^\mathrm{b}_{i} \,\,\mathrm{d} i. 
	\]
\end{itemize}
\end{definition}

The CE conditions are conceptually straightforward: the resulting market outcome, comprising the equilibrium price and traded quantities, must be stable. In other words, no buyer or seller should have an incentive to unilaterally adjust their quantity, given that individual actions do not influence the market price. This price-taking assumption is consistent with our non-atomic model, where each agent is infinitesimal and cannot affect the price unilaterally.
\brown{
	\begin{remark}[Price discovery process]\label{price:discovery}
		We intentionally omit the price discovery process and focus on the CE, as our interest lies in the outcomes rather than the mechanism used to obtain them. Such outcomes can arise, for instance, from buyers and sellers submitting bids to a market operator who clears the market. \textcolor{black}{In our \orange{non-atomic} game model, where all agents are infinitesimally small and act as price takers, we show that an NE of such a bidding process corresponds to \orange{a} CE that we consider here (see Appendix~\ref{appendix:equiv}).} 
	\end{remark}
}

With the equilibrium price $\pi$ and the cleared quantities $q^\mathrm{s}_i$ for sellers in the product-differentiated real-time market, the \rev{present value of the expected total revenue for sellers over the panel lifespan} with an aggregate panel capacity $c$ is given by
\begin{equation}\label{prt:total:rev}
	\Pi^\mathrm{s}_\mathrm{prt}(c) = \widetilde T \,\mathbb E\left[\int_{\mathcal I_\mathrm{s}} \Pi^\mathrm{s}_i (q_i^\mathrm{s},\pi)\,\, \mathrm{d}i\right],
\end{equation}
where $\widetilde T$, as defined earlier, scales the expected revenue from the representative operation period to the \rev{present value of the expected revenue over the panel lifespan}.
We next describe the single-product real-time market, which also trades electricity.

Unlike the product-differentiated real-time market, in the single-product real-time market both solar energy and utility-supplied energy are pooled together as a single product, regardless of the source of generation. Thus, consumers cannot express a preference or pay a premium for solar, even if they assign it a higher value. In fact, it can be shown that the outcome of this market coincides with that of the product-differentiated real-time market in the special case where $v_i \equiv 0$. We denote the \rev{present value of the expected total revenue for sellers over the panel lifespan} in the single-product real-time market as $\Pi^\mathrm{s}_\mathrm{srt}(c)$.

\subsection{Contract-based Market} 
\mehdi{Unlike the real-time solar markets, the \emph{contract-based market} (\texttt{cb}) operates well in advance of the actual delivery period. Consequently, service contracts are traded based on the expected, rather than realized, solar generation. For concreteness, we follow the terminology in \cite{10.1145/3538637.3538846} and focus on a setting where the contract is tied to \emph{panel capacity}, meaning a panel owner rents out units of capacity and the buyer receives the realized solar output associated with that rented capacity over the duration of the contract. Since both the traded quantity and price are determined prior to the realization of generation, this market differs fundamentally from real-time pricing mechanisms. Nevertheless, for notational consistency, we will reuse the notation from the previous subsection to describe outcomes in the contract-based setting.}
\begin{remark}[Time-scale of the contract-based market]
		In this section, we assume that contracts cover a single operation period. Although contracts are often longer-term in practice, this simplification is consistent with our assumption that operation periods are identically distributed. Under this assumption, each period has the same expected market environment, so focusing on one representative operation period is sufficient for characterizing the \rev{present value of expected revenue over the panel lifespan}. Importantly, contracts are still set \emph{ex ante}, before the realization of solar generation. Section~\ref{heterg:time} relaxes this assumption and considers heterogeneous operation periods and multi-period contracts.
\end{remark}

Since seller $i \in \mathcal I_\mathrm{s}$ can rent out any portion of \purple{their} panel capacity $c$, the set of feasible quantities to sell is $\mathcal Q_i^\mathrm{s} =[0, c]$. Meanwhile, as buyer $i$ can in principle lease any panel capacity, we have $\mathcal Q_i^\mathrm{b} = \reals_+$.  
Given the market equilibrium price $\pi$ for panel capacities, the payoff for seller $i \in \mathcal I_\mathrm{s}$ is $\Pi_i^\mathrm{s}(q_i^\mathrm{s},\pi) = \pi q^\mathrm{s}_i$, 
and the payoff for buyer $i \in \mathcal I_\mathrm{b}$ is 
\begin{align}\label{eq:payoff:cb}
	\Pi_i^\mathrm{b}(q_i^\mathrm{b},\pi) = v_i \mathbb E\min\{q^\mathrm{b}_i  G, L \}-\!\pi q^\mathrm{b}_i 
	 -\!\pi_\mathrm{u} \mathbb E (L\!-\!q^\mathrm{b}_i G)_+,
\end{align}
\purple{where $(z)_+:= \max(z,0)$}, and $q_i^\mathrm{s}$ and $q_i^\mathrm{b}$ are the cleared quantities (i.e., panel capacities) for the seller and buyer, respectively. \textcolor{black}{In~\eqref{eq:payoff:cb}, the first two terms capture the buyer’s expected surplus in the market, while the last term is the expected payment to the utility for any residual demand.}

\purple{Adapting the same CE concept as in Definition~\ref{def:ce}, with the traded product and payoff functions updated, and given the equilibrium price $\pi$ and sellers' cleared quantity $q^\mathrm{s}$, we can express the \rev{present value of total seller revenue over the panel lifespan} for the contract-based market as}
\begin{equation}\label{cb:total:rev}
	\Pi^\mathrm{s}_\mathrm{cb}(c) =\widetilde T \int_{\mathcal I_\mathrm{s}} \Pi^\mathrm{s}_i (q_i^\mathrm{s},\pi)\,\, \mathrm{d}i.
\end{equation}

\begin{remark}[Unified modeling framework]\label{remark2}
	\mehdi{The proposed market model serves as a general framework capable of representing a variety of solar market mechanisms, provided that the traded product and the payoff structures for buyers and sellers are appropriately defined \textcolor{black}{such that a CE exists}. For example, both real-time pricing and contract-based arrangements can be formulated within this framework, with key distinctions captured through the specification of the traded product, the feasible sets of buyers and sellers, and the corresponding payoff functions. We abstract away additional implementation-specific features, as our primary objective is to investigate how different market structures shape long-term investment outcomes. While these mechanisms may differ in operational frequency, often leading to variations in price volatility as discussed in prior work \cite{6197252,10.1145/3538637.3538846}, such factors do not directly affect the investment equilibrium when only expected returns matter.}
\end{remark}

Now that we have completed the modeling of these market mechanisms, we proceed to analyze how the corresponding CE can be derived for each.

\section{Competitive Equilibria of Short-Term Markets} \label{section3}

\brown{In this section, we analytically characterize the CE of each short-term market mechanism and, based on the resulting equilibrium prices and allocations, derive the \rev{present value of the expected total revenue of panel owners over the panel lifespan} as a function of panel capacity \(c\).}

\begin{lemma}[CE of real-time market mechanisms]\label{prop:eq:rt}
	A CE for the product-differentiated real-time market (\texttt{prt}) and the single-product real-time market (\texttt{srt}), is listed in Table~\ref{tab:rt}, where $\overline{F}_{V}(\cdot):=1-F_{V}(\cdot)$ is the complementary cumulative distribution function of $v_i$, and $\overline \pi^\mathrm{b}_i:=\pi_\mathrm{u}+v_{i}$. \purple{Furthermore, the \rev{present value of the expected total revenue for \textcolor{black}{panel} owners over the panel lifespan} under these two market mechanisms \purple{is}}
		\begin{align}
		&\Pi^\mathrm{s}_\mathrm{prt}(c) = \widetilde T \, \mathbb E\left[\left(\pi_{\mathrm{u}}+\overline{F}^{-1}_V\left(\frac{cG}{L}\right)\right)\,cG\,\mathbb{1} \bigl\{cG \leq L \bigl\} \right],\label{prt:rev}\\
		&\Pi^\mathrm{s}_\mathrm{srt}(c) = \widetilde T \, \mathbb E\left[\pi_{\mathrm{u}}\,cG\,\mathbb{1} \bigl\{cG \leq L \bigl\}\right]\label{srt:rev}.
	\end{align}
\end{lemma}

\begin{table}[h]
\centering
\caption{CE outcomes for real-time market mechanisms}\label{tab:rt}
\begin{tabular}{@{}cc|ccc@{}}
\toprule
 &
  Market &
  $\pi$ &
  $q_i^\mathrm{s}$ &
  $q_i^\mathrm{b}$ \\ \midrule
\multicolumn{1}{c|}{\multirow{2}{*}{\begin{tabular}[c]{@{}c@{}}Abundant supply\\ $cG > L$\end{tabular}}} &
  $\texttt{prt}$ &
  0 &
  $L$ &
  $L$ \\ \cmidrule(l){2-5} 
\multicolumn{1}{c|}{} &
  $\texttt{srt}$ &
  0 &
  $L$ &
  $L$ \\ \midrule
\multicolumn{1}{c|}{\multirow{2}{*}{\begin{tabular}[c]{@{}c@{}}Limited supply\\ $cG \leq L$\end{tabular}}} &
  $\texttt{prt}$ &
   $\pi_\mathrm{u}+\overline F_V ^{-1}\left(\frac{cG}{L}\right)$ &
  $cG$ &
  $L\mathbb 1\{\overline \pi^\mathrm{b}_i\ge \pi\}$ \\ \cmidrule(l){2-5}
\multicolumn{1}{c|}{} &
  $\texttt{srt}$ &
  $\pi_\mathrm{u}$ &
  $cG$ &
  $cG$ \\ \bottomrule
\end{tabular}
\end{table}

\mehdi{The CE characterization in Lemma~\ref{prop:eq:rt} shows that, when supply is abundant, i.e., the total realized solar generation exceeds demand, the resulting prices and allocations are identical under both types of real-time markets. This result follows from the fact that solar supply fully covers demand, rendering utility support unnecessary. As a consequence, the zero marginal cost of solar generation results in a zero equilibrium price.}
 When the solar supply is insufficient to cover the load, the market clearing outcomes of the two market types become different. Since the utility supply is required in this case, the market price for the single-product real-time market will be $\pi_\mathrm{u}$. When solar is traded as a separate product, however, we see that the scarce solar production $cG$ is allocated among the buyers with higher \mehdi{solar premiums} first. The solar market price turns out to have a premium over the \purple{utility-supplied} energy,  driven by the competition among the buyers. \textcolor{black}{Accordingly, substituting $\pi$ and $q_i^{\mathrm{s}}$ from Table~\ref{tab:rt} into~\eqref{seller:payoff:real} yields individual seller payoffs; aggregating via~\eqref{prt:total:rev} then gives~\eqref{prt:rev}. The expression in~\eqref{srt:rev} follows by a similar calculation under the single-product market model.}

\mehdi{We now proceed to characterize the CE for the contract-based market. For each seller \( i \in \mathcal{I}_\mathrm{s} \), the utility-maximizing choice within the feasible set for \( q_i^\mathrm{s} \) is \( q_i^\mathrm{s} = c \). On the buyer side, for a given market price \( \pi \), let \( d_i^\star(\pi) \) denote the capacity that maximizes \( \Pi_i^\mathrm{b}(q_i^\mathrm{b}, \pi) \), as defined in~\eqref{eq:payoff:cb}.
}

\revise{To analytically characterize $d_i^\star(\pi)$, we first define
	\begin{equation}\label{eq:funcg}
		\mu_{G_\mathrm{tr}}(d_i)
		:=
		\mathbb E\!\left[
		G\,\mathbb{1}\{d_iG\le L\}
		\right].
	\end{equation}
	The function $\mu_{G_\mathrm{tr}}(\cdot)$ is non-increasing in $d_i$, because increasing the contracted capacity shrinks the event $\{d_iG\le L\}$. It represents the expected per-unit generation that remains below the buyer's load limit.
}
\revise{
	To define capacity demand without imposing global strict invertibility of $\mu_{G_\mathrm{tr}}(\cdot)$, we use the generalized inverse
	\begin{equation}\label{eq:funcg_inverse}
		\mu_{G_\mathrm{tr}}^{-1}(z)
		:=
		\sup\{d_i\ge0 \mid \mu_{G_\mathrm{tr}}(d_i)=z\}.
	\end{equation}
	This definition is useful when $\mu_{G_\mathrm{tr}}(\cdot)$ has flat portions, as may occur under bounded support of $G$. It selects the largest capacity level corresponding to a given value of $z$ and avoids imposing a direct invertibility assumption on $\mu_{G_\mathrm{tr}}(\cdot)$. Thus, when multiple capacity levels correspond to the same value of $z$, the generalized inverse serves as a selection rule and does not require capacity demand to be unique at every price.
}

\revise{
	We then define the extended generalized inverse as
	\begin{equation}\label{eq:funcg_inverse_tilde}
		\tilde{\mu}_{G_\mathrm{tr}}^{-1}(z)
		:=
		\begin{cases}
			\mu_{G_\mathrm{tr}}^{-1}(z), & \text{if } 0\le z\le \mathbb E[G],\\
			0, & \text{if } z>\mathbb E[G].
		\end{cases}
	\end{equation}
	When $z>\mathbb E[G]$, the required value is above the maximum attainable value of $\mu_{G_\mathrm{tr}}(\cdot)$, so the associated capacity demand is set to zero.
}

We can now characterize $d_i^\star(\pi)$ in the following lemma.
\purple{\begin{lemma}[Individual and aggregate demand functions under contract-based market]\label{demand:cb}
The individual demand function of a buyer $i \in \mathcal{I}_{\mathrm{b}}$ is given by:
\begin{equation}\label{eq:prop1:dinverse}
	d_i^\star(\pi) := \tilde {\mu}_{G_\mathrm{tr}}^{-1}\left(\frac{\pi}{\pi_{\mathrm{u}} + v_i}\right).
\end{equation}
Accordingly, the aggregate market demand function is 
\begin{equation} \label{d:star:eq}
	\widehat d^\star(\pi) := \int d_{i}^\star(\pi)\,\, \mathrm{d}i=\int \tilde {\mu}_{G_\mathrm{tr}}^{-1}\left(\frac{\pi}{\pi_{\mathrm{u}}+v_{i}}\right)\,\, \mathrm{d}F_V(v_i).
\end{equation}
\end{lemma}}

This \purple{lemma} allows us to obtain the CE for the contract-based market participation game.

\begin{lemma}[CE of contract-based market] \label{lemma:ne:cb}
The following is a CE for the contract-based market:
\begin{equation}
	q_i^\mathrm{s} = c, \quad   i \in \mathcal I_\mathrm{s};  \quad q_i^\mathrm{b}=d_{i}^\star(\pi),	 \quad i \in \mathcal I_\mathrm{b}, \nonumber
\end{equation}
and $\pi$ is the solution to 
$
c = \widehat d^\star(\pi).$ \mehdi{Moreover, the \rev{present value of total revenue for \textcolor{black}{panel} owners over the panel lifespan} under the contract-based market is 
\begin{equation}\label{cb:rev}
\Pi^\mathrm{s}_\mathrm{cb}(c) = \widetilde T c \pi.
\end{equation}} 
\end{lemma}

\revise{\begin{remark}[Price-responsive demand]\label{rem:price_responsive_demand}
		The baseline short-term market analysis assumes an inelastic aggregate load $L$. This fixed-load formulation serves as a benchmark for isolating the effect of market design on PV investment incentives. This benchmark is consistent with the fact that short-run electricity demand is often relatively price-inelastic~\cite{burke2018price,labandeira2017meta}. Nevertheless, we acknowledge that demand-response programs can create price responsiveness in some settings~\cite{albadi2008summary}.
		In the single-product real-time market, price-responsive demand can be incorporated by replacing the fixed load with a continuous and strictly decreasing aggregate demand function $D(\pi)$. The \rev{present value of the expected total seller revenue over the panel lifespan} then becomes
		\begin{align}
			\Pi_{\mathrm{srt}}^{\mathrm{s,el}}(c)
			=
			\widetilde T\,
			\mathbb E
			\bigg[
			&
			\pi_{\mathrm u}cG\,
			\mathbb{1}\{cG<D(\pi_{\mathrm u})\}
			\nonumber\\
			&
			+
			D^{-1}(cG)cG\,
			\mathbb{1}\{D(\pi_{\mathrm u})\le cG\le D(0)\}
			\bigg].
		\end{align}
		Here, utility supply remains marginal when $cG<D(\pi_{\mathrm u})$, while the market clears at the inverse-demand price $D^{-1}(cG)$ when $D(\pi_{\mathrm u})\le cG\le D(0)$; for more abundant PV generation, the market price is zero.
		Extending price-responsive demand to the product-differentiated and contract-based markets requires additional modeling. In the former, consumption may respond differently to solar and utility-supplied electricity, while in the latter, capacity is contracted ex ante but electricity consumption occurs ex post after solar generation is realized. A full treatment would therefore require additional assumptions on consumer preferences and demand response. We retain the fixed-load formulation as our benchmark and leave these extensions for future work.
	\end{remark}
}

\section{Equilibrium Panel Capacities}\label{section4}
\textcolor{black}{Given the short-term market outcomes derived in the previous section,
we next apply the investment model to characterize the resulting equilibrium panel investment capacities.}

\begin{lemma}[NE panel capacities]\label{lemma:nec}
\purple{An} NE aggregate panel capacity $c_m^\mathrm{ne}$ for each market $m \in \{\mathrm{prt}, \mathrm{srt},\mathrm{cb}\}$ is \purple{a} solution to the following equation: 
\begin{equation}\label{eq:nec:gen}
\Pi_m^\mathrm{s}(c) = \pi_0  c,
\end{equation}
which can be specialized to individual markets as follows:
\begin{enumerate}
 \item For \purple{the} product-differentiated real-time market, $c_\mathrm{prt}^\mathrm{ne}$ is the solution \purple{to}
    \begin{equation}\label{eq:eqc:prt}
    \widetilde T \, \mathbb E \left [ \left(\pi_{\mathrm{u}}+\overline{F}_V^{-1}\left(\frac{cG}{L}\right)\right)\,G\,\mathbb{1} \bigl\{cG \leq L \bigl\}\right]=\pi_0  .
    \end{equation}

    \item For \purple{the} single-product real-time market,  $c_\mathrm{srt}^\mathrm{ne}$ is the solution \purple{to}
    \begin{equation}\label{eq:eqc:srt}
    \widetilde T \, \mathbb E \left [\pi_{\mathrm{u}}\,G\,\mathbb{1} \bigl\{cG \leq L \bigl\}\right ]=\pi_0.
    \end{equation}
       \item For \purple{the} contract-based market, 
    \begin{equation}\label{eq:eqc:cb}
    c_\mathrm{cb}^\mathrm{ne} = \widehat d^\star\left(\frac{\pi_0}{\widetilde T}\right).
    \end{equation}
\end{enumerate}
\end{lemma}

\begin{figure}[t]
	\includegraphics[width=0.48\textwidth]{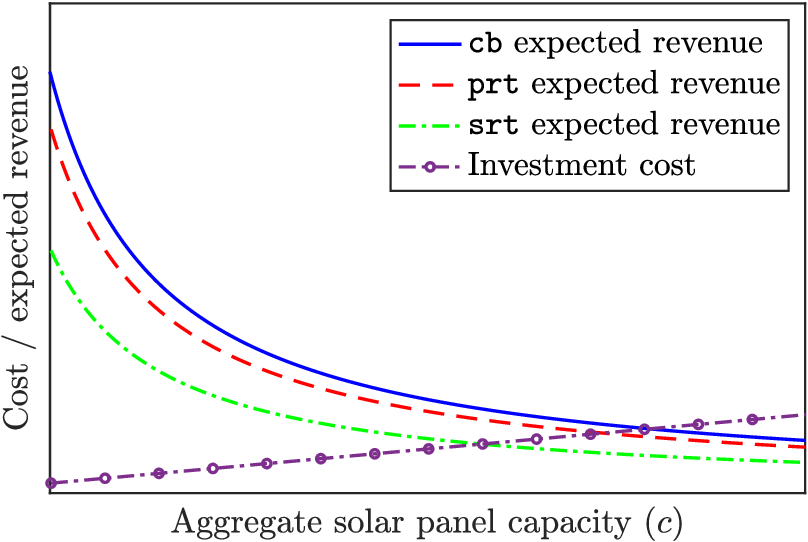}	
	\caption{Illustration of the equilibrium condition for the panel investment problem. The relative position and shape of the curves depend on the problem parameters. }\label{fig:eq}
\end{figure}

The underlying idea behind \eqref{eq:nec:gen} is straightforward: regardless of the specific market mechanism, the aggregate capacity reaches equilibrium when the last infinitesimal investor who decides to invest earns zero net profit, meaning $\Pi^\mathrm{inv}_{m,i}(x_i, c)$ defined in~\eqref{eq:inv:po} is zero for that investor. In other words, the equilibrium capacity can be found at the point where the investment cost curve intersects the expected revenue curve, both expressed as functions of capacity \( c \). This relationship is illustrated in Fig.~\ref{fig:eq}.

\revise{The existence and uniqueness of the positive solutions to~\eqref{eq:eqc:prt} and~\eqref{eq:eqc:srt} follow from the properties of the corresponding expected per-unit revenue functions. For the single-product real-time market, the left-hand side of~\eqref{eq:eqc:srt} is continuous in $c$. Under the bounded-support assumption $G\in[0,\overline g]$, if $c\le L/\overline g$, then $cG\le L$ for all realizations of $G$, so available solar generation never exceeds the load. In this region, all solar generation is sold, and the left-hand side of~\eqref{eq:eqc:srt} is constant and strictly above $\pi_0$ by condition~\eqref{eq:pi0cond}. Hence, the strict inequality in~\eqref{eq:pi0cond} rules out an equilibrium in this flat region. The solution to~\eqref{eq:eqc:srt} therefore lies in the region $c>L/\overline g$, where the left-hand side is strictly decreasing in $c$ and approaches zero as $c$ becomes large. Therefore,~\eqref{eq:eqc:srt} admits a unique positive solution. This uniqueness corresponds to the single intersection between the expected revenue curve and the investment cost curve illustrated in Fig.~\ref{fig:eq}. The same argument applies to the product-differentiated real-time market, since its expected per-unit revenue is continuous and strictly decreasing at the relevant solution, with the solar-premium term increasing expected revenue relative to the single-product market.
For the contract-based market, the aggregate capacity $c_\mathrm{cb}^\mathrm{ne}$ is also uniquely determined by \eqref{eq:eqc:cb}. At the equilibrium contract price $\pi_0/\widetilde T$, the argument of $\tilde{\mu}_{G_\mathrm{tr}}^{-1}(\cdot)$ is strictly below the maximum value $\mathbb E[G]$. Therefore, the generalized inverse is evaluated in the strictly decreasing region of $\mu_{G_\mathrm{tr}}(\cdot)$, where it coincides with the ordinary inverse. Hence, the individual capacity demands, and consequently the aggregate capacity $c_\mathrm{cb}^\mathrm{ne}$, are uniquely determined.}

With this characterization in place, Lemma~\ref{lemma:nec} enables us to compare the equilibrium capacities across market mechanisms via numerical simulations. However, deeper insight comes from a theoretical comparison of the resulting aggregate capacities. To enable a more meaningful comparison, we first define a benchmark capacity in
the next subsection.

\revise{
	\begin{remark}[Regulated compensation]
		Feed-in tariffs and other administratively determined export rates can be viewed as a special case of our framework in which the selling price is fixed by regulation rather than determined by interactions among sellers and buyers in a short-term market. For instance, suppose that each unit of PV generation is compensated at a fixed rate $\pi_{\mathrm{reg}}$. Then the \rev{present value of the expected total revenue of panel owners over the panel lifespan} is
		\begin{equation}
			\Pi^\mathrm{s}_{\mathrm{reg}}(c)
			=
			\widetilde T \mathbb{E}\left[\pi_{\mathrm{reg}} cG\right]
			=
			\widetilde T \pi_{\mathrm{reg}} c \mathbb{E}[G].
		\end{equation}
		Substituting this expression into~\eqref{eq:nec:gen} yields the zero-profit condition
		\begin{equation}\label{eq:regulation}
			\widetilde T \pi_{\mathrm{reg}}\mathbb{E}[G]=\pi_0.
		\end{equation}
		Unlike the market-based mechanisms studied in this paper, this condition does not determine a unique interior aggregate capacity because the per-unit expected revenue is independent of $c$. If $\widetilde T \pi_{\mathrm{reg}}\mathbb{E}[G]>\pi_0$, all potential investors have an incentive to invest; if $\widetilde T \pi_{\mathrm{reg}}\mathbb{E}[G]<\pi_0$, no investor has an incentive to invest; and if~\eqref{eq:regulation} holds, investors are indifferent. Thus, fixed regulated compensation does not generate the capacity-dependent feedback mechanism analyzed in this paper.
	\end{remark}
}

\subsection{Benchmark: Socially Optimal Panel Investment}
In addition to the market-driven investment outcomes, we also consider the \emph{socially optimal panel investment} capacity as a benchmark, defined as the aggregate capacity that maximizes social welfare, i.e., the total payoff of consumers and investors.

To this end, we first need to define how to optimally allocate solar energy, when it is scarce, among the collection of consumers \blue{in an operation period}. Given the realization of solar generation $G$, we can characterize \purple{an} optimal allocation as \purple{a} solution to the following optimization problem:

\begin{subequations}\label{opt:swo}
	\begin{align}
		\!\!\max_{\sigma:[0,\overline v]\mapsto \{0,1\}}  \quad& \int_{0}^{\overline{v}} v_i \sigma(v_i)\,\, \mathrm{d}F_V(v_i) \label{opt:swo:obj}\\ 
		\mbox{s.t.}\quad\quad & \int_{0}^{\overline{v}} \sigma(v_i)\,\, \mathrm{d} F_V(v_i) \le\min\{cG/L, 1\},\label{opt:swo:cond}
	\end{align}
\end{subequations}
\purple{where, for each $v_i$, the binary variable $\sigma(v_i)$ indicates whether the consumer's load is fully served by solar ($\sigma(v_i)=1$) or not ($\sigma(v_i)=0$).}
\textcolor{black}{The optimal value} $v^\star(c,G)$ \textcolor{black}{then} characterizes the \emph{maximum average \mehdi{solar premium}} that can be achieved given the solar capacity $c$ and the realized $G$.

The socially optimal panel capacity is then given by a solution to the following optimization problem:
\begin{align}\label{social:cap}
	\max_{c\in \reals_+} \enspace & \widetilde T\,\mathbb E \left[v^\star(c,G) L-\pi_\mathrm{u} (L-cG)_+\right]- \pi_0 c,
\end{align}
where the objective in \eqref{social:cap} represents the \rev{present value of social welfare over the panel lifespan}, i.e., the sum of consumers' and investors' payoffs, with internal monetary transfers between them canceling out.
\purple{The following result provides an analytical characterization of this capacity.}

\begin{proposition}[Socially optimal capacity]\label{prop:optc}
	The socially optimal panel capacity, $c_{\mathrm{opt}}$, is the solution \purple{to}
	\begin{equation}\label{eq:opt}
		\widetilde T \, \mathbb E \left [ \left(\pi_{\mathrm{u}}+\overline{F}_V^{-1}\left(\frac{cG}{L}\right)\right)\,G\,\mathbb{1} \bigl\{cG \leq L \bigl\}\right]=\pi_0.
	\end{equation}
\end{proposition}
\rev{
	\begin{remark}[Scope of the socially optimal capacity]
		The socially optimal capacity characterized above is defined with respect to the welfare components considered in our model, consistent with our focus on how short-term market mechanisms shape long-term PV investment incentives. In practice, additional physical and operational considerations may affect the socially desirable level of PV investment, including distribution-network capacity constraints, voltage constraints, and network-induced solar curtailment. These considerations are outside the scope of the present model and, if incorporated, could alter the resulting socially optimal capacity. 
	\end{remark}
}
\subsection{Analytical Comparison of Equilibrium Capacities}\label{epsilon:method}
\revise{As~\eqref{eq:eqc:prt},~\eqref{eq:eqc:srt}, and~\eqref{eq:opt} have similar structures, comparison among $c^\mathrm{ne}_\mathrm{srt}$, $c^\mathrm{ne}_\mathrm{prt}$, and $c_{\mathrm{opt}}$ is not difficult. However, comparing $c^\mathrm{ne}_\mathrm{cb}$ with other equilibrium capacities turns out to be challenging in general. For this purpose, we re-parametrize the solar premiums of consumers and introduce a sequence of scaled versions of the problem as follows.}

\revise{For each $i\in \mathcal I_\mathrm{b}$, fix $\tilde v_i$ and let $v_i=\epsilon\tilde v_i$ for $\epsilon\ge 0$. This type of scaling is a standard device for comparative-static analysis~\cite{milgrom1994monotone}: rather than changing each consumer's premium independently, it allows us to study a particular region of the parameter space by shifting the overall level of solar premiums while preserving the relative heterogeneity across consumers. In this parametrization, the baseline values $\tilde v_i$ capture the relative differences in consumers' environmental preferences and willingness to pay for solar energy. Such a baseline distribution can be interpreted as being obtained from empirical information, such as survey data, stated-preference studies, or observed participation in renewable-energy programs. The scalar $\epsilon$ then controls the overall strength of these preferences in the population.}

\revise{Economically, a small value of $\epsilon$ corresponds to a market setting in which consumers have relatively weak environmental preferences or limited willingness to pay a premium for solar energy over utility electricity. This may represent an early-stage retail market, a setting with low consumer awareness of renewable-energy products, limited policy support, or a market in which consumers are primarily price-sensitive. A large value of $\epsilon$ corresponds to a setting in which consumers are more environmentally conscious or have stronger willingness to pay for renewable electricity. Such a setting may arise from stronger environmental preferences, public information campaigns, green-energy programs, or policy incentives that increase consumers' valuation of solar energy.}

\revise{Thus, changing $\epsilon$ shifts the distribution of solar premiums for the population as a whole, while maintaining heterogeneity across buyers through the fixed baseline values $\tilde v_i$. These scaled versions of the problem allow us to examine how the relative equilibrium capacities change as the overall strength of consumers' solar preferences varies. This analysis leads to the main results of this section:}

\begin{theorem}[Comparison of NE capacities]\label{thm:main}
	The following relations hold for the equilibrium aggregate panel investment capacities induced by diverse market mechanisms.
	\begin{enumerate}
		\item \emph{General case}: For any 
		$\epsilon \ge 0$, 
		\begin{equation}
			c^\mathrm{ne}_\mathrm{srt} \le c^\mathrm{ne}_\mathrm{prt} = c_\mathrm{opt},
			\qquad \text{and} \qquad
			c^\mathrm{ne}_\mathrm{srt} \le c^\mathrm{ne}_\mathrm{cb}.
			\vspace{-0.25 em}\label{thm:main:general}
		\end{equation}
		\item \emph{Small \purple{solar premium}}: If $\epsilon$ is in a neighborhood of $0$ and $c^\mathrm{ne}_\mathrm{srt}$ is large such that
		\begin{equation}\label{fg:constant}
			(1-\delta) r_0 \le f_G(g) \le (1+\delta) r_{0},\vspace{-0.15 em}
		\end{equation}
		holds   
		for all $g \le L/c^\mathrm{ne}_\mathrm{srt}$ with some $r_0\in \reals_+$ and $\delta>0$ sufficiently small, then
		\begin{equation}
			c^\mathrm{ne}_\mathrm{srt} \le c^\mathrm{ne}_\mathrm{prt} = c_\mathrm{opt} \lesssim c^\mathrm{ne}_\mathrm{cb},\vspace{-0.25 em}\label{thm:main:case2} 
		\end{equation}
		where the last inequality holds in the sense that
		\begin{equation}
			c^\mathrm{ne}_\mathrm{cb} - c^\mathrm{ne}_\mathrm{prt} \ge \beta \epsilon - O\left(\epsilon^2\right),\vspace{-0.25 em}\label{thm:main:case2:detail} 
		\end{equation}
		with $\beta$ being a positive constant. 
		\item \emph{No \purple{solar premium}}: If $\epsilon =0$, 
		\begin{equation}\label{eq:thm:p3}
			c^\mathrm{ne}_\mathrm{srt} =c^\mathrm{ne}_\mathrm{prt} = c_\mathrm{opt} = c^\mathrm{ne}_\mathrm{cb}. 
		\end{equation}
	\end{enumerate}
\end{theorem}

The first interesting result \brown{from~\eqref{thm:main:general}} in Theorem~\ref{thm:main} is that the single-product real-time market leads to the lowest investment signal among the mechanisms considered here. In particular, it weakly under-invests relative to the socially optimal panel capacity, and it also induces a weakly smaller equilibrium capacity than the contract-based market. In fact, with $\epsilon>0$, it is easy to identify a distribution $F_{\widetilde V}$ such that $c^\mathrm{ne}_\mathrm{srt}< c_\mathrm{opt}$. 
While one could argue that the single-product real-time market is easier to implement, particularly because it avoids product differentiation and may involve simpler market rules and settlement procedures, it leads to a loss in social welfare when viewed through the lens of long-term investment equilibrium.
The second observation\footnote{In this case, a general analytical ordering between $c_{\mathrm{prt}}^{\mathrm{ne}}$ and $c_{\mathrm{cb}}^{\mathrm{ne}}$ is not available. Nevertheless, our numerical results indicate that the contract-based market yields a larger equilibrium capacity than the product-differentiated market in most realistic setups.} \brown{\textcolor{black}{from}~\eqref{thm:main:general}} is that the product-differentiated real-time market is guaranteed to \rev{achieve the benchmark socially optimal panel investment capacity $c_{\mathrm{opt}}$ under the welfare framework defined in~\eqref{social:cap}}. This stems from the fact that the product-differentiated market allocates solar energy in the same way as the socially optimal allocation in~\eqref{opt:swo}.

When electricity users are unwilling to pay much more for solar energy (i.e., when $\epsilon$ is small), we \purple{establish} a qualitative relation between the contract-based market equilibrium capacity, $ c^\mathrm{ne}_\mathrm{cb}$, and the socially optimal capacity, $c_\mathrm{opt}$. With a typical value of $\pi_0$ so that $c^\mathrm{ne}_\mathrm{srt}$ is relatively large,\brown{~\eqref{thm:main:case2}} suggests that the contract-based market consistently leads to over-investment. Although this analytical result is established only for small $\epsilon$, our numerical results suggest that the over-investment effect extends to large values of $\epsilon$ (see Section~\ref{constant}). From a static perspective, this may be viewed as suboptimal for the setting considered in this paper. However, if \purple{we incorporate} potential \blue{\emph{economies of scale}} in panel manufacturing, over-investment may reduce $\pi_0$ in the future and thus benefit society. \mehdi{Therefore, under a dynamic investment model, a policymaker may favor the contract-based market.}

\mehdi{Finally, and perhaps most surprisingly, \brown{~\eqref{eq:thm:p3} indicates that} all market mechanisms yield the same equilibrium panel capacity when users place no additional value on solar energy relative to utility-supplied electricity. This scenario may arise if \purple{utility-supplied} energy is also fully renewable, or if users exhibit no specific preference for solar generation. However, given current levels of renewable penetration, non-zero values of \( v_i \) are expected for a considerable portion of users \cite{farhar1999willingness}. As a result, the under- and over-investment effects that market mechanisms have on long-term solar equilibrium capacities are likely to remain substantial in the near future.}

\subsection{Trade-offs Between Investment Outcomes and Market Implementation}\label{sec:implementation}

Theorem~\ref{thm:main} shows that the product-differentiated real-time market supports the socially optimal investment capacity in the baseline analytical model. However, this result should be interpreted as a benchmark within the framework considered in this paper, since the model abstracts from practical implementation frictions. We therefore briefly discuss some practical factors that may affect this conclusion.

A first practical issue is metering. Importantly, the product-differentiated market does not require physical tracing of electricity from individual PV sellers to buyers, since electricity is physically pooled and product differentiation can be implemented through financial settlement. The relevant physical measurement is seller-side PV generation or grid injection, represented by $cG$ in the model. Such measurement is also required under the single-product and contract-based markets, so common PV installation, interconnection, and metering costs can be included in the investment cost $\pi_0$. Moreover, meter-related costs are relatively small compared with PV investment costs. For example, using a conservative AMI-related cost of \$600 per meter~\cite{doe_ami_summary_2016} and a residential PV benchmark of \$2.95/Wdc for an 8-kWdc system~\cite{nrel_panel}, the meter-related cost is approximately $600/(8000\times2.95)\approx2.5\%$ of the upfront PV system cost. Thus, although the precise metering requirements may differ across market mechanisms, the associated cost differences are likely to be modest relative to the overall PV investment cost, supporting the use of a common $\pi_0$ as a reasonable benchmark approximation.

Beyond metering, there are additional practical implementation considerations that differ across the three market mechanisms. In particular, the mechanisms differ in how the traded product is defined, allocated, and settled. The single-product real-time market has the simplest product definition but still requires real-time clearing and settlement. The product-differentiated market additionally requires financial differentiation of solar electricity, allocation of scarce solar generation according to buyers' solar premiums, and corresponding billing procedures. In contrast, the contract-based market shifts much of the market-clearing complexity to an ex-ante contracting stage. Physical measurement of realized solar generation is still needed to determine the residual demand that must be supplied by the utility; importantly, however, realized generation does not directly affect payments to panel owners or contract settlement, which are determined ex ante based on the contracted capacity.

With these considerations in mind, the finding that the product-differentiated real-time market achieves the socially optimal investment capacity \rev{under the welfare framework considered in this paper} should be interpreted as a benchmark efficiency result, not as a claim that it is the most practical market design in every setting. The single-product and contract-based markets may be simpler in some implementation dimensions, and such practical considerations may influence their attractiveness to policymakers or market operators.

\section{Extensions}\label{extensions}

\revise{In this section, we extend the baseline model in two directions. First, we incorporate heterogeneity across operation periods and analyze how time-varying solar generation, load, and utility prices affect the equilibrium characterization. Second, we consider the role of revenue risk in long-term PV investment decisions and examine how risk-sensitive investor behavior can change the investment equilibrium and the comparison across market mechanisms. 
}
\subsection{Heterogeneous Operation Periods}\label{heterg:time}

So far, we have assumed that the values/distributions of key parameters, such as loads and solar irradiation, are identical across operation periods within the base planning period. \textcolor{black}{In practice, however, these values/distributions may vary over time (e.g., by season, time of day, and day of week)}. In this section, we extend our analysis to incorporate this temporal heterogeneity.

\brown{In our extended model}, each period \purple{$t$} is characterized by a distinct set of parameters: the total consumer load $L_t$, the solar generation per unit panel capacity $G_t$ with associated probability density function $f_{G_t}$, and the utility supply price $\pi_{\mathrm{u}_t}$. 
\revise{Consistent with the bounded-support assumption in the baseline model, we assume that each $G_t$ is supported on $[0,\overline g_t]$, has finite first and second moments, and admits a density $f_{G_t}$ that is continuous on $[0,\overline g_t]$ and positive on $(0,\overline g_t)$.}

In this model, real-time markets are cleared independently for each $t\in\mathcal{T}$, and the expected revenues of \textcolor{black}{panel} investors may therefore vary across the operation periods of the base planning period. In contrast, the contract-based market involves a single ex-ante contract covering the entire base planning period. For consistency across the market mechanisms, we express the corresponding revenues on an average per-operation-period basis and then scale this average by $\widetilde T$ to obtain the \rev{present value of the expected total revenue over the panel lifespan}, $\Pi_m^{\mathrm{s}}(c)$.

We also impose the heterogeneous analogue of~\eqref{eq:pi0cond}:
\begin{equation}\label{eq:pi0cond:tv}
	\blue{c}\pi_0 <
	\sum_{t\in\mathcal{T}} \frac{\widetilde T}{T}
	\pi_{\mathrm{u}_t}\mathbb{E}[\blue{c}G_t].
\end{equation}
This condition means that, after accounting for the heterogeneous operation periods and the present-value scaling over the panel lifespan, the value of generated solar energy at the corresponding utility supply prices exceeds the investment cost.

\mehdi{We start by} characterizing the CE of short-term markets and associated revenue for investors under this new setup.
\subsubsection{Short-Term Market Equilibria}
\purple{The results from Lemma~\ref{prop:eq:rt} and Table~\ref{tab:rt}, with parameters $G$, $\pi_{\mathrm{u}}$, and $L$ indexed by $t$, apply to the real-time markets in each operation period $t \in \mathcal{T}$}. As a result, the \rev{present value of the expected total revenue for investors over the panel lifespan} \purple{is}
\begin{align}
	&\Pi^\mathrm{s}_\mathrm{prt} = \sum_{t\in\mathcal{T}} \frac{\widetilde T}{T} \mathbb E \left [ \left(\pi_{\mathrm{u}_{t}}+\overline{F}_V^{-1}\left(\frac{cG_{t}}{L_{t}}\right)\right)\,c G_{t}\,\mathbb{1} \bigl\{cG_{t} \leq L_{t} \bigl\}\right],\label{eq:eqc:prt:tv}\\
	&\Pi^\mathrm{s}_\mathrm{srt} = \sum_{t\in\mathcal{T}}\frac{\widetilde T}{T} \mathbb E \left [\pi_{\mathrm{u}_{t}}\,c G_{t}\,\mathbb{1} \bigl\{cG_{t} \leq L_{t} \bigl\}\right ].\label{eq:eqc:srt:tv}
\end{align}

\purple{In the contract-based market, 
	$d_i^\star(\pi)$ is a solution to}
\purple{\begin{equation}
		\!\!\max_{d_i \ge 0} \!\!\enspace \sum_{t\in\mathcal{T}}\left[v_i\mathbb E\min\{d_i G_{t}, L_{t} \}\!-\! \pi_{\mathrm{u}_{t}} \mathbb E (L_{t}-d_i G_{t})_+\right]\!-\! \pi d_i.
\end{equation}} 

Similar to~\eqref{eq:funcg}, for each \(t \in \mathcal{T}\), we first define
\begin{equation}\label{mug:time_varying}
	\mu_{G_{t,\mathrm{tr}}}(d_i)
	:=
	\mathbb E \left[
	G_t \mathbb{1}\bigl\{d_iG_t \le L_t\bigr\}
	\right].
\end{equation}
\revise{For the small-solar-premium comparative-static analysis in this heterogeneous-period setting, we assume that each function
\(\mu_{G_{t,\mathrm{tr}}}(\cdot)\) is locally differentiable around the zero-premium equilibrium capacity $c_{\mathrm{srt}}^{\mathrm{ne}}$.}
Using these functions, we can analytically characterize \(d_i^\star(\pi)\). To this end, define
\begin{equation}
	w_i(d_i):=\sum_{t\in\mathcal{T}} \left(\pi_{\mathrm{u}_t}+v_i\right)\mu_{G_{t,\mathrm{tr}}}(d_i).
\end{equation}
The function \(w_i(d_i)\) represents the consumer's expected total valuation for one unit of rented panel capacity over the entire planning period. Since each $\mu_{G_{t,\mathrm{tr}}}(d_i)$ is non-increasing in $d_i$, $w_i(d_i)$ is also non-increasing in $d_i$. Hence, \(d_i^\star(\pi)\) is characterized as a solution to
\begin{equation}
	w_i(d_i)=\pi.
\end{equation}
We then define the inverse
\begin{equation}
	w_i^{-1}(z):=\sup\{d_i\ge 0 \mid w_i(d_i)\ge z\},
\end{equation}
and the following extension:
\begin{equation}
	\widetilde w_i^{-1}(z):=
	\begin{cases}
		w_i^{-1}(z), & \mbox{if } 0 < z \le \sum_{t\in\mathcal{T}} (\pi_{\mathrm{u}_t}+v_i)\mathbb E[G_t],\\
		0, & \mbox{if } z > \sum_{t\in\mathcal{T}} (\pi_{\mathrm{u}_t}+v_i)\mathbb E[G_t].
	\end{cases}
\end{equation}
Thus, we obtain
\begin{equation}
	d_i^\star(\pi)=\widetilde w_i^{-1}(\pi),
\end{equation}
which implies
\begin{equation} \label{d:star:tv}
	\widehat d^\star(\pi)=\int d_i^\star(\pi)\,\,\mathrm{d}i
	=\int \widetilde w_i^{-1}(\pi)\,\,\mathrm{d}i.
\end{equation}

Given this result, the conditions for the CE of the contract-based market given in Lemma~\ref{lemma:ne:cb} apply to this new setup. Thus, the \rev{present value of total investors' revenue over the panel lifespan} is $\Pi^\mathrm{s}_\mathrm{cb} = \frac{\widetilde T}{T} c\pi$.

\subsubsection{Equilibrium Panel Capacities}

\purple{Lemma~\ref{lemma:nec} continues to hold under this setup. Substituting the heterogeneous-period revenue expressions in~\eqref{eq:eqc:prt:tv} and~\eqref{eq:eqc:srt:tv} into the equilibrium condition~\eqref{eq:nec:gen}, the equilibrium capacities for the real-time markets satisfy
	\begin{align}
		&\sum_{t\in\mathcal{T}}
		\frac{\widetilde T}{T}
		\mathbb E\left[
		\left(
		\pi_{\mathrm{u}_t}
		+\overline F_V^{-1}\left(\frac{cG_t}{L_t}\right)
		\right)
		G_t
		\mathbb 1\{cG_t\le L_t\}
		\right]
		=\pi_0,
		\label{eq:eqc:prt:tv:ne}\\
		&\sum_{t\in\mathcal{T}}
		\frac{\widetilde T}{T}
		\mathbb E\left[
		\pi_{\mathrm{u}_t}G_t
		\mathbb 1\{cG_t\le L_t\}
		\right]
		=\pi_0.
		\label{eq:eqc:srt:tv:ne}
	\end{align}
	For the contract-based market, the equilibrium capacity is
	\begin{equation}
		c_{\mathrm{cb}}^{\mathrm{ne}}
		=
		\widehat d^\star\left(
		\frac{\pi_0T}{\widetilde T}
		\right),
	\end{equation}
	where $\widehat d^\star(\cdot)$ is given by~\eqref{d:star:tv}.}

\subsubsection{Socially Optimal Investment}
Different operation periods may have non\brown{-}identical levels of load and expected solar generation per unit of panel capacity. Thus, for each $t \in \mathcal{T}$, the optimal allocation of solar energy among consumers, denoted by $\sigma_t:\reals_+ \mapsto \{0,1\}$, and the corresponding maximum average \mehdi{solar premium}, denoted by $v_t^\star(c,G_t)$, should be obtained separately by solving~\eqref{opt:swo} with the parameters $L$, $G$, and the function $\sigma$ replaced by their period-\(t\) counterparts. Once $v_t^\star(c,G_t)$ is obtained for all operation periods, the socially optimal capacity can be characterized as follows:

\begin{proposition}[Socially optimal capacity with heterogeneous periods]\label{prop:optc:tv}
	The socially optimal capacity, $c_\mathrm{opt}$, is \purple{a solution to}
	\begin{align} \label{social:cap:tv}
		\max_{c\in \reals_+} \enspace & \sum_{t\in\mathcal{T}} \frac{\widetilde T}{T} \mathbb E\!\left[v_t^\star(c,G_t)L_t-\pi_{\mathrm{u}_t}(L_t-cG_t)_+\right]-\pi_0 c.
	\end{align}
	An analytical characterization of $c_{\mathrm{opt}}$ is obtained by solving
	\begin{equation}\label{eq:opt:tv}
		\sum_{t\in\mathcal{T}} \frac{\widetilde T}{T} \mathbb E \left[\left(\pi_{\mathrm{u}_t}+\overline{F}_V^{-1}\left(\frac{cG_t}{L_t}\right)\right)G_t\,\mathbb{1}\bigl\{cG_t \le L_t\bigr\}\right]=\pi_0.
	\end{equation}
\end{proposition}

\subsubsection{Analytical Comparison of Equilibrium Capacities}
Building on these results, we now extend the main findings of Theorem~\ref{thm:main} to this new setting.
\begin{theorem}[Comparison of NE capacities with heterogeneous periods]\label{thm:main:tv}
		The results in~\eqref{thm:main:general} and~\eqref{eq:thm:p3}, corresponding to the \emph{general case} and the \purple{\emph{no solar premium}} case of Theorem~\ref{thm:main}, remain valid in this setting. \textcolor{black}{Moreover, if there exist constants $R>0$ and a sufficiently small $\delta > 0$ such that, for all $p\in[0,1]$, the aggregate density condition
		\begin{equation}\label{fg:constant:tv}
			(1-\delta)R
			\le
			\sum_{t\in\mathcal{T}}
			\left(\frac{L_t}{c^\mathrm{ne}_\mathrm{srt}}\right)^2
			\!\!\! f_{G_t}\!\left(\frac{L_t p}{c^\mathrm{ne}_\mathrm{srt}}\right)
			\!\mathbb{1}\left\{\frac{L_t p}{c^\mathrm{ne}_\mathrm{srt}}\le \overline g_t\right\}
			\le
			(1+\delta)R
		\end{equation}
		holds, then the capacity ordering in~\eqref{thm:main:case2}, corresponding to the \purple{\emph{small solar premium}} case of Theorem~\ref{thm:main}, continues to hold under this setting, with the corresponding first-order bound characterized by a positive constant $\beta_{\mathrm{tv}}$.}
	\end{theorem}

\subsection{Incorporating Risk Sensitivity into the Investment Model}\label{risk:analysis:subsection}

\revise{The investment model developed so far assumes risk-neutral investors, meaning that investors make long-term PV investment decisions by comparing the \rev{present value of expected revenue over the panel lifespan} with the investment cost. However, in practice, investors may also be sensitive to downside revenue risk, especially because PV generation and market-clearing outcomes depend on uncertain operating conditions. In this subsection, to capture risk-sensitive investment behavior at the aggregate level, we allow investors to share a common degree of risk aversion.}

\revise{There are several ways to incorporate risk sensitivity into the investment model, including mean-variance risk adjustment, conditional value-at-risk (CVaR), chance-constrained models, and robust optimization approaches \cite{markowitz2000mean,ruszczynski2009stochastic,ben2009robust}. In this subsection, we adopt a CVaR approach, which has been widely used in energy-sector problems involving investment, planning, and risk management \cite{xuan2021conditional,szolgayova2011dynamic,pan2024multi}. To isolate the effect of risk-sensitive investor behavior from the time-heterogeneity effects discussed in Section~\ref{heterg:time}, we maintain homogeneous operation periods while introducing risk sensitivity.}

\revise{\subsubsection{CVaR Definition}
	
	Before specifying the risk-adjusted investment equilibrium, we first introduce the CVaR measure used to quantify downside losses. A loss random variable represents the monetary loss associated with uncertain outcomes. In our setting below, this loss is induced by uncertainty in realized PV generation and the resulting market revenue.}

\revise{Let $X$ be a nonnegative integrable loss random variable, and let $\mathbb{P}$ denote the underlying probability measure. For $\alpha\in(0,1)$, the $\alpha$-quantile of $X$ is defined as follows~\cite{pavlikov2018cvar}:
	\begin{equation}\label{eq:quantile_xi}
		\xi_{\alpha}(X)
		=
		\inf\left\{
		\ell\in\mathbb{R}:
		\mathbb{P}(X>\ell)\le 1-\alpha
		\right\}.
	\end{equation}
}

\revise{For general loss distributions, including continuous, discrete, and mixed distributions, CVaR can be defined using the standard optimization representation \cite{rockafellar2000optimization}:
	\begin{equation}\label{eq:cvar_general}
		\mathrm{CVaR}_{\alpha}(X)
		:=
		\min_{\zeta\in\mathbb{R}}
		\left\{
		\zeta
		+
		\frac{1}{1-\alpha}
		\mathbb{E}
		\left[
		(X-\zeta)_+
		\right]
		\right\}.
	\end{equation}
 Intuitively, $\mathrm{CVaR}_{\alpha}(X)$ measures the average loss in the upper $(1-\alpha)$-tail of the loss distribution. For example, if $\alpha=0.95$, then CVaR captures the average loss among the worst $5\%$ of loss realizations.
}

\revise{
	\begin{proposition}[Continuous-distribution CVaR representation]\label{prop:cvar_continuous}
		If $X$ has a continuous distribution, then the CVaR definition in~\eqref{eq:cvar_general} reduces to the conditional-expectation form
		\begin{equation}
			\mathrm{CVaR}_{\alpha}(X)
			=
			\mathbb{E}
			\left[
			X
			\mid
			X\ge \xi_{\alpha}(X)
			\right].
		\end{equation}
	\end{proposition}
}

\subsubsection{Impact of Risk-Averse Investors on Equilibrium Capacity}

\revise{The short-term market-clearing outcomes remain unchanged, since risk sensitivity affects investors' long-term investment decisions rather than the operation of the short-term markets. Therefore, the expected-revenue characterizations in~\eqref{prt:rev},~\eqref{srt:rev}, and~\eqref{cb:rev} remain valid. However, the valuation used in the investment stage is modified to account for downside revenue risk.}

\revise{Consistent with the representative-operation-period structure of the baseline model, let $R_m^{\mathrm{s}}(c,g)$ denote the realized aggregate seller revenue in the representative operation period under market mechanism $m\in\{\mathrm{prt},\mathrm{srt},\mathrm{cb}\}$, when the realized solar generation per unit panel capacity is $G=g$. Specifically,
	\begin{equation}\label{eq:realized_aggregate_seller_revenue}
		R_m^{\mathrm{s}}(c,g)
		:=
		\int_{\mathcal I_{\mathrm{s}}}
		\Pi_i^{\mathrm{s}}
		\left(
		q_i^{\mathrm{s}},
		\pi
		\right)
		\,\mathrm{d}i,
	\end{equation}
	where $q_i^{\mathrm{s}}$ and $\pi$ are the equilibrium seller allocation and market price, respectively, as defined for each market mechanism in Section~\ref{sec:mechanisms}, evaluated at aggregate capacity $c$ and solar realization $G=g$. Thus, $R_m^{\mathrm{s}}(c,G)$ is the corresponding random representative-period aggregate seller revenue, and the \rev{present value of the expected total seller revenue over the panel lifespan} can be written as
	\begin{equation}
		\Pi_m^{\mathrm{s}}(c)
		=
		\widetilde T
		\mathbb{E}
		\left[
		R_m^{\mathrm{s}}(c,G)
		\right].
	\end{equation}}

\revise{To capture downside risk, we define the representative-period shortfall loss relative to the per-period equivalent investment cost as
	\begin{equation}\label{eq:shortfall_loss}
		X_m(c,G)
		:=
		\left(
		\frac{\pi_0 c}{\widetilde T}
		-
		R_m^{\mathrm{s}}(c,G)
		\right)_+.
	\end{equation}
	 Hence, $X_m(c,G)$ is positive only when the realized representative-period seller revenue falls below the \emph{per-period} equivalent investment cost.
}

\revise{Using the CVaR measure introduced above, we define the \rev{risk-adjusted present value of aggregate seller revenue over the panel lifespan} under market mechanism $m\in\{\mathrm{prt},\mathrm{srt},\mathrm{cb}\}$ as
	\begin{equation}\label{eq:risk_adjusted_revenue}
		\Pi_{m,\eta}^{\mathrm{s}}(c)
		:=
		\widetilde T
		\left[
		\mathbb{E}
		\left[
		R_m^{\mathrm{s}}(c,G)
		\right]
		-
		\eta
		\mathrm{CVaR}_{\alpha}
		\left(
		X_m(c,G)
		\right)
		\right],
	\end{equation}
	where $\eta\ge 0$ is a common risk-sensitivity parameter, $\alpha\in(0,1)$ is the CVaR confidence level, and $\mathrm{CVaR}_{\alpha}$ follows the definition in~\eqref{eq:cvar_general}. When $\eta=0$, the valuation in~\eqref{eq:risk_adjusted_revenue} reduces to the risk-neutral expected-revenue formulation in the baseline model.
}

\revise{
	The corresponding risk-adjusted investment equilibrium is obtained by replacing $\Pi_m^{\mathrm{s}}(c)$ in~\eqref{eq:nec:gen} with $\Pi_{m,\eta}^{\mathrm{s}}(c)$, as follows:
	\begin{equation}\label{eq:risk_adjusted_equilibrium}
		\Pi_{m,\eta}^{\mathrm{s}}(c)
		=
		\pi_0 c.
	\end{equation}
	For each market mechanism $m\in\{\mathrm{prt},\mathrm{srt},\mathrm{cb}\}$ and risk-sensitivity parameter $\eta\ge 0$, let $c_{m,\eta}^{\mathrm{ne}}$ denote the solution of~\eqref{eq:risk_adjusted_equilibrium}. In particular, $c_{m,0}^{\mathrm{ne}}$ corresponds to the risk-neutral equilibrium capacity. To simplify notation, we write $c_m^{\mathrm{ne}}:=c_{m,0}^{\mathrm{ne}}$, consistent with the equilibrium capacities characterized in Lemma~\ref{lemma:nec}. The following lemma summarizes the effect of risk sensitivity on the within-market equilibrium capacities.
}

\revise{
	\begin{lemma}[Effect of risk aversion on within-market equilibrium capacity]\label{lem:risk_aversion_capacity}
		For any $\eta\ge 0$, the risk-adjusted equilibrium capacities satisfy
		\begin{equation}
			c_{\mathrm{srt},\eta}^{\mathrm{ne}}
			\le
			c_{\mathrm{srt}}^{\mathrm{ne}},
		\end{equation}
		\begin{equation}
			c_{\mathrm{prt},\eta}^{\mathrm{ne}}
			\le
			c_{\mathrm{prt}}^{\mathrm{ne}},
		\end{equation}
		and
		\begin{equation}
			c_{\mathrm{cb},\eta}^{\mathrm{ne}}
			=
			c_{\mathrm{cb}}^{\mathrm{ne}}.
		\end{equation}
	\end{lemma}
}

\revise{
	Lemma~\ref{lem:risk_aversion_capacity} shows that CVaR-based risk aversion affects the real-time markets and the contract-based market differently. For the single-product and product-differentiated real-time markets, adding the shortfall-risk penalty can weakly reduce the equilibrium investment capacity. This is because real-time market revenues depend on realized solar generation and are therefore exposed to revenue shortfall risk. In contrast, \rev{under the contract-based market considered in this paper,} the equilibrium capacity is unchanged, since sellers receive an ex-ante capacity payment and their revenue is not directly tied to the realization of $G$.
}

\rev{
	\begin{remark}[Scope of the contract-based risk result]
			We note that the invariance of the contract-based equilibrium capacity to the CVaR penalty is based on the capacity-based contract considered in this paper, under which seller payments are determined ex ante and do not depend on realized PV output. Other contract designs may link seller payments to realized generation, for example through minimum-generation guarantees, shortfall penalties, or performance-based adjustments. Under such contracts, seller revenue would generally depend on the realization of $G$, and the CVaR penalty could therefore affect the contract-based investment equilibrium. Hence, the invariance result above need not hold for these alternative contract structures.
	\end{remark}
}

\revise{
	Equipped with Lemma~\ref{lem:risk_aversion_capacity}, we next compare the risk-adjusted equilibrium capacities across market mechanisms and with the socially optimal capacity $c_{\mathrm{opt}}$. In this extension, $c_{\mathrm{opt}}$ remains unchanged from the baseline model because risk sensitivity affects only investors' private valuation of uncertain revenues through the CVaR penalty, while the social planner's objective, the underlying market surplus, and the investment cost remain unchanged. To facilitate this comparison, we follow the same approach as in Section~\ref{epsilon:method} to construct problem instances by scaling users' valuations through $v_i=\epsilon\tilde v_i$ for all $i\in\mathcal I_{\mathrm b}$. The main results are summarized in the following theorem.
}

\revise{
	\begin{theorem}[Risk-sensitive equilibrium capacity comparison]\label{thm:risk_sensitive_comparison}
		For any $\eta\ge 0$, the following relations hold for the risk-adjusted equilibrium aggregate panel capacities induced by the considered market mechanisms.
		
		\begin{enumerate}
			\item \emph{General case}: For any $\epsilon\ge 0$,
			\begin{equation}
				c_{\mathrm{srt},\eta}^{\mathrm{ne}}
				\le
				c_{\mathrm{prt},\eta}^{\mathrm{ne}}
				\le
				c_{\mathrm{opt}},
				\qquad \text{and} \qquad
				c_{\mathrm{srt},\eta}^{\mathrm{ne}}
				\le
				c_{\mathrm{cb},\eta}^{\mathrm{ne}}.
			\end{equation}
			
	\item \emph{Small solar premium}: If $\epsilon>0$ is sufficiently small and the assumption of the small-solar-premium case of Theorem~\ref{thm:main}, i.e.,~\eqref{fg:constant}, holds, then
			\begin{equation}
				c_{\mathrm{srt},\eta}^{\mathrm{ne}}
				\le
				c_{\mathrm{prt},\eta}^{\mathrm{ne}}
				\le
				c_{\mathrm{opt}}
				\lesssim
				c_{\mathrm{cb},\eta}^{\mathrm{ne}}.
			\end{equation}
			Moreover,
			\begin{equation}
				c_{\mathrm{cb},\eta}^{\mathrm{ne}}
				-
				c_{\mathrm{prt},\eta}^{\mathrm{ne}}
				\ge
				\beta\epsilon
				-
				O(\epsilon^2).
			\end{equation}
			
			\item \emph{No solar premium}: If $\epsilon=0$, then
			\begin{equation}
				c_{\mathrm{srt},\eta}^{\mathrm{ne}}
				=
				c_{\mathrm{prt},\eta}^{\mathrm{ne}}
				\le
				c_{\mathrm{opt}}
				=
				c_{\mathrm{cb},\eta}^{\mathrm{ne}}.
			\end{equation}
		\end{enumerate}
	\end{theorem}
}

\revise{
	Theorem~\ref{thm:risk_sensitive_comparison} shows that the main qualitative comparison in Theorem~\ref{thm:main} is preserved under risk aversion. The single-product real-time market still induces no more investment than the product-differentiated real-time market, and it also induces no more investment than the contract-based market. However, unlike the risk-neutral case, the product-differentiated real-time market is no longer guaranteed to attain the socially optimal capacity; instead, its risk-adjusted equilibrium capacity is weakly below $c_{\mathrm{opt}}$. This reflects the fact that risk-averse investors place a lower value on risky real-time revenues because of downside revenue risk.
}

\revise{
	The contract-based market behaves differently. Since its equilibrium capacity is unchanged by the CVaR penalty, it continues to induce weakly higher investment than the single-product real-time market for any $\epsilon\ge0$, and the asymptotic separation between the contract-based market and the product-differentiated real-time market is preserved in the small-solar-premium regime. In particular, the gap between $c_{\mathrm{cb},\eta}^{\mathrm{ne}}$ and $c_{\mathrm{prt},\eta}^{\mathrm{ne}}$ is at least as large as the corresponding risk-neutral first-order gap. Thus, risk aversion strengthens the relative tendency of real-time markets to support lower investment compared with the contract-based market, while leaving the contract-based investment signal unchanged.
}

\section{Numerical Experiments}\label{Numerical}

\rev{In this section, we numerically evaluate the theoretical findings and examine their robustness using calibrated short-term electricity market parameters and benchmark investment and financial assumptions.}
\subsection{Case Study Setup}\label{data}

Motivated by California's leading solar adoption, we model a large population of potential investors in San Francisco, California, deciding at the beginning of 2025 whether to install PV. To simplify the analysis of the short-term markets, we treat San Francisco as a single pooled distribution-level market and consider the three representative market frameworks introduced earlier.

We consider a 25-year panel lifespan with hourly operation periods. \rev{To implement the present-value adjustment described in Section~\ref{lifecycle:financial}, we use a nominal annual discount rate of $\rho=5.64\%$ and an annual inflation rate of $2.5\%$, consistent with the financial assumptions used in a recent NREL techno-economic analysis~\cite{bb462edd3ab04760b9d614d0ec967e43}. Accordingly, we set $\kappa_y=(1.025)^{y-1}$ for $y\in\{1,\ldots,25\}$.}

Although our analysis allows for arbitrarily many heterogeneous periods, we group all hours into 16 representative periods defined by season and time of day. Specifically, we let $\mathcal{T}=\{1,\ldots,16\}$, where each $t\in\mathcal{T}$ corresponds to one season--time-of-day pair in 
$
\{\text{Winter},\text{Spring},\text{Summer},\text{Fall}\}
\times
\{\text{Morning},\text{Midday},\text{Evening},\text{Night}\}.
$
The periods are ordered first by season and then by time of day: Winter Morning, Winter Midday, Winter Evening, Winter Night, followed by the corresponding Spring, Summer, and Fall periods. Morning corresponds to 6:00--10:59, midday to 11:00--15:59, evening to 16:00--20:59, and night to 21:00--5:59. Each period is weighted in the expected-revenue calculations by the relative frequency of the hours it represents.

\begin{figure}[t]
	\centering
	\includegraphics[width=0.5\textwidth]{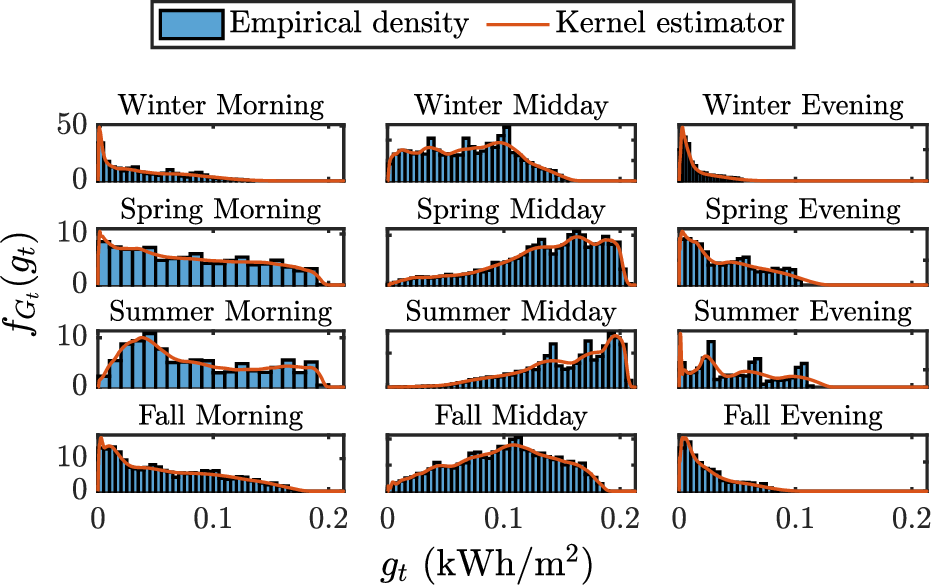}	
	\caption{Empirical and kernel density estimates of effective solar output for 12 non-night periods.}
	\label{fg:merged}
\end{figure}

We use hourly global horizontal irradiance data from 2000 to 2022 for San Francisco, California, to model the distribution of $G_t$ for all $t\in\mathcal T$~\cite{nrel_data}. To account for panel efficiency, we multiply the hourly global horizontal irradiance data by a factor of $0.2$~\cite{955fd4dbf3e34e6e92a6dcaf57db598f} and, for each one-hour observation, convert the resulting values to effective solar energy output in $\mathrm{kWh}/\mathrm{m}^2$. Thus, when $c$ denotes aggregate installed panel area in $\mathrm{m}^2$, $cG_t$ is measured in kWh. For morning, midday, and evening periods, we use a kernel density estimator to fit the distribution of effective solar output. Since night periods naturally have no solar irradiation, we model them separately by setting $G_t=0$ with probability one. The empirical densities and fitted kernel estimators for all 12 non-night season--time-of-day periods are shown in Fig.~\ref{fg:merged}.

We use survey data from \cite{farhar1999willingness} on consumers' willingness to pay a premium for renewable electricity and convert these values into the per-kWh solar premium $v_i$ used in our model. The original survey values are reported as additional monthly charges on consumers' electricity bills, measured in 1997 dollars. Since our model uses a solar premium in $\$/\mathrm{kWh}$, we first divide these monthly charges by the average California residential monthly electricity consumption in 1997, which is $542~\mathrm{kWh}$ \cite{eia1998electric}, to express the premium on a per-kWh basis. We then convert the resulting values to 2024 dollars by adjusting for inflation using the Consumer Price Index (CPI). Finally, we fit a truncated exponential distribution to model $F_V(v_i)$. The fitted distribution is a truncated exponential distribution with support $[0,20.27]$~\cent/kWh and rate parameter $\lambda=0.2856~(\cent/\mathrm{kWh})^{-1}$, yielding a mean solar premium of $3.44$~\cent/kWh. \rev{Given the age of the underlying survey data and the conversion of monthly stated willingness to pay into a per-kWh premium, we treat the fitted distribution as a calibration benchmark rather than a direct empirical estimate of current consumer solar premiums.}

The load parameter $L_t$ is calibrated using 2024 electricity-demand data from CAISO and the California Energy Commission, representing the most recent historical load information available to investors at the beginning of 2025. The CAISO historical hourly load data report demand for several load areas, including the PG\&E service territory \cite{caiso_hourly_load}, but not directly for San Francisco. We therefore use the PG\&E hourly load profile to capture the within-month load shape and scale it using San Francisco's monthly electricity consumption reported by the California Energy Commission \cite{cec_consumption_data}. Specifically, for each month, the PG\&E hourly load is rescaled so that its monthly total equals San Francisco's monthly electricity consumption. The resulting hourly San Francisco load series is then averaged within each of the 16 representative periods to obtain $L_t$, where the entries in each vector correspond to $[\text{morning},\text{midday},\text{evening},\text{night}]$. The calibrated load values are $[0.623,0.580,0.672,0.575]$ for winter, $[0.598,0.512,0.644,0.586]$ for spring, $[0.529,0.545,0.681,0.544]$ for summer, and $[0.563,0.551,0.667,0.539]$ for fall, all in $\mathrm{GWh}$. 

For electricity prices, we use PG\&E Residential Inclu TOU tariff sheets effective during 2024 \cite{pge_res_inclu_tou_rates}, \rev{consistent with the interpretation of $\pi_{\mathrm{u}_t}$ in our model as the per-unit electricity price faced by consumers when purchasing electricity supplied by the utility.} We use the E-TOU-C rate schedule and assign each hour the corresponding retail rate from the tariff sheet effective for that date. We then average the resulting hourly prices within each of the 16 representative periods to obtain $\pi_{\mathrm{u}_t}$. With entries ordered as $[\text{morning},\text{midday},\text{evening},\text{night}]$, the calibrated price values are $[47.91,47.91,50.80,47.91]$ for winter, $[49.31,49.31,52.14,49.31]$ for spring, $[50.35,50.35,60.65,50.35]$ for summer, and $[47.26,47.26,52.66,47.26]$ for fall, all in \cent/kWh.

Finally, we use a PV capital and installation cost of \$2.95/W$_{\mathrm{dc}}$ \rev{and an annual O\&M cost of \$34/kW$_{\mathrm{dc}}$-year for the base planning period (i.e., the first year)}, based on the 2025Q1 modeled market price benchmark for residential PV-only systems \cite{nrel_panel}. \rev{The annual O\&M cost in subsequent years is obtained by applying the inflation adjustment specified in Section~\ref{lifecycle:financial}, and the resulting costs are discounted when computing the present-value investment cost over the 25-year panel lifespan.}

\rev{Using these case-study parameters, we first examine the equilibrium investment outcomes under the baseline setting before considering their robustness to alternative assumptions.}

\subsection{\rev{Baseline Investment Outcomes}}\label{constant}

The baseline setting uses the calibrated San Francisco case-study parameters and benchmark financial assumptions described above and considers homogeneous risk-neutral investors.

Given the case-study setup described in Section~\ref{data}, we compute panel investors' expected revenues under the real-time markets using~\eqref{eq:eqc:prt:tv} and~\eqref{eq:eqc:srt:tv}. For the contract-based market, we first obtain the aggregate demand function, $\widehat{d}^\star(\pi)$, from~\eqref{d:star:tv} and then compute investors' revenue. We then combine these revenue calculations with the investment equilibrium condition to derive the NE aggregate investment capacities under the three short-term markets. We consider two values of the solar-premium scaling parameter introduced in Section~\ref{epsilon:method}: the no-premium case $\epsilon=0$ and the calibrated-premium case $\epsilon=1$, which uses the fitted solar-premium distribution without scaling. The resulting equilibrium capacities are summarized in Table~\ref{tab:agg-capacity}. For ease of interpretation, all capacities are reported in $\mathrm{GW}_{\mathrm{dc}}$, i.e., gigawatts of installed direct-current PV capacity. \rev{To do so, note that since the model represents panel capacity $c$ by area in $\mathrm{m}^2$, the numerical calculations use consistent units: load values are converted to kWh, while the PV capital and O\&M cost inputs are converted to a per-$\mathrm{m}^2$ basis using a capacity density of $0.2~\mathrm{kW}_{\mathrm{dc}}/\mathrm{m}^2$, based on the assumed panel efficiency of $0.2$~\cite{955fd4dbf3e34e6e92a6dcaf57db598f}. The resulting equilibrium panel areas are then converted to installed capacity in $\mathrm{GW}_{\mathrm{dc}}$ for reporting.}
\begin{table}[h]
	\centering
	\caption{NE aggregate investment capacity (GW$_{\mathrm{dc}}$)}
	\label{tab:agg-capacity}
	\begin{tabular}{c|c c c c}
		\toprule
		\text{Solar-premium scaling parameter} 
		& $c_{\mathrm{srt}}^{\mathrm{ne}}$ 
		& $c_{\mathrm{prt}}^{\mathrm{ne}}$ 
		& $c_{\mathrm{cb}}^{\mathrm{ne}}$ 
		& $c_{\mathrm{opt}}$ \\
		\addlinespace[2pt]
		\toprule
		$\epsilon=0$ & \rev{1.410} & \rev{1.410} & \rev{1.410} & \rev{1.410} \\
		\midrule
		$\epsilon=1$ & \rev{1.410} & \rev{1.444} & \rev{1.466} & \rev{1.444} \\
		\bottomrule
	\end{tabular}
\end{table}

We first consider the no-premium case $\epsilon=0$, where consumers assign no additional value to solar energy. In this case, all market mechanisms yield the same aggregate capacity, as shown in Table~\ref{tab:agg-capacity}. This is consistent with~\eqref{eq:thm:p3} under case 3 (\emph{no solar premium}) of Theorems~\ref{thm:main} and~\ref{thm:main:tv}.

We then consider the calibrated-premium case $\epsilon=1$, which uses the fitted solar-premium distribution without scaling. Table~\ref{tab:agg-capacity} shows that the single-product real-time market yields the lowest equilibrium capacity, while the product-differentiated real-time market yields the same capacity as the social optimum. This is consistent with the comparison outcome established in~\eqref{thm:main:general} for case 1 (\emph{general case}) in Theorem~\ref{thm:main}, and likewise in Theorem~\ref{thm:main:tv}. The contract-based market yields a larger capacity than the social optimum in this numerical instance, i.e., $c_{\mathrm{opt}}<c_{\mathrm{cb}}^{\mathrm{ne}}$.

We also numerically checked the aggregate density condition in~\eqref{fg:constant:tv}. The corresponding aggregate density term varies substantially over $p\in[0,1]$ rather than remaining approximately flat, so the sufficient density condition used for the small-premium analytical comparison is not satisfied in this calibration. Nevertheless, the numerical result still exhibits the same qualitative over-investment pattern. This suggests that the sufficient condition in case 2 (\emph{small solar premium}) of Theorems~\ref{thm:main} and~\ref{thm:main:tv} is not necessary for the ordering $c_{\mathrm{opt}}<c_{\mathrm{cb}}^{\mathrm{ne}}$ to arise.

The relatively small differences in the resulting equilibrium aggregate capacities across market mechanisms in the calibrated-premium case reflect the fitted premium distribution: the mean solar premium is $3.44$~\cent/kWh, which is small relative to the retail electricity prices used in the numerical analysis. This observation motivates the sensitivity analyses below, where we examine how the equilibrium aggregate capacities change with the solar-premium level and other key market-side and long-term investment parameters.

\subsection{\rev{Robustness to Market-Side Parameters}}

\subsubsection{\rev{Solar-Premium Sensitivity}}

\begin{figure}[t]
	\includegraphics[width=0.5\textwidth]{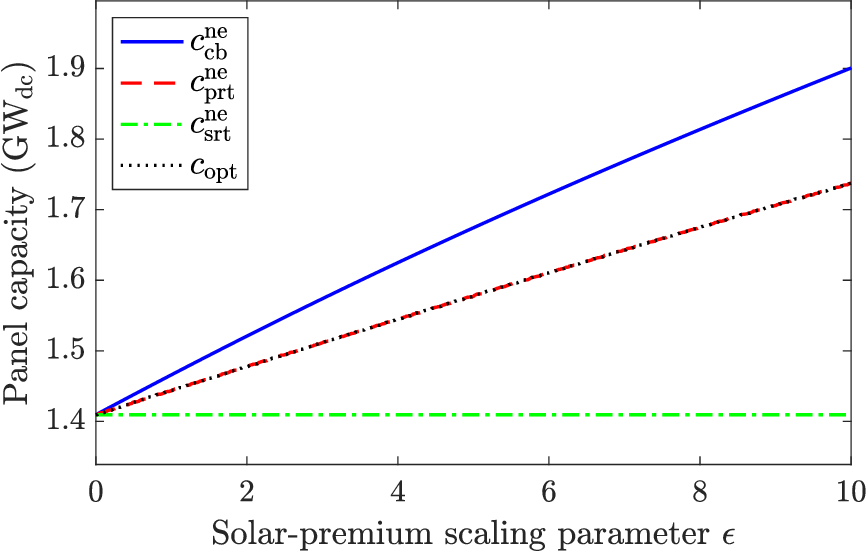}	
	\caption{Equilibrium aggregate PV investment capacities under different solar-premium scaling factors.}
	\label{fig:premium}
\end{figure}

Beyond the specific cases of $\epsilon=0$ and $\epsilon=1$, we vary the scaling factor applied to consumers' solar premiums and examine the resulting equilibrium investment capacities. The results are shown in Fig.~\ref{fig:premium}.

As consumers' solar premiums increase, Fig.~\ref{fig:premium} shows that the three market mechanisms respond differently. The equilibrium capacity under the single-product real-time market remains essentially unchanged because this market does not distinguish solar generation from conventional generation and therefore does not reflect consumers' additional willingness to pay for solar energy. Consequently, the gap between $c_{\mathrm{srt}}^{\mathrm{ne}}$ and the socially optimal capacity increases as $\epsilon$ increases.

In contrast, the product-differentiated real-time market tracks the socially optimal capacity throughout the sensitivity range. As the solar premium increases, both $c_{\mathrm{prt}}^{\mathrm{ne}}$ and $c_{\mathrm{opt}}$ increase, reflecting the additional value that consumers assign to solar electricity under the welfare framework considered in this paper.

The contract-based market also induces greater investment as consumers' solar premiums increase, but its equilibrium capacity increases more rapidly than the socially optimal capacity. Consequently, the gap between $c_{\mathrm{cb}}^{\mathrm{ne}}$ and $c_{\mathrm{opt}}$ widens as $\epsilon$ increases. This numerical pattern illustrates how heterogeneous consumer valuations can lead to higher aggregate investment under the contract-based market than under the socially optimal benchmark considered in this paper.

\begin{figure}[t]
	\includegraphics[width=0.5\textwidth]{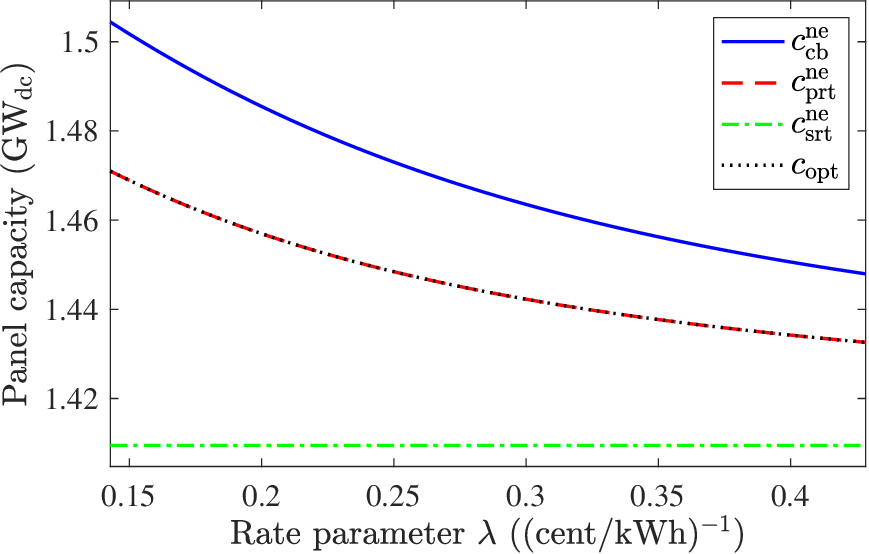}
	\caption{\rev{Equilibrium aggregate PV investment capacities under different values of the fitted solar-premium rate parameter $\lambda$.}}
	\label{fig:premium_lambda}
\end{figure}

\rev{The analysis above varies $\epsilon$ to examine how investment outcomes change as consumers place more or less additional value on solar electricity, with larger values of $\epsilon$ representing stronger preferences and willingness to pay for solar energy. In addition to this variation in the overall level of solar premiums, we examine the sensitivity of the results to the fitted premium distribution itself. Specifically, we vary the truncated-exponential rate parameter $\lambda$ around its calibrated value of $0.2856~(\cent/\mathrm{kWh})^{-1}$ while holding $\epsilon=1$ and the upper support $\overline{v}=20.27$~\cent/kWh fixed. A smaller value of $\lambda$ places relatively greater probability on higher solar premiums, whereas a larger value concentrates more probability near lower premiums. Because changing $\lambda$ shifts the overall truncated distribution, including its mean, this analysis captures sensitivity to the fitted solar-premium distribution through its rate parameter. The resulting equilibrium investment capacities are shown in Fig.~\ref{fig:premium_lambda}.}

\rev{As $\lambda$ increases, the truncated-exponential distribution places more probability on lower solar premiums, reducing consumers' average additional valuation of solar electricity. Consistent with this shift, Fig.~\ref{fig:premium_lambda} shows that $c_{\mathrm{prt}}^{\mathrm{ne}}$, $c_{\mathrm{cb}}^{\mathrm{ne}}$, and $c_{\mathrm{opt}}$ all decrease with $\lambda$, while $c_{\mathrm{srt}}^{\mathrm{ne}}$ remains unchanged because the single-product market does not depend on consumers' solar-premium distribution. Importantly, the qualitative capacity ordering is preserved throughout the considered range: the product-differentiated market continues to track the socially optimal capacity, whereas the contract-based market yields a higher equilibrium capacity. Thus, the main comparison across market mechanisms is robust to substantial variation in the fitted rate parameter of the solar-premium distribution.}

Having examined the sensitivity to consumers' solar valuations, we next consider the utility-supply price $\pi_{\mathrm{u}_t}$, another key short-term market parameter affecting investment incentives in the model.
\subsubsection{\rev{Utility-Supply Price Sensitivity}}

\rev{We next examine the sensitivity of the equilibrium capacity comparisons to the utility-supply price $\pi_{\mathrm{u}_t}$. Recall that our baseline analysis uses PG\&E residential TOU rates as $\pi_{\mathrm{u}_{t}}$, consistent with its interpretation in our model as the per-unit electricity price faced by consumers when purchasing electricity supplied by the utility. 
	
	To examine the robustness of the market-mechanism comparisons across different price levels, we introduce a utility-price scaling parameter $\gamma$ and define
	\begin{equation}
		\pi_{\mathrm{u}_t}(\gamma)
		=
		\gamma \pi_{\mathrm{u}_t}^{\mathrm{base}},
	\end{equation}
	where $\pi_{\mathrm{u}_t}^{\mathrm{base}}$ denotes the period-specific PG\&E TOU price used in the baseline analysis. We vary $\gamma$ over $[0.1,2]$ while holding all other calibrated parameters fixed. Thus, $\gamma=1$ corresponds to the baseline retail tariff, whereas values below and above one represent proportionally lower and higher utility-supply price levels, respectively. At $\gamma=0.1$, the resulting prices range approximately from $4.7$ to $6.1$~\cent/kWh, which is comparable in magnitude to wholesale electricity-price levels observed in California~\cite{caiso_2024_market_report}. The upper portion of the range represents substantially higher electricity-price levels than the calibrated baseline.}

\begin{figure}[t]
	\includegraphics[width=0.5\textwidth]{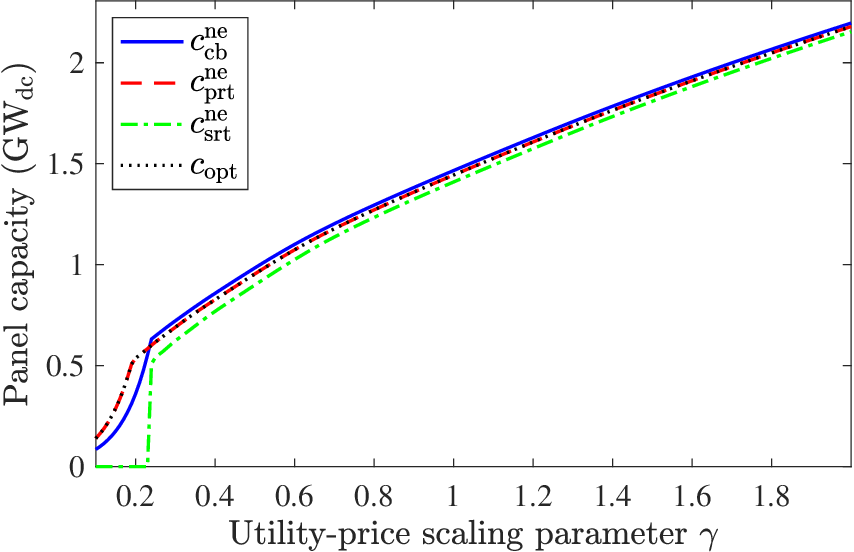}
	\caption{\rev{Equilibrium aggregate PV investment capacities under different utility-price scaling factors $\gamma$.}}
	\label{fig:utility_price}
\end{figure}

\rev{Figure~\ref{fig:utility_price} shows that, over most of the considered price range, the capacity ordering established by our theoretical analysis is preserved: $c_{\mathrm{srt}}^{\mathrm{ne}}<c_{\mathrm{prt}}^{\mathrm{ne}}=c_{\mathrm{opt}}<c_{\mathrm{cb}}^{\mathrm{ne}}$. As $\gamma$ increases, equilibrium capacities increase because higher utility-supply prices raise market-clearing prices and thereby increase the expected revenue from PV investment. At sufficiently low values of $\gamma$, however, condition~\eqref{eq:pi0cond:tv} is no longer satisfied, and the theoretical capacity ordering need not hold; in this region, the single-product equilibrium capacity falls to zero and the ordering between the product-differentiated and contract-based markets changes. The latter two markets nevertheless sustain positive investment because consumers' additional willingness to pay for solar continues to provide value even when utility-supply prices are low. Importantly, this behavior occurs only when price levels are far below the calibrated distribution-level retail prices and are closer in magnitude to wholesale electricity prices.}

\subsection{\rev{Robustness to Long-Term Investment Parameters}}

\subsubsection{\rev{Investment-Cost Sensitivity}}

\begin{figure}[t]
	\includegraphics[width=0.5\textwidth]{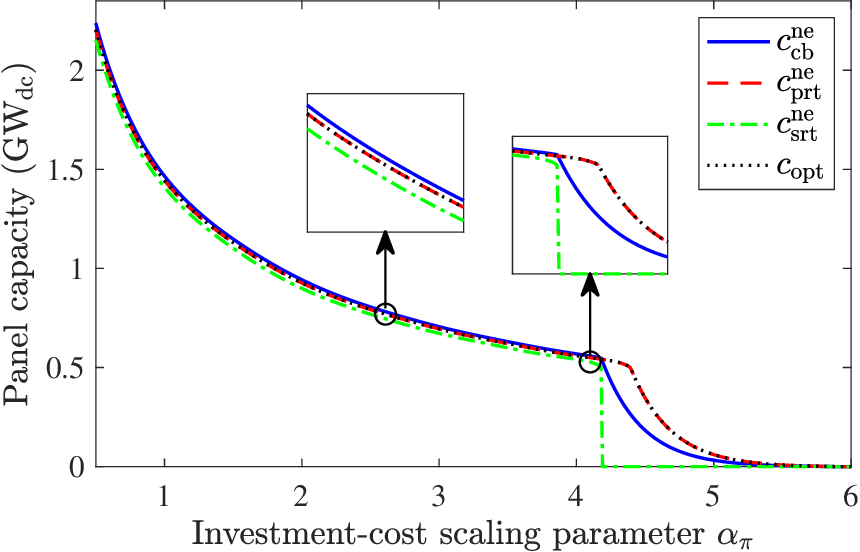}	
	\caption{\rev{Equilibrium aggregate PV investment capacities under different investment-cost scaling factors $\alpha_{\pi}$.}}
	\label{fig:comp}
\end{figure}

\rev{We next examine the sensitivity of the equilibrium capacities to the total present-value investment cost $\pi_0$. We introduce an investment-cost scaling parameter $\alpha_{\pi}$ and set
	\begin{equation}
		\pi_0(\alpha_{\pi})
		=
		\alpha_{\pi}\pi_0^{\mathrm{base}},
	\end{equation}
	where $\pi_0^{\mathrm{base}}$ is the benchmark investment cost used in the baseline analysis. We vary $\alpha_{\pi}$ over a broad range while holding the remaining calibrated parameters fixed. Thus, $\alpha_{\pi}=1$ corresponds to the baseline investment cost, whereas values below and above one represent proportionally lower and higher investment costs, respectively. The resulting equilibrium capacities are shown in Fig.~\ref{fig:comp}.}

\rev{As expected, increasing the investment cost reduces equilibrium capacity under all three market mechanisms. Over most of the considered range, the numerical results preserve the ordering
	\begin{equation}
		c_{\mathrm{srt}}^{\mathrm{ne}}
		<
		c_{\mathrm{prt}}^{\mathrm{ne}}
		=
		c_{\mathrm{opt}}
		<
		c_{\mathrm{cb}}^{\mathrm{ne}}.
		\nonumber
	\end{equation}
	Thus, the main market comparison remains robust to substantial variation in the investment cost.}

\rev{At sufficiently high investment-cost levels, however, condition~\eqref{eq:pi0cond:tv} is violated. The single-product market then reaches the corner solution $c_{\mathrm{srt}}^{\mathrm{ne}}=0$, and the theoretical capacity ordering need not continue to hold. The product-differentiated and contract-based markets can sustain positive investment over a wider cost range because consumers' additional willingness to pay for solar provides an additional source of revenue. In this extreme high-cost region, the ordering between the product-differentiated and contract-based markets may therefore change. This behavior occurs only when investment costs are several times larger than the calibrated benchmark and therefore well outside the baseline cost range. Moreover, such high cost levels are unlikely to be practically relevant if ongoing improvements in PV technology and associated cost reductions continue.}

\subsubsection{\rev{Discount-Rate Sensitivity}}

\rev{We further examine the sensitivity of the investment outcomes to the nominal annual discount rate $\rho$, which directly affects the present-value adjustment over the 25-year panel lifespan. We vary $\rho$ over $[0,35\%]$ while holding inflation and the remaining calibrated parameters fixed, with $\rho=5.64\%$ corresponding to the baseline value. For each value of $\rho$, we recompute $\widetilde T$, the present value of expected revenues and O\&M costs, and the resulting investment cost to determine the associated equilibrium capacities.}

\begin{figure}[t]
	\includegraphics[width=0.5\textwidth]{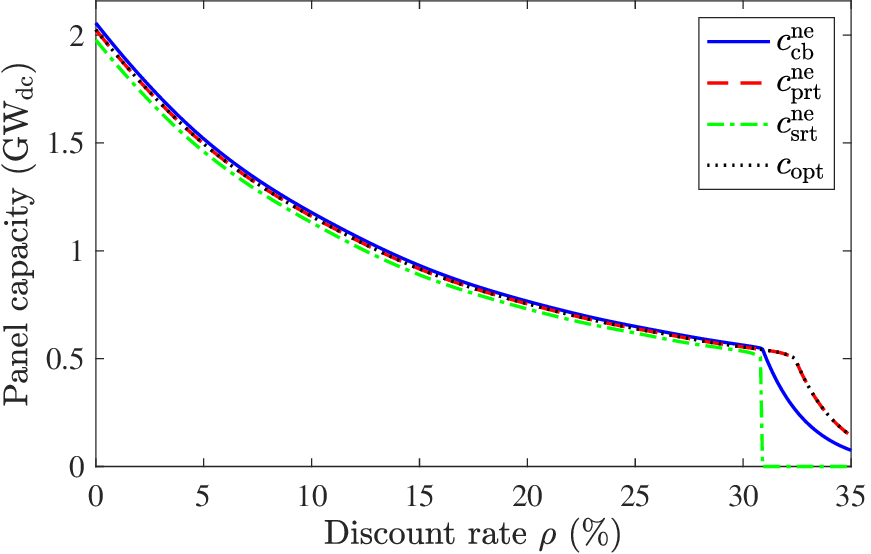}
	\caption{\rev{Equilibrium aggregate PV investment capacities under different annual discount rates $\rho$.}}
	\label{fig:discount_rate}
\end{figure}

\rev{As shown in Fig.~\ref{fig:discount_rate}, increasing the discount rate reduces equilibrium investment under all market mechanisms because future PV revenues receive less weight in present-value terms. Over most of the considered range, the qualitative ordering established by the theoretical analysis is preserved, with $c_{\mathrm{srt}}^{\mathrm{ne}}<c_{\mathrm{prt}}^{\mathrm{ne}}=c_{\mathrm{opt}}<c_{\mathrm{cb}}^{\mathrm{ne}}$. At very high discount rates, however, the single-product equilibrium reaches zero and the ordering can change. This is the same boundary effect observed previously when $\pi_0$ becomes sufficiently large or when $\pi_{\mathrm{u}_t}$ falls sufficiently low: condition~\eqref{eq:pi0cond:tv} is no longer satisfied, so the assumptions underlying the theoretical comparison no longer hold. \rev{However, these extreme discount rates are far above the baseline value. For more practically relevant discount-rate levels, the numerical results continue to support the qualitative market ordering established by our theoretical analysis.}}

\subsection{\rev{Robustness to Investor Heterogeneity}}

\rev{Our theoretical analysis assumes homogeneous investors with a common investment cost $\pi_0$ and a binary installation decision. A general treatment of investor heterogeneity could incorporate investor-specific installation costs, available rooftop capacity, financing conditions, and technology characteristics, and would require a more detailed investment-side formulation and calibration. Such an extension is beyond the scope of this paper, whose primary focus is on how alternative short-term market mechanisms affect long-term PV investment incentives. Nevertheless, to examine whether the main capacity comparison is sensitive to investor heterogeneity, we focus on one representative dimension and consider a simplified numerical extension in which investors differ in their investment costs.}

\rev{Specifically, we assume that present-value investment costs vary uniformly across investors:
	\begin{equation}
		\pi_{0,i}
		\sim
		U\!\left[
		(1-\delta_{\pi})\pi_0,\,
		(1+\delta_{\pi})\pi_0
		\right],
		\qquad i\in\mathcal I_{\mathrm{inv}},
	\end{equation}
	where $\pi_0$ is the baseline investment cost and $\delta_{\pi}$ controls the degree of investment-cost heterogeneity. We vary $\delta_{\pi}$ from $0$ to $0.8$, with $\delta_{\pi}=0$ corresponding to the homogeneous-investor benchmark and the upper end of the range representing substantial dispersion in investor costs. Under positive cost heterogeneity, investors whose costs do not exceed the market-specific expected revenue per unit of installed capacity choose to invest, while higher-cost investors do not.}

\rev{For $\delta_\pi>0$, let $F_{\pi_0}$ denote the CDF of investor-specific investment costs among potential investors. Given aggregate capacity $c$, the equilibrium aggregate capacity satisfies
	\begin{equation}
		c_m^{\mathrm{ne}}
		=
		C_{\max}
		F_{\pi_0}\!\left(
		\frac{\Pi_m^{\mathrm{s}}(c_m^{\mathrm{ne}})}
		{c_m^{\mathrm{ne}}}
		\right),
	\end{equation}
	where $C_{\max}$ denotes the aggregate capacity that would be installed if all potential investors participated. In the homogeneous-investor benchmark, the interior equilibrium is determined directly by the zero-profit condition $\Pi_m^{\mathrm{s}}(c)/c=\pi_0$, so $C_{\max}$ does not affect the equilibrium as long as it is sufficiently large and nonbinding. With heterogeneous investment costs, however, aggregate capacity is determined through the cost distribution $F_{\pi_0}$, and therefore also depends on $C_{\max}$.}

\rev{A full heterogeneous-investor calibration would require detailed information on the distribution of investor characteristics and the scale of the potential investor population, which is beyond the scope of this paper. We therefore examine robustness by varying $\delta_\pi$, which controls investment-cost heterogeneity. Since the heterogeneous equilibrium also depends on $C_{\max}$, we vary this parameter as well according to
	\begin{equation}
		C_{\max}
		=
		\chi c_{\mathrm{ref}},
	\end{equation}
	where $c_{\mathrm{ref}}$ is the baseline product-differentiated equilibrium capacity, which coincides with the baseline socially optimal capacity under the welfare framework considered in this paper, and $\chi$ ranges from $1.25$ to $4$.}

\begin{figure}[t]
	\centering
	\includegraphics[width=0.5\textwidth]{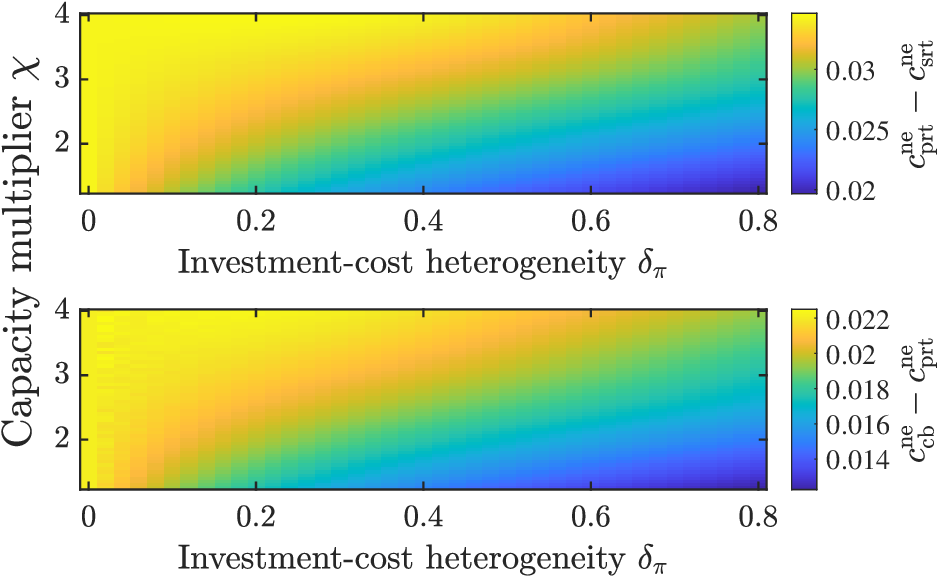}
	\caption{\rev{Robustness of the equilibrium-capacity ordering to investment-cost heterogeneity and the potential-capacity multiplier $\chi$. The plotted capacity differences are reported in GW$_{\mathrm{dc}}$, with positive values indicating preservation of the corresponding capacity ordering.}}
	\label{fig:investor_heterogeneity}
\end{figure}

\rev{In Fig.~\ref{fig:investor_heterogeneity}, we vary both the investment-cost heterogeneity parameter $\delta_{\pi}$ and the potential-capacity multiplier $\chi$. For each parameter combination, the upper panel plots $c_{\mathrm{prt}}^{\mathrm{ne}}-c_{\mathrm{srt}}^{\mathrm{ne}}$, while the lower panel plots $c_{\mathrm{cb}}^{\mathrm{ne}}-c_{\mathrm{prt}}^{\mathrm{ne}}$. As shown in the figure, both differences remain positive over the entire parameter ranges considered. This provides numerical evidence that the relative ranking of equilibrium capacities across the three market mechanisms is robust over the considered ranges of investment-cost heterogeneity and potential-capacity scale. However, this result should not be generalized to arbitrary heterogeneous-investor settings, which would require a richer investment-side formulation and additional calibration.}
\subsection{Policy Implications and Practical Pathways}\label{sec:policy-implementation}

\revise{The numerical results reinforce the central policy implication of our theoretical analysis: allowing consumers' heterogeneous valuations for solar energy to affect PV revenues can materially change long-term investment incentives. In the calibrated benchmark and across the main sensitivity ranges in the baseline risk-neutral setting, the single-product real-time market yields the lowest aggregate investment because consumers' additional willingness to pay for solar energy is not transmitted to panel investors. In contrast, both the product-differentiated real-time market and the contract-based market transmit these valuations to investors, although through different mechanisms and with different investment consequences.}

\revise{These two mechanisms also differ in their resulting investment levels. The product-differentiated real-time market aligns decentralized investment with the \rev{benchmark socially optimal capacity under the welfare framework considered in this paper}, whereas the contract-based market can create stronger investment incentives and induce a larger aggregate PV capacity. Within this static setting, the additional capacity induced by the contract-based market therefore represents over-investment relative to the benchmark social optimum. Contract-based procurement may nevertheless support greater solar deployment, and higher deployment could generate longer-run benefits through economies of scale, supply-chain development, or improved installation practices that reduce future costs. Such dynamic feedbacks are outside the scope of the present model, so our results should be interpreted as a static efficiency comparison within the framework considered here.}

\revise{A related practical question is how consumers' heterogeneous valuations for renewable electricity could be reflected in PV investment incentives through existing electricity-sector programs and market arrangements. Feed-in tariffs already provide structured revenue channels for distributed PV investment, but generally rely on administratively determined compensation rather than directly transmitting consumers' heterogeneous solar valuations. More advanced distribution-level or product-differentiated markets may require substantial institutional and technological development before large-scale implementation. Community choice aggregation (CCA), on the other hand, provides a useful practical analogue: local governments procure electricity on behalf of residents, businesses, and municipal accounts, while transmission and distribution service remains with the existing utility~\cite{epa_cca,cpuc_cca_regulatory}. CCA is not itself an investment mechanism, but it can serve as a procurement layer that aggregates consumer preferences for renewable electricity and translates them into renewable-energy portfolios or long-term contracts. If procurement decisions and contracts with PV owners or developers reflect these preferences, differentiated demand for renewable electricity can create revenue streams that support additional PV development. Thus, CCA provides one practical pathway for connecting consumers' renewable-energy valuations to the PV investment incentives studied in our model.}

\section{Concluding Remarks}

This paper studies how short-term market mechanisms shape long-term distributed PV investment, where investors' decisions are driven by expected lifetime revenue from participating in these markets. To analyze this relationship, we develop an investment model that links short-term market equilibria to the resulting long-term investment outcomes. As a key tool, we propose a unified framework that yields closed-form expressions for short-term equilibrium outcomes across a broad class of mechanisms; these expressions feed into the investment model. We then analyze three representative mechanisms: a standard single-product real-time market, a new product-differentiated real-time market, and a new contract-based panel market, each capturing a distinct approach to pricing and trading in distribution networks. For each mechanism, we apply the unified framework and the investment model to derive analytical expressions for the resulting long-term equilibrium panel capacities. We theoretically establish that, in the baseline risk-neutral setting, the product-differentiated real-time market yields \rev{the benchmark socially optimal investment capacity under the welfare framework considered in this paper}, whereas the single-product market induces under-investment \rev{relative to this benchmark}. More generally, under both risk neutrality and risk aversion, the single-product market yields no greater equilibrium capacity than either the product-differentiated or contract-based market. Under risk aversion, the product-differentiated equilibrium capacity may fall below the socially optimal capacity but remains no smaller than the single-product equilibrium capacity. When consumers' solar premiums are small, we further show that, under both risk-neutral and risk-averse investment preferences, the contract-based market can lead to over-investment \rev{relative to this benchmark social optimum}, a pattern our numerical results confirm even for larger premiums.

\revise{Several directions for future work remain. One direction is to develop a full
	elastic-demand model for the product-differentiated and contract-based markets.
	Another direction is to relax assumptions such as uniform buy/sell pricing in
	short-term markets and to incorporate richer investor and seller heterogeneity,
	including cases where some investors are large enough to exercise market power.
	A further direction is to extend the investment model to incorporate additional
	factors, such as bounded rationality, subjective risk perceptions, and adoption
	inertia.
}

\bibliographystyle{IEEEtran}
\bibliography{bib}
\appendices

\section{Competitive Equilibrium and Nash Equilibrium Equivalence in Short-Term Markets}\label{appendix:equiv}

As noted in Remark~\ref{price:discovery}, rather than detailing the price-discovery process, we focus on the CE concept for the short-term markets, yet claim that it is equivalent to an NE of a bidding game between sellers and buyers. We establish this equivalence in this section. 

We consider a standard double-sided auction process for our unified solar market participation game.  
In this process, each seller $i \in \mathcal I_\mathrm{s}$ submits a price-quantity bid $(\underline \pi_i^\mathrm{s}, \overline q^\mathrm{s}_i)$, where $\underline \pi_i^\mathrm{s} \ge 0$ is the minimum price that the seller is willing to accept, and $\overline q^\mathrm{s}_i \in \mathcal Q^\mathrm{s}_i \subset \reals_{+}$ is the maximum quantity that the seller can provide. Each buyer $i\in \mathcal I_\mathrm{b}$ submits a price-quantity offer $(\overline \pi_i^\mathrm{b}, \overline q^\mathrm{b}_i)$, where $\overline \pi_i^\mathrm{b} \ge 0$ is the maximum willingness to pay for buyer $i$, and $\overline q_i^\mathrm{b}  \in \mathcal Q^\mathrm{b}_i \subset \reals_{+}$ is the maximum quantity to be consumed by the buyer. We will refer to the collection of quantities for all sellers or buyers by omitting the subscript $i$, e.g., $\overline q^\mathrm{s} := \{\overline q_i^\mathrm{s}\}_{i \in \mathcal I_\mathrm{s}}$.  To further simplify our notation, we denote the collection of submitted market data by $\mathbf{D}= (\underline \pi^\mathrm{s}, \overline q^\mathrm{s}, \overline \pi^\mathrm{b}, \overline q^\mathrm{b})$; we also use $\mathbf{D}_i$ to denote the data submitted by supplier $i\in\mathcal I_\mathrm{s}$ (or consumer $i\in \mathcal I_\mathrm{b}$), and $\mathbf{D}_{-i}$ to denote the data submitted by everyone else\footnote{We will always make clear whether $i$ refers to a supplier or a consumer to avoid ambiguity in this notation.}. We slightly overload notation and write the payoff of seller $i\in\mathcal I_{\mathrm s}$ as
$\Pi_i^{\mathrm s}(\underline{\pi}_i^{\mathrm s},\overline{q}_i^{\mathrm s},\mathbf D_{-i})$,
using the same payoff function $\Pi_i^{\mathrm s}(q_i^{\mathrm s},\pi)$ defined in Section~\ref{sec:mechanisms} for the corresponding market. 
Similarly, we denote the payoff of buyer $i\in\mathcal I_{\mathrm b}$ by
$\Pi_i^{\mathrm b}(\overline{\pi}_i^{\mathrm b},\overline{q}_i^{\mathrm b},\mathbf D_{-i})$,
which is identical to the buyer payoff function $\Pi_i^{\mathrm b}(q_i^{\mathrm b},\pi)$ in Section~\ref{sec:mechanisms}.

The market operator will then collect all these bids and offers, and clear the market by identifying the intersection of the supply curve and demand curve obtained by integrating the bids and offers from individual participants.  
The market clearing process determines a market price $\pi$, and the allocation for all bids and offers, i.e., $q_i^\mathrm{s} \in [0,  \overline q_i^\mathrm{s}]$ for all $i \in \mathcal I_\mathrm{s}$, and $q_i^\mathrm{b} \in [0,  \overline q_i^\mathrm{b}]$ for all $i \in \mathcal I_\mathrm{b}$. 
In other words, 
\begin{equation}
	(\pi, q^\mathrm{s}, q^\mathrm{b}) = \mathsf{Market\,Clearing}(\mathbf{D}), \nonumber
\end{equation}
where the mathematical details for the market clearing process are provided as follows. 

\subsection{Market Clearing Process}
Given the bids and offers $\mathbf{D}= (\underline \pi^\mathrm{s}, \overline q^\mathrm{s}, \overline \pi^\mathrm{b}, \overline q^\mathrm{b})$, the market operator clears the market by identifying the intersection of the supply and demand curves. It is easy to show that the following optimization is an equivalent formulation:
\begin{subequations}\label{opt:mc}
	\begin{align}
		\max_{q^\mathrm{s}, q^\mathrm{b}}\quad & \int_{\mathcal I_\mathrm{b}} \overline \pi^\mathrm{b}_i q_i^\mathrm{b}\,\, \mathrm{d}i - \int_{\mathcal I_\mathrm{s}} \underline \pi^\mathrm{s}_i q_i^\mathrm{s}\,\, \mathrm{d}i \label{eq:obj:mc}\\
		\mbox{s.t.}\quad & \pi: \int_{\mathcal I_\mathrm{s}} q_i^\mathrm{s}\,\, \mathrm{d}i =\int_{\mathcal I_\mathrm{b}}  q_i^\mathrm{b}\,\, \mathrm{d}i, \label{eq:con:mc}\\
		& \quad \,\,\,0 \le q^\mathrm{s}_i\le \overline q^\mathrm{s}_i, \quad i \in \mathcal I_\mathrm{s}\label{eq:mu:mc},\\
		& \quad \,\,\, 0 \le q^\mathrm{b}_i\le \overline q^\mathrm{b}_i, \quad i\in \mathcal I_\mathrm{b}\label{eq:nu:mc},
	\end{align}
\end{subequations}
where $\pi$ is the Lagrange multiplier associated with the market clearing constraint \eqref{eq:con:mc}. Optimization~\eqref{opt:mc} is an infinite dimensional linear program, which has many solutions. To avoid the potential non-uniqueness issue in both the primal and dual solutions, we adapt the following rules:
\begin{enumerate}
	\item If the dual solution of constraint~\eqref{eq:con:mc}  is not unique, denote the set of such dual solutions by $\Omega^\mathrm{eq}$. We set the market price to $\pi := \min\{\pi': \pi'\in \Omega^\mathrm{eq}\}$. 
	\item For each buyer $i \in \mathcal I_\mathrm{b}$, 
	\[
	q_i^\mathrm{b} = \begin{cases}
		\overline q^\mathrm{b}_i, & \mbox{ if }	\overline \pi^\mathrm{b}_i > \pi, \\
		0, & \mbox{ if }	\overline \pi^\mathrm{b}_i < \pi.
	\end{cases}
	\]
	If both $i,j \in \mathcal I_\mathrm{b}$ such that $\overline \pi^\mathrm{b}_i= \overline \pi^\mathrm{b}_j= \pi$, then $q^\mathrm{b}_i/\overline q^\mathrm{b}_i =q^\mathrm{b}_j/\overline q^\mathrm{b}_j $. If $i \in \mathcal I_\mathrm{b}$ is the only buyer with $\overline \pi^\mathrm{b}_i=  \pi$, then $q_i^\mathrm{b} = 
	\overline q^\mathrm{b}_i$. 
	\item For each seller $i \in \mathcal I_\mathrm{s}$, 
	\[
	q_i^\mathrm{s} = \begin{cases}
		\overline q^\mathrm{s}_i, & \mbox{ if }	\underline \pi^\mathrm{s}_i < \pi, \\
		0, & \mbox{ if }	\underline \pi^\mathrm{s}_i > \pi.
	\end{cases}
	\]
	If both $i,j \in \mathcal I_\mathrm{s}$ such that $\underline \pi^\mathrm{s}_i= \underline \pi^\mathrm{s}_j= \pi$, then $q^\mathrm{s}_i/\overline q^\mathrm{s}_i =q^\mathrm{s}_j/\overline q^\mathrm{s}_j $. If $i \in \mathcal I_\mathrm{s}$ is the only seller with $\underline \pi^\mathrm{s}_i=  \pi$, then $q_i^\mathrm{s} = 
	\overline q^\mathrm{s}_i$. 
\end{enumerate} 
One can verify that the proposed rules above correspond to a solution of optimization~\eqref{opt:mc}. 

In the following two subsections, we characterize an NE of the buyer-seller bidding process under each of the three short-term markets, yielding a stable $\mathbf{D}$ for the market-clearing problem~\eqref{opt:mc}.

\subsection{Nash Equilibrium of Real-Time Market Mechanisms and Its Equivalence to the Associated CE of Lemma~\ref{prop:eq:rt}}

\begin{lemma}[NE of real-time market mechanisms]\label{prop:eq:rt:bid}
	For product-differentiated real-time market (\texttt{prt}) and single-product real-time market (\texttt{srt}), an NE market participation strategy and the corresponding market clearing outcomes are listed in Table~\ref{tab:rt:bid}. 
\end{lemma}
\begin{table*}[t]
	\centering
	\caption{NE market participation strategy and market clearing outcomes for real-time market mechanisms}\label{tab:rt:bid}
	\begin{tabular}{@{}cc|cccc|ccc@{}}
		\toprule
		\multicolumn{1}{l}{} &
		\multicolumn{1}{l|}{} &
		\multicolumn{4}{c|}{NE participation strategy} &
		\multicolumn{3}{c}{Market clearing outcome} \\ \midrule
		\multicolumn{1}{l|}{} &
		Market &
		$\underline \pi_i^\mathrm{s}$ &
		$\overline q_i^\mathrm{s}$ &
		$\overline \pi_i^\mathrm{b}$ &
		$\overline q_i^\mathrm{b}$ &
		$\pi$ &
		$q_i^\mathrm{s}$ &
		$q_i^\mathrm{b}$ \\ \midrule
		\multicolumn{1}{c|}{\multirow{2}{*}{\begin{tabular}[c]{@{}c@{}}Abundant supply\\ $cG>L$\end{tabular}}} &
		$\texttt{prt}$ &
		0 &
		$cG$ &
		$\pi_\mathrm{u}+v_i$ &
		$L$ &
		0 &
		$L$ &
		$L$ \\ \cmidrule(l){2-9} 
		\multicolumn{1}{c|}{} &
		$\texttt{srt}$ &
		0 &
		$cG$ &
		$\pi_\mathrm{u}$ &
		$L$ &
		0 &
		$L$ &
		$L$ \\ \midrule
		\multicolumn{1}{c|}{\multirow{2}{*}{\begin{tabular}[c]{@{}c@{}}Limited supply\\ $cG\le L$\end{tabular}}} &
		$\texttt{prt}$ &
		0 &
		$cG$ &
		$\pi_\mathrm{u}+v_i$ &
		$L$ &
		$\pi_\mathrm{u}+\overline F_V ^{-1}\left(\frac{cG}{L}\right)$ &
		$cG$ &
		$L\mathbb 1\{\overline \pi^\mathrm{b}_i\ge \pi\}$ \\ \cmidrule(l){2-9} 
		\multicolumn{1}{c|}{} &
		$\texttt{srt}$ &
		0 &
		$cG$ &
		$\pi_\mathrm{u}$ &
		$L$ &
		$\pi_\mathrm{u}$ &
		$cG$ &
		$cG$ \\ \bottomrule
	\end{tabular}
\end{table*}
\begin{IEEEproof}[Proof of Lemma~\ref{prop:eq:rt:bid}]
	We present the main proof for the product-differentiated real-time market; the single-product real-time market follows as a special case.
	We first show that, under the proposed participation strategies, the market outcomes in Table~\ref{tab:rt:bid} solve~\eqref{opt:mc}. We then verify that these strategies constitute a Nash equilibrium, i.e., no supplier or consumer can profitably deviate unilaterally. Following Table~\ref{tab:rt:bid}, we consider two cases.
	
	\emph{Abundant supply:} The outcomes in Table~\ref{tab:rt:bid} satisfy the constraints of~\eqref{opt:mc} and are therefore feasible. To establish optimality, note that since $0\le q_i^{\mathrm b}\le L$ for all $i\in\mathcal I_{\mathrm b}$, the objective in~\eqref{opt:mc} is bounded by
	\begin{equation}
		\int_{\mathcal I_\mathrm{b}} \left(\pi_{\mathrm{u}}+v_{i}\right) q_i^\mathrm{b}\, \mathrm{d}i
		\leq
		\int_{\mathcal I_\mathrm{b}} \left(\pi_{\mathrm{u}}+v_{i}\right) L\, \mathrm{d}i .\nonumber
	\end{equation}
	This upper bound is attained by the allocation $q_i^{\mathrm b}=L$ for all $i \in \mathcal{I}_{\mathrm{b}}$, hence the outcomes in Table~\ref{tab:rt:bid} are optimal for~\eqref{opt:mc}. To derive the associated clearing price $\pi=0$ (as in Table~\ref{tab:rt:bid}), we next apply the KKT conditions by forming the Lagrangian of~\eqref{opt:mc} as
	\begin{align}\label{eq:prt:lag}
		\mathcal{L} = &-\int_{\mathcal I_\mathrm{b}} \left(\pi_{\mathrm{u}}+v_{i}\right) q_i^\mathrm{b}\,\, \mathrm{d}i+\pi \left( \int_{\mathcal I_\mathrm{b}}  q_i^\mathrm{b}\,\, \mathrm{d}i -	\int_{\mathcal I_\mathrm{s}} q_i^\mathrm{s}\,\, \mathrm{d}i\right)\nonumber\\
		& +\int_{\mathcal I_\mathrm{s}} \mu_{i}^{\mathrm{s}}\left(q_i^\mathrm{s}-cG\right)\,\, \mathrm{d}i
		+\int_{\mathcal I_\mathrm{b}} \mu_{i}^{\mathrm{b}}\left(q_i^\mathrm{b}-L\right)\,\, \mathrm{d}i\nonumber\\
		& +\int_{\mathcal I_\mathrm{s}} \nu_{i}^{\mathrm{s}}\left(-q_i^\mathrm{s}\right)\,\, \mathrm{d}i+\int_{\mathcal I_\mathrm{b}} \nu_{i}^{\mathrm{b}}\left(-q_i^\mathrm{b}\right)\,\, \mathrm{d}i,
	\end{align}
	in which $\mu_{i}^{\mathrm{s}}$ and $\nu_{i}^{\mathrm{s}}$ are the Lagrangian multipliers associated with~\eqref{eq:mu:mc}, while  $\mu_{i}^{\mathrm{b}}$ and $\nu_{i}^{\mathrm{b}}$ are for~\eqref{eq:nu:mc}. Differentiating $\mathcal{L}$ with respect to $q_i^{\mathrm b}$ and $q_i^{\mathrm s}$ yields
	\begin{align}
		& -\pi_{\mathrm{u}}-v_i+\pi+\mu_i^{\mathrm b}-\nu_i^{\mathrm b}=0, && i\in\mathcal I_{\mathrm b}, \nonumber\\
		& -\pi+\mu_i^{\mathrm s}-\nu_i^{\mathrm s}=0, && i\in\mathcal I_{\mathrm s}. \nonumber
	\end{align}
	Given the clearing outcomes in Table~\ref{tab:rt:bid}, the constraints associated with $\mu_i^{\mathrm s}$, $\nu_i^{\mathrm s}$, and $\nu_i^{\mathrm b}$ are nonbinding; hence, by complementary slackness, $\mu_i^{\mathrm s}=\nu_i^{\mathrm s}=\nu_i^{\mathrm b}=0$.
	Therefore,
	\begin{align}
		& \pi=\pi_{\mathrm u}+v_i-\mu_i^{\mathrm b}, && i\in\mathcal I_{\mathrm b}, \nonumber\\
		& \pi=0, && i\in\mathcal I_{\mathrm s}, \nonumber
	\end{align}
	which implies $\pi=0$. Thus, the clearing outcomes in Table~\ref{tab:rt:bid} are optimal for~\eqref{opt:mc} and therefore constitute the market-clearing solution in this case. We next show that the corresponding submitted bids form an NE of the bidding process.
	
	Under the results summarized for this case in Table~\ref{tab:rt:bid}, each supplier earns $\Pi_i^{\mathrm s}(0,cG,\mathbf D_{-i})=0$. Suppose, for contradiction, that some $i\in\mathcal I_{\mathrm s}$ deviates to $(\underline{\pi}_i^{\mathrm{s}'},\overline q_i^{\mathrm{s}'})$ and obtains $\Pi_i^{\mathrm s}(\underline{\pi}_i^{\mathrm{s}'},\overline q_i^{\mathrm{s}'},\mathbf D_{-i})>0$, which would require a strictly positive clearing price $\pi'>0$. However, with all other bids fixed, a single seller's deviation does not affect the aggregate terms in~\eqref{eq:obj:mc}-\eqref{eq:con:mc}; hence the corresponding KKT multiplier of~\eqref{eq:con:mc} (the market price) remains unchanged. Therefore $\pi'=\pi=0$ and the deviating seller's payoff stays zero, a contradiction. 
	Similarly, suppose that there exists an $ i \in \mathcal I_\mathrm{b}$ with $\overline \pi_i^{\mathrm{b}'} \neq \pi_{\mathrm{u}}+v_i$, and $\Pi_i^\mathrm{b}(\overline \pi_i^{\mathrm{b}'},L, \mathbf{D}_{-i}) > \Pi_i^\mathrm{b}(\pi_{\mathrm{u}}+v_{i},L, \mathbf{D}_{-i}).$
	
	However, using~\eqref{cons:payoff} gives
	\begin{equation}
		\Pi_i^\mathrm{b}(\overline \pi_i^{\mathrm{b}'},L, \mathbf{D}_{-i}) \leq v_{i}q_i^{\mathrm{b}},\nonumber
	\end{equation}
	and
	\begin{equation}
		\Pi_i^\mathrm{b}(\pi_{\mathrm{u}}+v_{i},L, \mathbf{D}_{-i})=v_{i}q_i^{\mathrm{b}}.\nonumber
	\end{equation}
	Therefore, 
	$\Pi_i^\mathrm{b}(\overline \pi_i^{\mathrm{b}'},L, \mathbf{D}_{-i}) \leq \Pi_i^\mathrm{b}(\pi_{\mathrm{u}}+v_{i},L, \mathbf{D}_{-i})$ which contradicts the initial assumption. 
	Thus, no supplier or consumer has a profitable unilateral deviation, and the outcomes in Table~\ref{tab:rt:bid} constitute an NE for this case. The single-product real-time market follows by setting $v_{i} \equiv 0$.

	\emph{Limited supply}:
	To verify that the market outcomes in Table~\ref{tab:rt:bid} clear the market under the proposed bids, it suffices to show that they satisfy the KKT conditions of~\eqref{opt:mc}. We therefore form the Lagrangian as in~\eqref{eq:prt:lag} and obtain the stationarity conditions as follows:
	\begin{align}
		& v_i-\overline F_V^{-1}\!\left(\frac{cG}{L}\right)=\mu_i^{\mathrm b}-\nu_i^{\mathrm b}\triangleq \Upsilon_i^{\mathrm b}, 
		&& i\in\mathcal I_{\mathrm b}, \nonumber\\
		& \pi_{\mathrm u}+\overline F_V^{-1}\!\left(\frac{cG}{L}\right)=\mu_i^{\mathrm s}-\nu_i^{\mathrm s}, 
		&& i\in\mathcal I_{\mathrm s}, \label{prt:feas:case2}
	\end{align}
	where $\Upsilon_i^{\mathrm b}\in\reals$ is introduced for notational convenience.
	Moreover, the given bids also satisfy the primal feasibility conditions:
	\begin{align}
		& 0 \leq L\mathbb{1} \left\{v_{i} \geq \overline F_V ^{-1}\left(\frac{cG}{L}\right)\right\} \leq L,\nonumber \\ 
		& 0 \leq cG\leq cG,\nonumber \\
		& cG=\int_{\mathcal I_\mathrm{b}} L\mathbb{1} \left\{v_{i} \geq \overline F_V ^{-1}\left(\frac{cG}{L}\right)\right\}\,\, \mathrm{d}i. \nonumber
	\end{align}
	The complementary slackness conditions associated with~\eqref{eq:mu:mc} can be written as:
	\begin{align}
		& \nu_{i}^{\mathrm{s}}\bigl(-cG\bigl)=0 \Rightarrow \nu_{i}^{\mathrm{s}}=0, & \quad i \in \mathcal I_\mathrm{s},\nonumber \\ 
		&\mu_{i}^{\mathrm{s}} \bigl(cG-cG\bigl)=0 \Rightarrow \mu_{i}^{\mathrm{s}} \geq 0,& \quad i \in \mathcal I_\mathrm{s}. \nonumber 
	\end{align}
	Based on these results and according to~\eqref{prt:feas:case2}, it is immediate to get
	\begin{equation}
		\mu_{i}^{\mathrm{s}}=\pi_{\mathrm{u}} +\overline F_V ^{-1}\left(\frac{cG}{L}\right), \quad i \in \mathcal I_\mathrm{s}.\nonumber
	\end{equation}
	
	Similarly, the complementary slackness conditions associated with~\eqref{eq:nu:mc} can be written as:
	\begin{align}
		& \nu_{i}^{\mathrm{b}}\left(-L\mathbb{1} \left\{v_{i} \geq \overline F_V ^{-1}\left(\frac{cG}{L}\right)\right\}\right)=0, & \quad i \in \mathcal I_\mathrm{b},\nonumber \\ 
		&\mu_{i}^{\mathrm{b}} \left(L\mathbb{1} \left\{v_{i} \geq \overline F_V ^{-1}\left(\frac{cG}{L}\right)\right\}-L\right)=0,& \quad i \in \mathcal I_\mathrm{b}.\nonumber 
	\end{align}
	
	In particular, letting $\mu_i^{\mathrm b}=(\Upsilon_i^{\mathrm b})_+$ and $\nu_i^{\mathrm b}=(-\Upsilon_i^{\mathrm b})_+$ satisfies dual feasibility and complementary slackness. Therefore, under the proposed bids, the claimed market outcomes satisfy the KKT conditions and hence are feasible and optimal solutions of~\eqref{opt:mc}.

	To establish that the proposed strategies form a Nash equilibrium, suppose for contradiction that some $i\in\mathcal I_{\mathrm s}$ deviates to $(\underline{\pi}_i^{\mathrm{s}'},\overline q_i^{\mathrm{s}'})$ and attains
	\[
	\Pi_i^{\mathrm s}(\underline{\pi}_i^{\mathrm{s}'},\overline q_i^{\mathrm{s}'},\mathbf D_{-i})
	>
	\Pi_i^{\mathrm s}(0,cG,\mathbf D_{-i})
	=
	\Bigl(\pi_{\mathrm u}+\overline F_V^{-1}\!\Bigl(\tfrac{cG}{L}\Bigr)\Bigr)cG .
	\]
	Such a profitable deviation would require the clearing price to rise to some $\pi' > \pi_{\mathrm u}+\overline F_V^{-1}\!\bigl(\tfrac{cG}{L}\bigr)$, which is impossible since any individual participant has a negligible impact on the market price, as discussed in the previous case.
	
	Similarly, suppose for contradiction that there exists $i\in\mathcal I_{\mathrm b}$ with
	$\overline{\pi}_i^{\mathrm{b}'}\neq \pi_{\mathrm u}+v_i$ and
	$\Pi_i^{\mathrm b}(\overline{\pi}_i^{\mathrm{b}'},L,\mathbf D_{-i})>
	\Pi_i^{\mathrm b}(\pi_{\mathrm u}+v_i,L,\mathbf D_{-i})$.
	We distinguish two cases:
	
	\begin{enumerate}
		\item If $\overline\pi_i^{\mathrm b}\ge \pi$, then
		\begin{equation}
			\Pi_i^{\mathrm b}(\pi_{\mathrm u}+v_i,L,\mathbf D_{-i})
			=\Bigl(v_i-\pi_{\mathrm u}-\overline F_V^{-1}\!\Bigl(\tfrac{cG}{L}\Bigr)\Bigr)L. \nonumber
		\end{equation}
		If $\overline{\pi}_i^{\mathrm{b}'}\geq\overline\pi_i^{\mathrm b}$, the deviation does not affect the clearing price (as argued before), hence yields the same payoff:
		\begin{equation}
			\Pi_i^{\mathrm b}(\overline{\pi}_i^{\mathrm{b}'},L,\mathbf D_{-i})
			=\Bigl(v_i-\pi_{\mathrm u}-\overline F_V^{-1}\!\Bigl(\tfrac{cG}{L}\Bigr)\Bigr)L, \nonumber
		\end{equation}
		contradicting strict improvement. The same conclusion holds for $\pi<\overline{\pi}_i^{\mathrm{b}'}<\overline\pi_i^{\mathrm b}$. If instead $\overline{\pi}_i^{\mathrm{b}'}<\pi$, then
		$\Pi_i^{\mathrm b}(\overline{\pi}_i^{\mathrm{b}'},L,\mathbf D_{-i})=-\pi_{\mathrm u}L$,
		again contradicting the assumed profitable deviation.
		
		\item If $\overline\pi_i^{\mathrm b}<\pi$, then
		\begin{equation}
			\Pi_i^{\mathrm b}(\pi_{\mathrm u}+v_i,L,\mathbf D_{-i})=-\pi_{\mathrm u}L. \nonumber
		\end{equation}
		For any $\overline{\pi}_i^{\mathrm{b}'}<\pi$ (including $\overline{\pi}_i^{\mathrm{b}'}<\overline\pi_i^{\mathrm b}$ or $\overline\pi_i^{\mathrm b}<\overline{\pi}_i^{\mathrm{b}'}<\pi$), the payoff remains $-\pi_{\mathrm u}L$, contradicting strict improvement. If $\overline{\pi}_i^{\mathrm{b}'}>\pi$, then
		\begin{equation}
			\Pi_i^{\mathrm b}(\overline{\pi}_i^{\mathrm{b}'},L,\mathbf D_{-i})
			=\Bigl(v_i-\pi_{\mathrm u}-\overline F_V^{-1}\!\Bigl(\tfrac{cG}{L}\Bigr)\Bigr)L. \nonumber
		\end{equation}
		For this to exceed $-\pi_{\mathrm u}L$, it must hold that
		\begin{align}\label{eq:pr:case2}
			\Bigl(v_i-\pi_{\mathrm u}-\overline F_V^{-1}\!\Bigl(\tfrac{cG}{L}\Bigr)\Bigr)L+\pi_{\mathrm u}L>0,
		\end{align}
		which implies $v_i-\overline F_V^{-1}\!\bigl(\tfrac{cG}{L}\bigr)>0$. However, since we are in the case that $\overline\pi_i^{\mathrm b}<\pi$, it gives
		\[
		\pi_{\mathrm u}+v_i-\pi=\pi_{\mathrm u}+v_i-\Bigl(\pi_{\mathrm u}+\overline F_V^{-1}\!\Bigl(\tfrac{cG}{L}\Bigr)\Bigr)<0,
		\]
		and hence $v_i-\overline F_V^{-1}(\tfrac{cG}{L})<0$, contradicting~\eqref{eq:pr:case2}. Therefore, no buyer or seller admits a profitable unilateral deviation, and the strategies in Table~\ref{tab:rt:bid} form an NE in this case. The single-product real-time market follows by setting $v_i \equiv 0$.
	\end{enumerate}
\end{IEEEproof}

Having shown that the bidding outcomes in Table~\ref{tab:rt:bid} constitute an NE and clear the market, we compare them with the CE outcomes in Table~\ref{tab:rt}. Since the two coincide, the proposed NE of the bidding process is equivalent to the CE characterized in Lemma~\ref{prop:eq:rt} for the real-time mechanisms. We next establish the same equivalence for the contract-based market.

\subsection{Nash Equilibrium of the Contract-Based Market Mechanism and Its Equivalence to the Associated CE of Lemma~\ref{lemma:ne:cb}}
\begin{lemma}[NE for the contract-based market] \label{lemma:ne:cb:bid}
	The following states an NE for the contract-based market and the corresponding clearing outcomes:
	\begin{align}
		&(\underline \pi_i^\mathrm{s},\overline q_i^\mathrm{s}) = (0, c),& q_i^\mathrm{s} = c, \qquad & \qquad i \in \mathcal I_\mathrm{s}, \nonumber \\ 	 
		&(\overline \pi_i^\mathrm{b},\overline q_i^\mathrm{b}) = (\pi,d_{i}^\star(\pi)),&
		q_i^\mathrm{b}=d_{i}^\star(\pi),	 &\qquad i \in \mathcal I_\mathrm{b}, \nonumber
	\end{align}
	and $\pi$ is the solution to 
	$
	c = \widehat d^\star(\pi). 	
	$ 
\end{lemma}
\begin{IEEEproof}[Proof of Lemma~\ref{lemma:ne:cb:bid}]
	We first verify that, under the proposed strategies, the claimed market outcome is valid by showing it solves~\eqref{opt:mc}. To this end, note that the outcome clearly satisfies all constraints of~\eqref{opt:mc} and is therefore feasible. Moreover, under the assumed strategies, the objective in~\eqref{opt:mc} is bounded by
	\begin{equation}
		\int_{\mathcal I_\mathrm{b}} \pi q_i^\mathrm{b}\, \mathrm{d}i \leq \int_{\mathcal I_\mathrm{b}} \pi d_{i}^\star(\pi)\, \mathrm{d}i.\nonumber
	\end{equation}
	Since the claimed outcome attains this bound, it is optimal for~\eqref{opt:mc}.
	
	To establish that these strategies form an NE, we first show that no supplier has a profitable unilateral deviation, using the same argument as in the real-time market proof. Since each supplier is infinitesimal, a unilateral deviation cannot affect the clearing price. Therefore, increasing the submitted minimum price cannot increase the clearing price, while reducing the submitted quantity cannot increase the supplier's payoff.
	For consumers, fix any buyer $i\in\mathcal I_{\mathrm b}$. Since an individual buyer is infinitesimal, its unilateral deviation does not affect the market-clearing price $\pi$. Therefore, after any unilateral deviation, the buyer's payoff is evaluated at the same clearing price $\pi$. By definition, $d_i^\star(\pi)$ maximizes the buyer payoff in~\eqref{eq:payoff:cb} at price $\pi$ over all feasible quantities. Hence, for any quantity $q_i^{\mathrm b}\ge0$,
	\begin{equation}
		\Pi_i^{\mathrm b}(d_i^\star(\pi),\pi)
		\ge
		\Pi_i^{\mathrm b}(q_i^{\mathrm b},\pi).\nonumber
	\end{equation}
	Now consider any deviation $(\overline\pi_i^{\mathrm{b}'},\overline q_i^{\mathrm{b}'})$. If $\overline\pi_i^{\mathrm{b}'}<\pi$, then the buyer is not cleared and obtains the payoff associated with $q_i^{\mathrm b}=0$, which cannot exceed the payoff at $d_i^\star(\pi)$. If $\overline\pi_i^{\mathrm{b}'}>\pi$, then the buyer may be cleared up to some quantity $q_i^{\mathrm b}\le \overline q_i^{\mathrm{b}'}$, but the clearing price remains $\pi$; therefore, the resulting payoff is still bounded above by the payoff at $d_i^\star(\pi)$. If $\overline\pi_i^{\mathrm{b}'}=\pi$, the same conclusion follows because $d_i^\star(\pi)$ is already an optimal quantity at price $\pi$. Thus, no buyer has a profitable unilateral deviation. Therefore, no buyer or seller has a profitable unilateral deviation from the strategies given in Lemma~\ref{lemma:ne:cb:bid}, and they constitute an NE. 
\end{IEEEproof}

Comparing the market outcomes in Lemma~\ref{lemma:ne:cb:bid} with the CE characterization in Lemma~\ref{lemma:ne:cb} shows that the proposed NE coincides with the CE for the contract-based market.

\section{Proofs for Section~\ref{section3}} \label{appA}

\begin{IEEEproof}[Proof of Lemma~\ref{prop:eq:rt}]
	For each case presented in Table~\ref{tab:rt}, we demonstrate that the outcomes satisfy the CE conditions:
	\begin{enumerate}
		\item Abundant supply, \texttt{prt}:
		If $\pi = 0$, any feasible $q_i^{\mathrm{s}}$, including $q_i^{\mathrm{s}} = L$, maximizes $\Pi_i^\mathrm{s}(q, 0)$ for each solar owner $i \in \mathcal{I}_{\mathrm{s}}$. Moreover, $q_i^{\mathrm{b}} = L$ uniquely maximizes $\Pi_i^\mathrm{b}(q, 0)$ for each consumer $i \in \mathcal{I}_{\mathrm{b}}$, and together these choices satisfy the market clearing condition.
		
		\item Abundant supply, \texttt{srt}:
		Although $\Pi_i^\mathrm{b}(q,\pi)$ differs from the previous case, the market outcome remains the same, and the previous proof still applies.
		
		\item Limited supply, \texttt{prt}:
		Given $\pi = \pi_\mathrm{u}+\overline F_V ^{-1}\left(\frac{cG}{L}\right)$, $q_i^\mathrm{b}$ can be obtained as
		\begin{equation}
			q_i^\mathrm{b} \in \argmax_{q \in \mathcal Q^\mathrm{b}_i} \left(v_i- \pi_\mathrm{u}-\overline F_V ^{-1}\left(\frac{cG}{L}\right)\right) q - \pi_\mathrm{u} (L-q). \nonumber
		\end{equation}
		This is a linear optimization problem over \mehdi{the compact set $[0,L]$, so the optimum lies at either endpoint, determined by the slope:}
		\begin{equation}
			\frac{\mathrm{d} \, \Pi_i^\mathrm{b}(q,\pi_\mathrm{u}+\overline F_V ^{-1}\left(\frac{cG}{L}\right))} {\mathrm{d} q} = v_i-\overline F_V ^{-1}\left(\frac{cG}{L}\right). \nonumber  
		\end{equation}
		If $\overline{\pi}_i^\mathrm{b}<\pi$, then 
		$v_i<\overline{F}_V^{-1}\left(\frac{cG}{L}\right)$, so the slope is negative and the unique maximizer is $q_i^\mathrm{b}=0$. If $\overline{\pi}_i^\mathrm{b}>\pi$, then the slope is positive and the unique maximizer is $q_i^\mathrm{b}=L$. If $\overline{\pi}_i^\mathrm{b}=\pi$, then the slope is zero, so every $q_i^\mathrm{b}\in[0,L]$ is optimal; under the allocation rule used in Table~\ref{tab:rt}, we select $q_i^\mathrm{b}=L$. 
		At this price, the unique maximizer of $\Pi_i^\mathrm{s}(q,\pi)$ for each solar owner $i\in\mathcal I_\mathrm{s}$ is $q_i^\mathrm{s}=cG$. Given the resulting supply and demand decisions, the market clearing condition is satisfied:
		\[
		\int_{\mathcal I_\mathrm{s}} cG \, \mathrm{d}i
		=
		\int_{\mathcal I_\mathrm{b}} L\,\mathbb{1}\left\{v_i\geq \overline F_V^{-1}\left(\frac{cG}{L}\right)\right\}\,\mathrm{d}i.
		\]
		The equality follows from the definition of $\overline F_V^{-1}(\cdot)$ and the continuity of $F_V$.
		
		\item Limited supply, \texttt{srt}: Given $\pi = \pi_\mathrm{u}$, the payoff for each consumer $i \in \mathcal{I}_\mathrm{b}$ is $\Pi_i^\mathrm{b}(q_i^\mathrm{b}, \pi) = -\pi_\mathrm{u}L$, which is independent of $q_i^\mathrm{b}$. Hence, any $q_i^\mathrm{b} \in \mathcal{Q}^\mathrm{b}_i$, including $q_i^\mathrm{b} = cG$, maximizes the consumer's payoff. Moreover, for each solar owner $i \in \mathcal{I}_\mathrm{s}$, the unique maximizer of $\Pi_i^\mathrm{s}(q, \pi)$ is $q_i^\mathrm{s} = cG$. Together, these choices also satisfy the market clearing condition.
	\end{enumerate}
	
	\mehdi{The results in Table~\ref{tab:rt} yield $\Pi^\mathrm{s}_i(q_i^\mathrm{s}, \pi)$ for all $i \in \mathcal{I}_{\mathrm{s}}$, which can then be used in~\eqref{prt:total:rev} to derive~\eqref{prt:rev} and its special case~\eqref{srt:rev}.}
\end{IEEEproof} 

\begin{IEEEproof}[Proof of Lemma~\ref{demand:cb}]
	We begin by observing that in~\eqref{eq:payoff:cb}, the term 
	$v_i\,\mathbb {E}\,\min\{q_i^\mathrm{b}G,L\}$ can be rewritten as
	$v_iL-v_i\,\mathbb E(L-q_i^\mathrm{b}G)_+$. Therefore, up to the constant term $v_iL$, the buyer's payoff can be written as
	\begin{equation}
		-(\pi_{\mathrm u}+v_i)\mathbb E(L-q_i^\mathrm{b}G)_+-\pi q_i^\mathrm{b}. \nonumber
	\end{equation}
	Since $(L-q_i^\mathrm{b}G)_+$ is convex in $q_i^\mathrm{b}$, the buyer's payoff is concave in $q_i^\mathrm{b}$. Hence, the first-order optimality condition, interpreted using the generalized inverse in~\eqref{eq:funcg_inverse}, is necessary and sufficient for optimality.
	
	For $q_i^\mathrm{b}>0$, using the continuity of the distribution of $G$, differentiating $\Pi_i^\mathrm{b}(q_i^\mathrm{b},\pi)$ with respect to $q_i^\mathrm{b}$ gives
	\begin{equation}
		\frac{\partial \Pi_i^\mathrm{b}(q_i^\mathrm{b},\pi)}{\partial q_i^\mathrm{b}}
		=
		(\pi_{\mathrm u}+v_i)\mu_{G_{\mathrm{tr}}}(q_i^\mathrm{b})-\pi,
		\nonumber
	\end{equation}
	where $\mu_{G_{\mathrm{tr}}}(\cdot)$ is defined in~\eqref{eq:funcg}. If
	$\pi/(\pi_{\mathrm u}+v_i)>\mathbb E[G]$, this derivative is negative at $q_i^\mathrm{b}=0$, and hence the optimal demand is $q_i^\mathrm{b}=0$. Otherwise, the optimal demand is characterized by
	\begin{equation}
		\mu_{G_{\mathrm{tr}}}(q_i^\mathrm{b})
		=
		\frac{\pi}{\pi_{\mathrm u}+v_i}.
		\nonumber
	\end{equation}
	Using the extended generalized inverse in~\eqref{eq:funcg_inverse_tilde}, this gives
	\begin{equation}
		d_i^\star(\pi)
		=
		\tilde{\mu}_{G_{\mathrm{tr}}}^{-1}
		\left(
		\frac{\pi}{\pi_{\mathrm u}+v_i}
		\right),
		\nonumber
	\end{equation}
	which yields~\eqref{eq:prop1:dinverse}. Integrating $d_i^\star(\pi)$ over $i\in\mathcal I_{\mathrm b}$ gives~\eqref{d:star:eq}.
\end{IEEEproof}

\begin{IEEEproof}[Proof of Lemma~\ref{lemma:ne:cb}]
	By Lemma~\ref{demand:cb}, $d_i^\star(\pi)$ is an optimizer of the buyer problem, i.e.,
	\[
	d_i^\star(\pi)\in \arg\max_{q\in\mathcal Q_i^\mathrm{b}}
	\Pi_i^\mathrm{b}(q,\pi).
	\]
	Moreover, for any $\pi>0$, $q_i^\mathrm{s}=c$ is the unique maximizer of $\Pi_i^\mathrm{s}(q_i^\mathrm{s},\pi)$ over the seller's feasible set, while for $\pi=0$, it is a maximizer. Given that $c=\widehat d^\star(\pi)$, the resulting supply and demand decisions satisfy both individual optimality and the market-clearing condition. These outcomes yield $\Pi_i^\mathrm{s}(q_i^\mathrm{s},\pi)$ for all $i\in\mathcal I_{\mathrm{s}}$, which can then be used in~\eqref{cb:total:rev} to derive~\eqref{cb:rev}.
\end{IEEEproof}

\section{Proofs for section~\ref{section4}}
\subsection{Proof of Lemma~\ref{lemma:nec} and Proposition~\ref{prop:optc}}

\begin{IEEEproof}[Proof of Lemma~\ref{lemma:nec}]
	For any positive feasible aggregate capacity $c$ satisfying~\eqref{eq:nec:gen}, substituting~\eqref{eq:nec:gen} into~\eqref{eq:inv:po} implies that an investor who invests earns zero net profit at the aggregate equilibrium capacity. Consider any investor $i\in\mathcal I_\mathrm{inv}$. Since individual investors are infinitesimal, a unilateral change in $x_i$ does not affect the aggregate capacity $c$. If $x_i=0$, then the investor obtains payoff zero, and deviating to $x_i'=1$ also yields zero payoff by~\eqref{eq:inv:po}. If $x_i=1$, then the investor obtains zero payoff, and deviating to $x_i'=0$ also yields zero payoff. Therefore, no investor has a profitable unilateral deviation, and~\eqref{eq:nec:gen} captures the NE condition of the investment game.
	
	Substituting the left-hand side of~\eqref{eq:nec:gen} with~\eqref{prt:rev} and~\eqref{srt:rev} yields~\eqref{eq:eqc:prt} and~\eqref{eq:eqc:srt}, respectively. Similarly, applying~\eqref{cb:rev} to the left-hand side of~\eqref{eq:nec:gen} gives
	\[
	\widetilde T \pi c=\pi_0 c.
	\]
	For a positive equilibrium capacity, this implies $\pi=\pi_0/\widetilde T$. Then, using Lemma~\ref{demand:cb}, market clearing requires $c=\widehat d^\star(\pi)$, and therefore
	\[
	c_\mathrm{cb}^\mathrm{ne}
	=
	\widehat d^\star\left(\frac{\pi_0}{\widetilde T}\right),
	\]
	which gives~\eqref{eq:eqc:cb} and completes the proof.
\end{IEEEproof}

\begin{IEEEproof}[Proof of Proposition~\ref{prop:optc}]
	First, note that \( v_i \) for all \( i \in \mathcal{I}_\mathrm{b} \) can be viewed as realizations of a random variable \( V \) with cumulative distribution function \( F_V \) and density \( f_V \), as described in Section~\ref{common:models}.
	
	To compute \( v^\star(c, G) \), we solve~\eqref{opt:swo}, which we divide into two cases:
	\begin{enumerate}
		\item \emph{Abundant supply $\left(cG> L\right)$}: Inequality~\eqref{opt:swo:cond} becomes
		$
		\int_{0}^{\overline{v}} \sigma(v_i)\,\, \mathrm{d} F_V(v_i) \le 1.  
		$
		Setting $\sigma(v_i) = 1$ for all $i \in \mathcal{I}_\mathrm{b}$ satisfies this constraint and maximizes the objective, since $v_{i}\geq 0$ for all $i \in \mathcal{I}_{\mathrm{b}}$, yielding
		\begin{equation}
			v^\star(c, G) = \int_0^{\overline{v}} v_i \, \,\mathrm{d}F_V(v_i) = \mathbb{E}[V].\nonumber
		\end{equation}
		
		\item \emph{Limited supply $\left(cG \leq L\right)$}: In this case,~\eqref{opt:swo:cond} becomes
		$
		\int_{0}^{\overline{v}} \sigma(v_i)\,\, \mathrm{d} F_V(v_i) \le cG/L 
		$. Given $v_{i} \geq 0$ for all $i \in \mathcal{I}_{\mathrm{b}}$, a $\sigma$ that maximizes the objective is obtained by allocating the limited supply to the portion of consumers with higher solar premiums, leading to an optimal solution
		\begin{equation*}
			\sigma(v_i) = \begin{cases}
				1, &\mbox{if } v_i \geq F_{V}^{-1}\left(1-\frac{cG}{L}\right),\\
				0, &\mbox{otherwise.}
			\end{cases}
		\end{equation*}
		This solution yields the corresponding optimal value of~\eqref{opt:swo:obj} as 
		\[
		v^\star(c,G)=\int_{F_{V}^{-1}\left(1-\frac{cG}{L}\right)}^{\overline{v}} v_i\,\, \mathrm{d}F_V(v_i).
		\]
	\end{enumerate}
	
	Substituting the expressions for $v^\star(c, G)$ into the optimization problem~\eqref{social:cap} yields: 
	\begin{align*}
		\max_{c \in \mathbb{R}_+} \quad \widetilde{T} \, \mathbb{E} \Big[\, &L \, \mathbb{E}[V] - L \, \mathbb{1}\{cG \leq L\} 
		\int_0^{F_{V}^{-1}\left(1-\frac{cG}{L}\right)} \!  v_i \,\, \mathrm{d}F_V(v_i) \nonumber \\
		&- \pi_\mathrm{u} (L - cG)_+ \,\Big] - \pi_0 c.
	\end{align*}
	
	Under our assumptions, we can verify that the second derivative of the objective with respect to $c$ is given by
	\begin{equation*}
		-\widetilde T \int_0^{\min\{L/c,\overline g\}} \frac{g^2}{L} \, 
		\frac{f_G(g)}
		{f_V\left(F_V^{-1}\left(1 - \tfrac{cg}{L}\right)\right)} \, \mathrm{d}g \leq 0,
	\end{equation*}
	which implies that the objective function is concave in $c$. Since the objective is concave in $c$ and the feasible set $\mathbb R_+$ is convex, any point satisfying the first-order optimality condition is globally optimal. Moreover, under~\eqref{eq:pi0cond}, the objective has a positive marginal value near $c=0$, while its marginal value becomes negative for sufficiently large $c$; hence the optimum is interior. Applying the first-order optimality condition and using the relation $F_V^{-1}(1-z)=\overline F_V^{-1}(z)$ under our inverse convention yield the following analytical characterization of \( c_{\mathrm{opt}} \) as the solution to
	\[
	\widetilde{T} \, \mathbb{E} \left[ \left(\pi_{\mathrm{u}} + \overline{F}_V^{-1}\left(\frac{cG}{L}\right)\right) G \, \mathbb{1}\{cG \leq L\} \right] = \pi_0.
	\]
\end{IEEEproof}

\subsection{Proof of Theorem~\ref{thm:main}}\label{app:ps}

For the remainder of the analysis, we introduce the notation $\widetilde V$ to denote the random variable associated with the distribution of the baseline values $\{\tilde v_i\}_{i\in\mathcal I_{\mathrm b}}$ across buyers, with CDF $F_{\widetilde V}$, complementary CDF $\overline F_{\widetilde V}=1-F_{\widetilde V}$, density $f_{\widetilde V}$, and mean $\mathbb E[\widetilde V]$. Under the scaling $v_i=\epsilon\tilde v_i$, the corresponding solar-premium random variable satisfies $V=\epsilon\widetilde V$. Hence, for $\epsilon>0$,
\[
\mathbb E[V]
=
\epsilon\mathbb E[\widetilde V],
\qquad
\overline F_V^{-1}(p)
=
\epsilon\overline F_{\widetilde V}^{-1}(p),
\quad p\in[0,1],
\]
while for $\epsilon=0$, $V\equiv0$.

We establish each case of Theorem~\ref{thm:main} individually.

\begin{IEEEproof}[General case]
	Since~\eqref{eq:eqc:prt} and~\eqref{eq:opt} are identical, we have
	\[
	c^\mathrm{ne}_\mathrm{prt} = c_\mathrm{opt}.
	\]
	Furthermore, for any $\epsilon \geq 0$, $c^\mathrm{ne}_\mathrm{prt}$ solves the modified version of~\eqref{eq:eqc:prt}:
	\begin{equation}\label{eq:prt:expect}
		\pi_0 - \widetilde{T} \, \mathbb{E}\left[\left(\pi_{\mathrm{u}} + \epsilon \, \overline{F}_{\widetilde{V}}^{-1}\left(\frac{cG}{L}\right)\right) G \, \mathbb{1}\{cG \leq L\} \right] = 0.
	\end{equation}
	Similarly, \( c^\mathrm{ne}_\mathrm{srt} \) is characterized by~\eqref{eq:eqc:srt}, or equivalently,
	\begin{equation}\label{eq:srt:expect}
		\widetilde T\pi_{\mathrm u}\mu_{G_{\mathrm{tr}}}(c^\mathrm{ne}_\mathrm{srt})=\pi_0.
	\end{equation}
Evaluating~\eqref{eq:prt:expect} at $c=c^\mathrm{ne}_\mathrm{prt}$ gives
\begin{equation}
	\widetilde T\,\mathbb{E}\left[\left(\pi_{\mathrm{u}} + \epsilon \, \overline{F}_{\widetilde{V}}^{-1}\left(\frac{c^\mathrm{ne}_\mathrm{prt}G}{L}\right)\right)G \mathbb{1}\{c^\mathrm{ne}_\mathrm{prt}G \leq L\} \right]
	=
	\pi_0.
	\label{eq:prt:eval}
\end{equation}
Moreover, since $\epsilon\ge0$ and $\overline{F}_{\widetilde V}^{-1}(\cdot)\ge0$, we have the pointwise inequality
\begin{equation}\label{prt:srt:integrand:homo}
\pi_{\mathrm u}G\mathbb 1\{c^\mathrm{ne}_\mathrm{prt}G\le L\}
\le
\left(\!\pi_{\mathrm{u}} \!+\! \epsilon \, \overline{F}_{\widetilde{V}}^{-1}\!\left(\frac{c^\mathrm{ne}_\mathrm{prt}G}{L}\right)\!\right)\!G \mathbb{1}\{c^\mathrm{ne}_\mathrm{prt}G \leq L\}\!.
\end{equation}
Taking expectations and multiplying by $\widetilde T$ yields
\begin{align}
	\widetilde T\pi_{\mathrm u}\mu_{G_{\mathrm{tr}}}(c^\mathrm{ne}_\mathrm{prt})
	&=
	\widetilde T\,\mathbb E\left[
	\pi_{\mathrm u}G\mathbb 1\{c^\mathrm{ne}_\mathrm{prt}G\le L\}
	\right]\nonumber\\
	&\le
	\!\widetilde T\,\mathbb{E}\left[\!\left(\!\pi_{\mathrm{u}} \!+\! \epsilon \, \overline{F}_{\widetilde{V}}^{-1}\!\!\left(\frac{c^\mathrm{ne}_\mathrm{prt}G}{L}\right)\!\right)\!\!G \mathbb{1}\{c^\mathrm{ne}_\mathrm{prt}G \leq L\} \!\right]\nonumber\\
	&=\pi_0,
\end{align}
where the last equality follows from~\eqref{eq:prt:eval}.
	Combining this inequality with~\eqref{eq:srt:expect} yields
	\[
	\mu_{G_{\mathrm{tr}}}(c^\mathrm{ne}_\mathrm{prt})
	\le
	\mu_{G_{\mathrm{tr}}}(c^\mathrm{ne}_\mathrm{srt}).
	\]
	As discussed in the uniqueness argument for Lemma~\ref{lemma:nec}, the relevant positive solutions lie in the region where $\mu_{G_{\mathrm{tr}}}(\cdot)$ is strictly decreasing. Therefore,
	\[
	c^\mathrm{ne}_\mathrm{srt}\le c^\mathrm{ne}_\mathrm{prt}.
	\]
	Hence,
	\[
	c^\mathrm{ne}_\mathrm{srt}\le c^\mathrm{ne}_\mathrm{prt}=c_\mathrm{opt}.
	\]
	
		It remains to show that $c^\mathrm{ne}_\mathrm{srt}\le c^\mathrm{ne}_\mathrm{cb}$. By Lemma~\ref{demand:cb}, the individual demand of buyer $i$ in the contract-based market at price $\pi_0/\widetilde T$ is characterized by the first-order condition
	\begin{equation}\label{cb:consumer:demand:epsilon}
		-\left(\pi_\mathrm{u}+\epsilon \tilde{v}_{i}\right) 
		\mathbb{E} \left[G \, \mathbb{1}\left\{d_i^\star\left(\frac{\pi_0}{\widetilde T}\right) G \leq L\right\} \right]
		+ \frac{\pi_0}{\widetilde{T}} = 0.
	\end{equation}
	Equivalently, using the generalized inverse representation in Lemma~\ref{demand:cb}, we have
	\[
	d_i^\star\left(\frac{\pi_0}{\widetilde T}\right)
	=
	\tilde{\mu}_{G_{\mathrm{tr}}}^{-1}
	\left(
	\frac{\pi_0/\widetilde T}{\pi_{\mathrm u}+\epsilon\tilde v_i}
	\right).
	\]
	Since $\epsilon\tilde v_i\ge0$, we have
	\[
	\frac{\pi_0/\widetilde T}{\pi_{\mathrm u}+\epsilon\tilde v_i}
	\le
	\frac{\pi_0}{\widetilde T\pi_{\mathrm u}}.
	\]
	Using~\eqref{eq:srt:expect}, this becomes
	\[
	\frac{\pi_0/\widetilde T}{\pi_{\mathrm u}+\epsilon\tilde v_i}
	\le
	\mu_{G_{\mathrm{tr}}}(c^\mathrm{ne}_\mathrm{srt}).
	\]
	Because $\mu_{G_{\mathrm{tr}}}(\cdot)$ is non-increasing, its generalized inverse is order-reversing. Therefore,
	\[
	d_i^\star\left(\frac{\pi_0}{\widetilde T}\right)
	\ge
	c^\mathrm{ne}_\mathrm{srt},
	\qquad i\in\mathcal I_{\mathrm b}.
	\]
	Integrating over all buyers gives
	\[
	c^\mathrm{ne}_\mathrm{cb}
	=
	\int_{\mathcal I_{\mathrm b}}
	d_i^\star\left(\frac{\pi_0}{\widetilde T}\right)\,\mathrm di
	\ge
	\int_{\mathcal I_{\mathrm b}}c^\mathrm{ne}_\mathrm{srt}\,\mathrm di
	=
	c^\mathrm{ne}_\mathrm{srt}.
	\]
	Thus, for any $\epsilon\ge0$,
	\[
	c^\mathrm{ne}_\mathrm{srt} \le c^\mathrm{ne}_\mathrm{prt}=c_\mathrm{opt},
	\qquad
	c^\mathrm{ne}_\mathrm{srt} \le c^\mathrm{ne}_\mathrm{cb}.
	\]
	\end{IEEEproof}

\begin{IEEEproof}[No solar premium]
	In this case, i.e., $\epsilon=0$,~\eqref{eq:prt:expect} and~\eqref{eq:eqc:srt} become identical. Therefore,
	\[
	c_{\mathrm{opt}}=c^\mathrm{ne}_\mathrm{prt}=c^\mathrm{ne}_\mathrm{srt}.
	\]
	It remains to show $c^\mathrm{ne}_\mathrm{cb}=c^\mathrm{ne}_\mathrm{srt}$. 
	
	To compute the equilibrium capacity for the contract-based market, as defined in~\eqref{eq:eqc:cb}, we use Lemma~\ref{demand:cb}. When $\epsilon=0$, the individual demand of buyer $i$ at price $\pi_0/\widetilde T$ becomes
	\[
	d_i^\star\left(\frac{\pi_0}{\widetilde T}\right)
	=
	\tilde{\mu}_{G_{\mathrm{tr}}}^{-1}
	\left(
	\frac{\pi_0}{\widetilde T\pi_{\mathrm u}}
	\right),
	\]
	which is identical for all $i\in\mathcal I_{\mathrm b}$. Hence,
	\[
	c^\mathrm{ne}_\mathrm{cb}
	=
	\widehat d^\star\left(\frac{\pi_0}{\widetilde T}\right)
	=
	d_i^\star\left(\frac{\pi_0}{\widetilde T}\right),
	\qquad i\in\mathcal I_{\mathrm b}.
	\]
	On the other hand,~\eqref{eq:eqc:srt} implies
	\[
	\mu_{G_{\mathrm{tr}}}(c^\mathrm{ne}_\mathrm{srt})
	=
	\frac{\pi_0}{\widetilde T\pi_{\mathrm u}}.
	\]
	Therefore, on the relevant strictly decreasing region of $\mu_{G_{\mathrm{tr}}}(\cdot)$,
	\[
	d_i^\star\left(\frac{\pi_0}{\widetilde T}\right)
	=
	c^\mathrm{ne}_\mathrm{srt}.
	\]
	Thus,
	\[
	c^\mathrm{ne}_\mathrm{cb}=c^\mathrm{ne}_\mathrm{srt}.
	\]
	Combining this with $c_{\mathrm{opt}}=c^\mathrm{ne}_\mathrm{prt}=c^\mathrm{ne}_\mathrm{srt}$ gives
	\[
	c^\mathrm{ne}_\mathrm{srt}
	=
	c^\mathrm{ne}_\mathrm{prt}
	=
	c_\mathrm{opt}
	=
	c^\mathrm{ne}_\mathrm{cb}.
	\]
\end{IEEEproof}

\begin{IEEEproof}[Small solar premium]
	Define $c_0 := c_{\mathrm{srt}}^{\mathrm{ne}}$. The main steps of the proof are summarized as follows: 1) we treat the left-hand sides of~\eqref{eq:prt:expect} and~\eqref{cb:consumer:demand:epsilon} as functions of both $\epsilon$ and $c$, where $\epsilon$ enters through the scaling $v_i=\epsilon\tilde v_i$; 2) we show that if $\epsilon$ lies in a neighborhood of $0$, then the corresponding capacities (i.e., the values of $c$ that solve~\eqref{eq:prt:expect} and~\eqref{cb:consumer:demand:epsilon}) must lie in a neighborhood of $c_0$; 3) we derive the first-order Taylor expansions of these functions at $(\epsilon, c) = (0, c_0)$, and provide explicit bounds on the higher-order terms; and 4) we compare the resulting Taylor expansions term by term.
	
	We defer the detailed proofs of the first three steps to Appendix~\ref{App:c}, and summarize their results below in the form of lemmas.
	\begin{lemma}[First-order expansion of $c_{\mathrm{prt}}^{\mathrm{ne}}$ around $\epsilon = 0$]\label{lemma4}
		For a sufficiently small $\epsilon>0$, we have
		\begin{equation}\label{prt:lemma4}
			c_\mathrm{prt}^{\mathrm{ne}}=c_{0}-\frac{\epsilon \, \mathbb E \left [\overline{F}_{\widetilde V}^{-1}\left(\frac{c_{0}G}{L}\right)\,G\,\mathbb{1} \bigl\{c_{0}G \leq L \bigl\}\right]}{\pi_\mathrm{u}\mu_{G_\mathrm{tr}}'(c_{0})}+O\left(\epsilon^2\right),
		\end{equation}
		\orange{where $\mu_{G_\mathrm{tr}}(\cdot)$ is defined in~\eqref{eq:funcg}.}
	\end{lemma}
	
	\begin{lemma}[First-order expansion of $c_{\mathrm{cb}}^{\mathrm{ne}}$ around $\epsilon = 0$]\label{lemma5}
		For a sufficiently small $\epsilon>0$, we have
		\begin{equation}\label{di:c_0:epsilon:lemma}
			d_i^\star\left(\frac{\pi_0}{\widetilde T}\right)=c_{0}-\frac {\epsilon \tilde v_{i}\mu_{G_{\mathrm{tr}}}(c_{0})} {\pi_\mathrm{u} \mu_{G_\mathrm{tr}}'\left(c_{0}\right)}+O\left(\epsilon^2\right),
		\end{equation}
		and consequently
		\begin{equation}\label{c_0:epsilon:lemma}
			c_\mathrm{cb}^{\mathrm{ne}}=c_{0}-\frac{\epsilon \mathbb E[\widetilde V] \mu_{G_{\mathrm{tr}}}(c_0)} {\pi_\mathrm{u} \mu_{G_\mathrm{tr}}'(c_0)}+O\left(\epsilon^2\right). 
		\end{equation}
	\end{lemma}  
	
	Equipped with these lemmas, we can compare $c_{\mathrm{cb}}^{\mathrm{ne}}$ and $c_{\mathrm{prt}}^{\mathrm{ne}}$ by isolating the non-identical terms in~\eqref{prt:lemma4} and~\eqref{c_0:epsilon:lemma}.
	
\revise{Note that, by the uniqueness argument for Lemma~\ref{lemma:nec}, the strict
inequality in~\eqref{eq:pi0cond} rules out the flat no-excess-generation
region and implies that $c_0>L/\overline g$. Hence, $L/c_0\in(0,\overline g)$,
and the same relation holds for all $c$ sufficiently close to $c_0$.
Therefore, for all such $c$, we can write
\[
\mu_{G_\mathrm{tr}}(c)
=
\int_0^{L/c} g f_G(g)\,\mathrm{d}g.
\]
Under the assumed local continuity of $f_G$ around $L/c_0$, this representation
implies that $\mu_{G_\mathrm{tr}}(\cdot)$ is differentiable at $c_0$.
Applying the Leibniz integral rule gives
\begin{equation}\label{mug_prime_homo}
	\mu_{G_\mathrm{tr}}'(c_0)
	=
	-\frac{L^2}{c_0^3} f_G\left(\frac{L}{c_0}\right)
	<0,
\end{equation}
where the strict inequality follows from the assumed positivity of
$f_G$ at $L/c_0$.
}

	Therefore, the proof is complete if we show
	\begin{equation} \label{50}
		\underbrace{ 
			\mathbb{E} \left[ \overline{F}_{\widetilde V}^{-1}\left( \frac{c_0 G}{L} \right) G \, \mathbb{1} \left\{ c_0 G \leq L \right\} \right]
		}_{:=A}
		<
		\underbrace{
			\mathbb{E}[\widetilde V] \, \mu_{G_{\mathrm{tr}}}(c_0)
		}_{:=B}.
	\end{equation}
	
For notational simplicity, let
\[
\tilde v(p):=\overline{F}_{\widetilde V}^{-1}(p), \qquad p\in[0,1],
\]
denote the decreasing inverse associated with $\overline F_{\widetilde V}$. By the definition of $\overline{F}_{\widetilde V}^{-1}(\cdot)$ and the bounded support $\widetilde V\in[0,\overline v]$, this function maps $[0,1]$ into $[0,\overline v]$ and is non-increasing in $p$.
	Thus,
	\[
	A = \mathbb{E}\left[ \tilde{v}\left( \frac{c_0 G}{L} \right) G \, \mathbb{1}\{ c_0 G \leq L \} \right].
	\]
	
	Since $\tilde{v}(1) = 0$, we have
	$
	\tilde{v}\left( \frac{c_0 G}{L} \right) = \int_{c_0 G / L}^{1} -\tilde{v}'(p) \, \mathrm{d}p,
	$
	which implies
	\begin{equation}\label{A}
		A= \int_{0}^{1} \left(-\tilde v'(p)\right) \, \mathbb E \left [G\,\mathbb{1} \bigl\{\frac{c_0G}{L} \leq p \bigl\}\right] \,\, \mathrm{d} p.    
	\end{equation}
	
	Observe that
	\begin{equation}\label{EV:orig}
		\mathbb{E}[\widetilde V]=-\int_{0}^{\overline{v}}\tilde v_i \, \mathrm{d}\overline{F}_{\widetilde V}(\tilde v_i)=\int_{0}^{\overline{v}} \overline{F}_{\widetilde V}(\tilde v_i) \, \,\mathrm{d}\tilde v_i, 
	\end{equation}
	where the first equality follows from the definition of expectation and the identity \( \mathrm{d}F_{\widetilde V} = -\mathrm{d}\overline{F}_{\widetilde V} \), and the second equality follows from integration by parts.
	
	Next, recall that by the integral identity for inverse functions, if \( f : [a, b] \to \mathbb{R} \) is strictly monotonic, then the following identity holds:
	\[
	\int_a^b f(x) \, dx = b f(b) - a f(a) - \int_{f(a)}^{f(b)} f^{-1}(y) \,\, dy.
	\]
	Since \( \overline{F}_{\widetilde{V}} \) is strictly monotonic, we can apply this identity and obtain
	\begin{align}
		\int_{0}^{\overline{v}} \overline{F}_{\widetilde V}(\tilde v_i) \, \mathrm{d}\tilde v_i
		= &\left(\overline{v}\right)
		\underset{0}{\underbrace{\overline{F}_{\widetilde V}(\overline{v})}} 
		- 0
		- \int_{\overline{F}_{\widetilde V}(0)}^{\overline{F}_{\widetilde V}(\overline{v})}  
		\overline{F}_{\widetilde V}^{-1}\left( p\right) \, \,\mathrm{d}p
		\nonumber \\
		&=\int_{\overline{F}_{\widetilde V}(\overline{v})}^{\overline{F}_{\widetilde V}(0)}  
		\underset{\tilde v(p)}{\underbrace{\overline{F}_{\widetilde V}^{-1}\left( p\right)}} \, \,\mathrm{d}p
		= \int_{0}^{1} \tilde v(p)\, \mathrm{d}p.\label{42}
	\end{align}
	By integration by parts, we have
	\begin{equation}\label{43}
		\int_{0}^{1} \tilde v(p) \, \mathrm{d}p = -\int_{0}^{1} \tilde v'(p)\, p \,\, \mathrm{d}p.
	\end{equation}
	
	Combining~\eqref{EV:orig},~\eqref{42}, and~\eqref{43}, we obtain
	\[
	\mathbb{E}[\widetilde V] = -\int_{0}^{1} \tilde v'(p)\, p \, \,\mathrm{d}p.
	\]
	Substituting this result into $B$ then gives
	\begin{equation}\label{B}
		B=\int_{0}^{1} \left(-\tilde v'(p)\right) \,\mathbb{E}\left[p\,G\, \mathbb{1} \bigl\{\frac{c_{0}G}{L} \leq 1 \bigl\}\right]\,\,\mathrm{d}p.
	\end{equation} 
	
	It follows from~\eqref{A} and~\eqref{B} that, since $-\tilde v'(p)>0$, establishing $A < B$ reduces to showing:
	\[
	\mathbb{E} \left[ G \, \mathbb{1} \left\{ \frac{c_0 G}{L} \leq p \right\} \right] 
	< 
	\mathbb{E} \left[ p\, G \, \mathbb{1} \left\{ \frac{c_0 G}{L} \leq 1 \right\} \right].
	\]
	Since $p\in[0,1]$ and $L/c_0<\overline g$, we have $Lp/c_0\leq L/c_0<\overline g$. Therefore,
	\begin{equation*}
		\mathbb{E} \left[G\, \mathbb{1} \left\{ \frac{c_0 G}{L} \leq p \right\} \right] = \int_0^{Lp / c_0} g \,f_G(g) \, \,\mathrm{d}g.
	\end{equation*}
	Furthermore, under assumption~\eqref{fg:constant}, we have  
	$\int_0^{Lp / c_0} g f_G(g) \, \,\mathrm{d}g \leq (1 + \delta) r_0 \frac{1}{2} \left( \frac{Lp}{c_0} \right)^2$,  
	which leads to the following upper bound for $A$:
	\begin{equation*}  
		A \leq (1 + \delta) r_0 \frac{1}{2} \left( \frac{L}{c_0} \right)^2 \int_0^1 -\tilde{v}'(p) \, p^2 \,\, \mathrm{d}p.
	\end{equation*}
	Similarly,  
	$\mathbb{E} \left[p\,G \, \mathbb{1} \left\{ \frac{c_0 G}{L} \leq 1 \right\} \right] = \int_0^{L / c_0} p \, g \,f_G(g) \, \,\mathrm{d}g$,  
	and applying assumption~\eqref{fg:constant} again, we obtain the lower bound for \( B \):
	\begin{equation*}
		B \geq (1 - \delta) r_0 \frac{1}{2} \left( \frac{L}{c_0} \right)^2 \int_0^1 -\tilde{v}'(p) \, p \, \,\mathrm{d}p.
	\end{equation*}
	
We can proceed by defining
\[
\theta
:=
\frac{
	\int_{0}^{1} -\tilde v'(p) \, p^2 \, \mathrm{d}p
}{
	\int_{0}^{1} -\tilde v'(p) \, p \, \mathrm{d}p
},
\]
where $0<\theta<1$ since $p\in[0,1]$. One can find a $\delta>0$ such that
\begin{equation}\label{A:B:theta}
	\frac{1+\delta}{1-\delta}\theta
	\leq
	\frac{\theta+1}{2}.
\end{equation}
Multiplying both sides of~\eqref{A:B:theta} by
$\frac{r_0}{2}\left(\frac{L}{c_0}\right)^2$
and applying the derived bounds for $A$ and $B$ yields
\begin{equation}
	A
	\leq
	\frac{\theta+1}{2}B
	<
	B,
	\nonumber
\end{equation}
which proves~\eqref{50}.
Thus, we conclude:
\begin{equation}
	\mathbb E[\widetilde V]\mu_{G_{\mathrm{tr}}}(c_0)
	\ge
	\frac{2}{\theta+1}
	\mathbb E\left[
	\overline{F}_{\widetilde V}^{-1}
	\left(\frac{c_0G}{L}\right)
	G\,
	\mathbb{1}\{c_0G\le L\}
	\right],
	\nonumber
\end{equation}
which can be used in~\eqref{prt:lemma4} and~\eqref{c_0:epsilon:lemma}
to derive~\eqref{thm:main:case2:detail}, where
\begin{equation*}
	\beta
	:=
	\frac{
		(1-\theta)
		\mathbb E\left[
		\overline{F}_{\widetilde V}^{-1}
		\left(\frac{c_0G}{L}\right)
		G\,
		\mathbb{1}\{c_0G\le L\}
		\right]
	}{
		-(1+\theta)\pi_\mathrm{u}
		\mu_{G_\mathrm{tr}}'(c_0)
	}
	>0.
\end{equation*}
\end{IEEEproof}

\section{Proof of Lemma~\ref{lemma4} and Lemma~\ref{lemma5}}\label{App:c}

\subsection{Proof of Lemma~\ref{lemma4}}

We can use~\eqref{eq:funcg} to reformulate~\eqref{eq:prt:expect}. Define
\begin{equation}\label{prt:eps:g}
	\phi_{\mathrm{prt}}(\epsilon,c)
	:=
	\frac{\pi_{0}}{\widetilde T}
	- \pi_{\mathrm{u}}\,\mu_{G_\mathrm{tr}}(c)
	- \epsilon\, k(c),
\end{equation}
where
\begin{equation}\label{k:def}
	k(c):=\mathbb E \left[ \overline{F}_{\widetilde V}^{-1}\left(\frac{cG}{L}\right) 
	\,G\,\mathbb{1} \bigl\{cG \leq L \bigl\}\right].
\end{equation}
For given values of $\pi_0$, $\pi_{\mathrm{u}}$, and $\epsilon$, the equilibrium capacity $c^\mathrm{ne}_\mathrm{prt}$ is characterized as the solution to
\[
\phi_{\mathrm{prt}}(\epsilon,c)=0.
\] We first show that if $\epsilon$ is sufficiently close to $0$, then the corresponding $c^\mathrm{ne}_\mathrm{prt}$ lies in a neighborhood of $c_0$.

To emphasize the dependence of the solution $c^\mathrm{ne}_\mathrm{prt}$ on $\epsilon$, we write $c^\mathrm{ne}_\mathrm{prt} = \Psi(\epsilon)$, where $\Psi: [0, 1] \to \mathbb{R}$. From~\eqref{eq:thm:p3}, it follows that $\Psi(0) = c_0$. 
Furthermore, the following lemma shows that $\Psi(\epsilon)$ is Lipschitz continuous with constant $L_{\Psi}$.
\begin{lemma}[Lipschitz continuity of $\Psi(\epsilon)$ near $\epsilon = 0$]\label{lemma:Psi:Lipschitz}
	In a neighborhood of $\epsilon = 0$, the function $\Psi(\epsilon)$ is Lipschitz continuous with Lipschitz constant $L_{\Psi}$.
\end{lemma}

Equipped with Lemma~\ref{lemma:Psi:Lipschitz}, we have
\[
|\Psi(\epsilon) - \Psi(0)| \leq L_{\Psi} |\epsilon - 0|,
\]
which implies
\begin{equation}\label{prt:error}
	|c^\mathrm{ne}_\mathrm{prt} - c_0| \leq L_{\Psi} \epsilon.
\end{equation}

In other words, if $\epsilon$ is sufficiently close to $0$, then $c^\mathrm{ne}_\mathrm{prt}$ remains close to $c_0$. This proximity allows us to expand $\phi_{\mathrm{prt}}(\epsilon, c^\mathrm{ne}_\mathrm{prt})$ in a Taylor series at $(0, c_0)$:
\begin{align}\label{prt:Taylor}
	\phi_{\mathrm{prt}}(\epsilon,c^\mathrm{ne}_\mathrm{prt})
	=&-\epsilon\,k\left(c_{0}\right)
	-\pi_{\mathrm{u}}\, (c^\mathrm{ne}_\mathrm{prt}-c_0)\,\mu_{G_\mathrm{tr}}'(c_0)\nonumber \\
	&+O(\lVert (\epsilon ,c^\mathrm{ne}_\mathrm{prt})- (0, c_0)\rVert ^2)=0.
\end{align}
Using~\eqref{prt:error}, we can bound the higher-order term as
\begin{equation}
	\lVert (\epsilon ,c^\mathrm{ne}_\mathrm{prt})- (0, c_0)\rVert ^2
	\leq \epsilon ^{2} \left(1+L_{\Psi}^2\right).
	\nonumber 
\end{equation}
Thus,~\eqref{prt:Taylor} can be written as
\begin{align}
	\phi_{\mathrm{prt}}(\epsilon,c^\mathrm{ne}_\mathrm{prt})
	=&-\epsilon\,k\left(c_{0}\right)
	-\pi_{\mathrm{u}}\, (c^\mathrm{ne}_\mathrm{prt}-c_0)\,\mu_{G_\mathrm{tr}}'(c_0)\nonumber \\
	&+O(\epsilon^{2})=0.\nonumber
\end{align}
Isolating $c^\mathrm{ne}_\mathrm{prt}$ gives
\[
c^\mathrm{ne}_\mathrm{prt}
=
c_0
-
\frac{\epsilon k(c_0)}
{\pi_{\mathrm u}\mu_{G_\mathrm{tr}}'(c_0)}
+
O(\epsilon^2).
\]
Substituting the definition of $k(c_0)$ from~\eqref{k:def} yields~\eqref{prt:lemma4}, which completes the proof of Lemma~\ref{lemma4}. It only remains to prove the Lipschitz-continuity result stated in Lemma~\ref{lemma:Psi:Lipschitz}. $\,\, \,\blacksquare$

\begin{IEEEproof}[Proof of Lemma~\ref{lemma:Psi:Lipschitz}]
	We first show that there exists a $\Delta > 0$, chosen sufficiently small so that for all $c_1, c_2 \in [c_0 - \Delta, c_0 + \Delta]$ with $c_1 > c_2$, the following holds \mehdi{for some $m, M \in \mathbb{R}_{++}$}:
	\begin{equation}\label{Tlim}
		0< m\leq \frac{\phi_{\mathrm{prt}}(\epsilon,c_{1})-\phi_{\mathrm{prt}}(\epsilon,c_{2})}{c_{1}-c_{2}} \leq M. 
	\end{equation}
	According to~\eqref{prt:eps:g}, this inequality is equivalent to:
	\begin{align}\label{63}
		0< m\leq &\pi_{\mathrm{u}} \left(\frac{ \mu_{G_\mathrm{tr}}(c_{2})-\mu_{G_\mathrm{tr}}(c_{1})}{c_{1}-c_{2}}\right)\nonumber\\
		&+\epsilon \left( \frac{k(c_{2})-k(c_{1})} {c_{1}-c_{2}}\right) \leq M.  
	\end{align}
	
	To prove the inequality, note that by the uniqueness argument for Lemma~\ref{lemma:nec}, the strict inequality in~\eqref{eq:pi0cond} implies that $c_0>L/\overline g$. Thus, by choosing $\Delta>0$ sufficiently small, we have $L/c\in(0,\overline g)$ for all $c \in [c_0 - \Delta, c_0 + \Delta]$. Hence, under the bounded-support assumption and the positivity of $f_G$ on $(0,\overline g)$, there exist \mehdi{some $m_1, M_1 \in \mathbb{R}_{++}$} such that
	\begin{equation}\label{gdot:lim}
		-M_1 \leq \pi_{\mathrm{u}} \mu_{G_\mathrm{tr}}'(c) = \pi_{\mathrm{u}} \frac{-L^2}{c^3} f_G\left( \frac{L}{c} \right) \leq -m_1 < 0.
	\end{equation}
	By the Mean Value Theorem \cite{rudin1976principles}, there exists a point $c_{3} \in (c_{2}, c_{1})$ such that:
	\begin{equation}
		\frac{\mu_{G_\mathrm{tr}}(c_{2})-\mu_{G_\mathrm{tr}}(c_{1})}{c_{1}-c_{2}}=-\frac{\mu_{G_\mathrm{tr}}(c_{1})-\mu_{G_\mathrm{tr}}(c_{2})}{c_{1}-c_{2}} = -\mu_{G_\mathrm{tr}}'(c_{3}).\nonumber
	\end{equation}
	Therefore, for all $c_1, c_2 \in [c_0 - \Delta, c_0 + \Delta]$ with $c_1 > c_2$, it follows that   
	\begin{equation}\label{gc:lim}
		0<m_{1}\leq \pi_{\mathrm{u}} \left(\frac{ \mu_{G_\mathrm{tr}}(c_{2})-\mu_{G_\mathrm{tr}}(c_{1})}{c_{1}-c_{2}}\right)\leq M_{1}.   
	\end{equation}
	
	Next, we consider the derivative of $k(c)$, which is given by:
	\begin{equation}
		k'(c)=-\mathbb E \left[\frac{G^{2}}{L} \frac{\mathbb{1} \bigl\{cG \leq L\bigl\}}{f_{\widetilde V}\left(\overline{F}_{\widetilde V}^{-1}\left(\frac{cG}{L}\right)\right)}\right]< 0.\nonumber
	\end{equation} 
	For all $c \in [c_0 - \Delta, c_0 + \Delta]$ and for each fixed $\epsilon > 0$, \mehdi{there exist constants $m_2, M_2 \in \mathbb{R}_{++}$ such that}:
	\begin{equation}\label{kdot:lim}
		- M_2 \leq \epsilon k'(c) \leq - m_2 < 0.
	\end{equation}
	Then, by the Mean Value Theorem, it follows that:
	\begin{equation}\label{kc:lim}
		0 < m_2 \leq \epsilon \left( \frac{k(c_2) - k(c_1)}{c_1 - c_2} \right) \leq M_2.
	\end{equation}
	
	Adding inequalities~\eqref{gc:lim} and~\eqref{kc:lim}, and letting $m = m_1 + m_2$ and $M = M_1 + M_2$, we obtain~\eqref{63}, which in turn implies~\eqref{Tlim}.

	Next, observe that $\phi_{\mathrm{prt}}(\epsilon, c): [0,1] \times \mathbb{R} \to \mathbb{R}$ is continuous, and $[0,1]$ is a compact set. These properties, together with the bounded derivative condition in~\eqref{Tlim}, ensure that \( \phi_{\mathrm{prt}}(\epsilon, c) \) satisfies the conditions of the Lipschitz Implicit Function Theorem (see Theorem 6 of~\cite{border2013notes}) in the considered neighborhood of \( c_0 \).
	Thus, there exists a \emph{unique} function \( \Psi: [0,1] \to \mathbb{R} \) such that $ \phi_{\mathrm{prt}}(\epsilon, \Psi(\epsilon)) = 0$, and \( \Psi \) is \emph{continuous}.  
	To establish that \( \Psi \) is Lipschitz continuous, it suffices to show that there exists a constant \( L_{\Psi} \geq 0 \) such that for all \( \epsilon_1, \epsilon_2 \in [0,1] \),
	\[
	\frac{|\Psi(\epsilon_2) - \Psi(\epsilon_1)|}{|\epsilon_2 - \epsilon_1|} \leq L_{\Psi}.
	\]

	Let $c_1 = \Psi(\epsilon_1)$ and $c_2 = \Psi(\epsilon_2)$. \mehdi{From~\eqref{prt:eps:g}, it follows that they satisfy} $\frac{\pi_0}{\widetilde{T}} - \pi_{\mathrm{u}}\, \mu_{G_\mathrm{tr}}(c_1) - \epsilon_1\, k(c_1) = 0$ and $\frac{\pi_0}{\widetilde{T}} - \pi_{\mathrm{u}}\, \mu_{G_\mathrm{tr}}(c_2) - \epsilon_2\, k(c_2) = 0$, respectively.
	Subtracting the second equation from the first gives:
	\begin{equation}\label{lhs_h}
		\pi_{\mathrm{u}} \left[\mu_{G_\mathrm{tr}}(c_{2})-\mu_{G_\mathrm{tr}}(c_{1})\right]+\epsilon_{2}k(c_{2})-\epsilon_{1}k(c_{1})=0.
	\end{equation}
	
	Define the function $H: [0, 1] \to \mathbb{R}$ as
	$
	H(\tau
	) :=\, \left(\epsilon_1 + \tau
	(\epsilon_2 - \epsilon_1)\right) k\left(c_1 + \tau
	(c_2 - c_1)\right) 
	+ \pi_{\mathrm{u}}\, \mu_{G_\mathrm{tr}}\left(c_1 + \tau
	(c_2 - c_1)\right).$
	Then,~\eqref{lhs_h} is equivalent to $H(1) - H(0) = 0$.
	
	By applying the Mean Value Theorem to the left-hand side of this equation, we obtain:
	\begin{equation}
		H(1) - H(0) = H'(\zeta) = 0, \nonumber
	\end{equation}
	for some $\zeta \in (0, 1)$. Moreover, we have
	\begin{align*}
		H'(\zeta) =\,& \pi_{\mathrm{u}}\, \mu_{G_\mathrm{tr}}'\left(c_1 + \zeta(c_2 - c_1)\right)(c_2 - c_1) \\
		&+ (\epsilon_2 - \epsilon_1) \, k\left(c_1 + \zeta(c_2 - c_1)\right) \\
		&+ \left(\epsilon_1 + \zeta(\epsilon_2 - \epsilon_1)\right) \, k'\left(c_1 + \zeta(c_2 - c_1)\right)(c_2 - c_1).
	\end{align*}
	
	Since $H'(\zeta) = 0$, it follows that
	\mehdi{\begin{equation}\label{upper:K&J}
			\frac{c_2 - c_1}{\epsilon_2 - \epsilon_1}=\frac{k\left(c_1 + \zeta(c_2 - c_1)\right)}{J},
		\end{equation}
		where
		\begin{align*}
			J:=&-\pi_{\mathrm{u}}\,\mu_{G_\mathrm{tr}}'\left(c_1 + \zeta(c_2 - c_1)\right) \nonumber\\
			&- \left(\epsilon_1 + \zeta(\epsilon_2 - \epsilon_1)\right) k'\left(c_1 + \zeta(c_2 - c_1)\right).
		\end{align*}
		
		It remains to upper bound the absolute value of the numerator and lower bound \( |J| \) on the right-hand side of~\eqref{upper:K&J}.
		To this end, note that  
		\begin{equation*}
			\left|k\left(c_1 + \zeta(c_2 - c_1)\right)\right| \leq \overline{v} \, \mathbb{E}[G].
		\end{equation*}
		Moreover, from~\eqref{gdot:lim}, the first term in $J$ is bounded below by a positive constant $m_1$, and from~\eqref{kdot:lim}, the second term is similarly bounded below by a positive constant $m_2$.
		Thus, we conclude:
		\begin{equation}
			\frac{|c_2 - c_1|}{|\epsilon_2 - \epsilon_1|}=\frac{\left | k\left(c_1 + \zeta(c_2 - c_1)\right)\right|}{|J|} \leq \frac{\overline{v} \, \mathbb{E}[G]}{\left(m_1 + m_2\right)} = L_{\Psi},\nonumber
		\end{equation}
		which completes the proof of Lemma~\ref{lemma:Psi:Lipschitz}.}
\end{IEEEproof}

\subsection{Proof of Lemma~\ref{lemma5}}

Using~\eqref{eq:funcg}, we can rewrite~\eqref{cb:consumer:demand:epsilon} as
\begin{equation}
	\phi_{\mathrm{cb},i}(\epsilon,d)
	:=
	\frac{\pi_{0}}{\widetilde T}
	-
	\left(\pi_{\mathrm{u}}+\epsilon\tilde v_i\right)
	\mu_{G_{\mathrm{tr}}}(d).\nonumber
\end{equation}
Then $d_i^\star\left(\frac{\pi_0}{\widetilde T}\right)$ is characterized by
\[
\phi_{\mathrm{cb},i}\left(\epsilon,d_i^\star\left(\frac{\pi_0}{\widetilde T}\right)\right)=0.
\]

The idea mirrors the previous case: if $\epsilon$ is close to $0$, then the solution $d_i^\star\left(\frac{\pi_0}{\widetilde T}\right)$ lies near $c_0$. To prove this, note from~\eqref{eq:prop1:dinverse} that
\[
d_i^\star \left(\frac{\pi_0}{\widetilde T}\right)
=
\tilde{\mu}_{G_\mathrm{tr}}^{-1}\left(\frac{\pi_0}{\widetilde T\left(\pi_{\mathrm{u}}+\epsilon \tilde v_i\right)}\right),
\]
and
\[
c_0
=
\tilde{\mu}_{G_\mathrm{tr}}^{-1}\left(\frac{\pi_0}{\widetilde T\, \pi_{\mathrm{u}}} \right).
\]
Thus,
\[
\left\lvert d_i^\star\left(\frac{\pi_0}{\widetilde T}\right) - c_{0} \right\rvert
=
\left\lvert  \Delta_{\mathrm{cb},i}\left(\epsilon\right)\right\rvert,
\]
where
\begin{equation}\label{Delta:cb:i}
	\Delta_{\mathrm{cb},i}\left(\epsilon\right)
	:=
	\tilde{\mu}_{G_\mathrm{tr}}^{-1}\left( \frac{\pi_0}{\widetilde T\left(\pi_{\mathrm{u}} + \epsilon \tilde{v}_i\right)} \right) 
	- 
	\tilde{\mu}_{G_\mathrm{tr}}^{-1}\left( \frac{\pi_0}{\widetilde T\, \pi_{\mathrm{u}}} \right).
\end{equation}
It remains to bound $\lvert \Delta_{\mathrm{cb},i}\left(\epsilon\right) \rvert$, which diminishes as $\epsilon \rightarrow 0$.

To proceed, define
\[
r_i(\epsilon)
:=
\frac{\pi_0}{\widetilde T\left(\pi_{\mathrm{u}}+\epsilon \tilde v_i\right)}.
\]
Then,
\[
\Delta_{\mathrm{cb},i}\left(\epsilon\right)
=
\tilde {\mu}_{G_\mathrm{tr}}^{-1}(r_i(\epsilon))
-
\tilde {\mu}_{G_\mathrm{tr}}^{-1}(r_i(0)).
\]
We now show that $\tilde {\mu}_{G_\mathrm{tr}}^{-1}(r_i(\cdot))$ is Lipschitz continuous by proving that both $r_i(\cdot)$ and $\tilde {\mu}_{G_\mathrm{tr}}^{-1}(\cdot)$ are individually Lipschitz continuous.

First, $r_i(\cdot)$ is decreasing, continuous, and differentiable. Since $\tilde v_i\leq \overline v$, we have the uniform bound
\[
\left| \frac{\mathrm{d}r_i(\epsilon)}{\mathrm{d}\epsilon}\right|
=
\frac{\pi_0\tilde v_i}{\widetilde T\left(\pi_{\mathrm u}+\epsilon\tilde v_i\right)^2}
\le
\frac{\pi_0\overline v}{\widetilde T\,\pi_{\mathrm u}^2}
=:L_r.
\]
Moreover, by the strict inequality in~\eqref{eq:pi0cond}, we have
\[
\frac{\pi_0}{\widetilde T\,\pi_{\mathrm u}}<\mathbb E[G].
\]
Therefore, for
\[
\frac{\pi_{0}}{\widetilde T \left(\pi_{\mathrm{u}}+\overline{v}\right)}
\leq y \leq
\frac{\pi_{0}}{\widetilde T \, \pi_{\mathrm{u}}},
\]
the extended generalized inverse $\tilde{\mu}_{G_\mathrm{tr}}^{-1}(y)$ coincides with the ordinary inverse of $\mu_{G_\mathrm{tr}}(\cdot)$ and is evaluated in the strictly decreasing region of $\mu_{G_\mathrm{tr}}(\cdot)$. Thus, the Lipschitz constant for $\tilde{\mu}_{G_\mathrm{tr}}^{-1}(\cdot)$ can be bounded as
\begin{equation}
	L_{\tilde{\mu}_{G_\mathrm{tr}}^{-1}}
	:=
	\max_y
	\left|
	\frac{\mathrm{d}\tilde{\mu}_{G_\mathrm{tr}}^{-1}(y)}{\mathrm{d}y}
	\right|
	=
	\max_y
	\left|
	\frac{\left(\tilde{\mu}_{G_\mathrm{tr}}^{-1}(y)\right)^3}
	{L^2 f_G\left(\frac{L}{\tilde{\mu}_{G_\mathrm{tr}}^{-1}(y)}\right)}
	\right|.
	\nonumber
\end{equation}
Here, $\tilde{\mu}_{G_\mathrm{tr}}^{-1}(y)$ is bounded as
\[
\tilde{\mu}_{G_\mathrm{tr}}^{-1}\left(\frac{\pi_{0}}{\widetilde T \, \pi_{\mathrm{u}}}\right)
\leq
\tilde{\mu}_{G_\mathrm{tr}}^{-1}(y)
\leq
\tilde{\mu}_{G_\mathrm{tr}}^{-1}\left(\frac{\pi_{0}}{\widetilde T \left(\pi_{\mathrm{u}}+\overline{v}\right)}\right).
\]
Since the composition of two Lipschitz continuous functions is Lipschitz continuous, the function $\tilde{\mu}_{G_\mathrm{tr}}^{-1}(r_i(\epsilon))$ is Lipschitz continuous with the uniform constant
\[
L_{\mathrm{cb}}
:=
L_{\tilde{\mu}_{G_\mathrm{tr}}^{-1}}L_r.
\]
Therefore,
\[
\frac{
	\left|\tilde{\mu}_{G_\mathrm{tr}}^{-1}(r_i(\epsilon))
	-
	\tilde{\mu}_{G_\mathrm{tr}}^{-1}(r_i(0))\right|
}{\epsilon}
\leq
L_{\mathrm{cb}}.
\]
Combining this with~\eqref{Delta:cb:i}, we obtain
\begin{equation}\label{epsilon_cb:lim}
	0\leq \Delta_{\mathrm{cb},i}(\epsilon)
	=
	\left|\Delta_{\mathrm{cb},i}(\epsilon)\right|
	\leq
	\epsilon L_{\mathrm{cb}},
\end{equation}
where the first inequality holds because $r_i(\epsilon)\le r_i(0)$ for $\epsilon\ge0$ and $\tilde{\mu}_{G_\mathrm{tr}}^{-1}(\cdot)$ is decreasing.

Equipped with this result, we derive the Taylor expansion of 
$\phi_{\mathrm{cb},i}\left(\epsilon, d_i^\star\left(\frac{\pi_0}{\widetilde T}\right)\right)$
around $(0,c_0)$. Since 
$d_i^\star\left(\frac{\pi_0}{\widetilde T}\right)$ solves
$\phi_{\mathrm{cb},i}(\epsilon,d)=0$, we have
\begin{align}\label{h:Taylor}
	0
	&=
	\phi_{\mathrm{cb},i}\left(\epsilon, d_i^\star\left(\frac{\pi_0}{\widetilde T}\right)\right)
	\nonumber\\
	&=
	-\left(d_i^\star\left(\frac{\pi_{0}}{\widetilde T}\right)-c_0\right)
	\pi_{\mathrm{u}}\mu_{G_\mathrm{tr}}'(c_0)  \nonumber \\
	&\quad
	-\epsilon \tilde v_{i} \mu_{G_\mathrm{tr}}(c_0) \nonumber \\
	&\quad
	+O\left(\left\lVert \left(\epsilon ,d_i^\star\left(\frac{\pi_{0}}{\widetilde T}\right)\right)- (0, c_0)\right\rVert ^2\right).
\end{align}
Then, based on~\eqref{epsilon_cb:lim}, we obtain the uniform bound
\begin{equation}
	\left\lVert \left(\epsilon ,d_i^\star\left(\frac{\pi_{0}}{\widetilde T}\right)\right)- (0, c_0)\right\rVert ^2
	\leq
	\epsilon ^2 \left(1+L_{\mathrm{cb}}^{2}\right),
	\nonumber     
\end{equation}
where the right-hand side is independent of $i$. Therefore, the remainder term in~\eqref{h:Taylor} is $O(\epsilon^2)$ uniformly over $i\in\mathcal I_{\mathrm b}$, and we can write
\begin{align}
	0
	=
	\phi_{\mathrm{cb},i}\left(\epsilon, d_i^\star\left(\frac{\pi_0}{\widetilde T}\right)\right)
	= &-\left(d_i^\star\left(\frac{\pi_{0}}{\widetilde T}\right)-c_0\right)
	\pi_{\mathrm{u}}\mu_{G_\mathrm{tr}}'(c_0)  \nonumber \\
	&-\epsilon \tilde v_{i} \mu_{G_\mathrm{tr}}(c_0) 
	+O\!\left(\epsilon^{2}\right).\nonumber
\end{align}
Isolating $d_i^\star\left(\frac{\pi_{0}}{\widetilde T}\right)$ gives
\begin{equation}
	d_i^\star\left(\frac{\pi_{0}}{\widetilde T}\right)
	=
	c_0
	-
	\frac{\epsilon \tilde v_i\mu_{G_\mathrm{tr}}(c_0)}
	{\pi_{\mathrm u}\mu_{G_\mathrm{tr}}'(c_0)}
	+
	O(\epsilon^2),
	\nonumber
\end{equation}
which yields~\eqref{di:c_0:epsilon:lemma}. Finally, applying~\eqref{d:star:eq} and using the fact that the $O(\epsilon^2)$ term is uniform in $i$, we obtain
\begin{align}
	c_\mathrm{cb}^{\mathrm{ne}}
	&=
	\int_{\mathcal I_\mathrm{b}}
	d_i^\star\left(\frac{\pi_{0}}{\widetilde T}\right)\,\mathrm di
	\nonumber\\
	&=
	c_0
	-
	\frac{\epsilon \mathbb E[\widetilde V]\mu_{G_\mathrm{tr}}(c_0)}
	{\pi_{\mathrm u}\mu_{G_\mathrm{tr}}'(c_0)}
	+
	O(\epsilon^2),
	\nonumber
\end{align}
which yields~\eqref{c_0:epsilon:lemma} and completes the proof. $\qquad \quad \quad \qquad  \quad \, \blacksquare$

\section{Proofs for Subsection~\ref{heterg:time}}

\subsection{Proof of Proposition~\ref{prop:optc:tv}}

The expression for $v_t^\star(c, G_t)$ is derived individually for each $t \in \mathcal{T}$ using the same method as in the proof of Proposition~\ref{prop:optc}. Substituting these expressions into~\eqref{social:cap:tv} yields
\begin{align}
	\max_{c\in \reals_+} \enspace \sum_{t\in\mathcal{T}} \! \frac{\widetilde T}{T} \mathbb E \Biggr[
	&\! L_{t}\mathbb E[V]
	\!-\!\!
	L_{t}\mathbb{1}\!\left\{cG_{t} \leq L_{t} \right\}\!\!
	\int_{0}^{F_{V}^{-1}\left(1-\frac{c G_{t}}{L_{t}}\right)} \!\! \! \! \! \! v_i \, \mathrm{d}F_{V}(\! v_{i}\!)
	\nonumber\\
	&-\pi_{\mathrm {u}_t}(L_{t}-cG_{t})_+
	\Biggr] - \pi_0 c. \nonumber
\end{align}

The objective is concave in \( c \) since, under the bounded-support assumption, its second derivative can be written as
\begin{equation*}
	-\sum_{t\in\mathcal{T}}
	\frac{\widetilde T}{T}
	\int_0^{\min\{L_{t}/c,\overline g_t\}}
	\frac{g_{t}^{2}}{L_{t}} \,
	\frac{f_{G_t}\left(g_{t}\right)}
	{f_V\left(F_V^{-1}\left(1 - \tfrac{cg_{t}}{L_{t}}\right)\right)}
	\,\mathrm{d}g_{t}
	\leq 0.
\end{equation*}
Since the objective is concave in $c$ and the feasible set is convex, the first-order optimality condition is sufficient to characterize the global optimum. Applying this condition yields~\eqref{eq:opt:tv}, completing the proof.
\quad  \quad \qquad  \quad \quad \quad \,\, \qquad\quad\qquad\qquad \,\,\,\, \scalebox{0.9}{$\blacksquare$}

\subsection{Proof of Theorem~\ref{thm:main:tv}}
We prove each case individually:

\begin{IEEEproof}[General case]
	Since the characterizing equation for $c_{\mathrm{prt}}^{\mathrm{ne}}$ is identical to that of~\eqref{eq:opt:tv}, it follows that
	\[
	c_{\mathrm{prt}}^{\mathrm{ne}} = c_{\mathrm{opt}}.
	\]
	
	We next show that $c_{\mathrm{srt}}^{\mathrm{ne}}\leq c_{\mathrm{prt}}^{\mathrm{ne}}$. By the investment equilibrium condition, the positive equilibrium capacities under the product-differentiated and single-product real-time markets satisfy
	\begin{equation}\label{cprt:e:eq}
		\frac{\Pi^\mathrm{s}_{\mathrm{prt}}(c_{\mathrm{prt}}^{\mathrm{ne}})}
		{c_{\mathrm{prt}}^{\mathrm{ne}}}
		=
		\pi_0,
	\end{equation}
	and
	\begin{equation}\label{eq:srt:eq:tv}
		\frac{\Pi^\mathrm{s}_{\mathrm{srt}}(c_{\mathrm{srt}}^{\mathrm{ne}})}
		{c_{\mathrm{srt}}^{\mathrm{ne}}}
		=
		\pi_0.
	\end{equation}
	For each operation period $t\in\mathcal T$, the pointwise comparison in~\eqref{prt:srt:integrand:homo} applies with $G$, $L$, and $\pi_{\mathrm u}$ replaced by $G_t$, $L_t$, and $\pi_{\mathrm {u}_t}$, respectively. Hence, for every $c>0$,
	\begin{equation}\label{term:integrand:time}
		\pi_{\mathrm {u}_t}G_t\mathbb 1\{cG_t\le L_t\}
		\!\le
		\!\left(\!\pi_{\mathrm {u}_t}
		+
		\epsilon \overline F_{\widetilde V}^{-1}
		\left(\frac{cG_t}{L_t}\right)\!
		\right)
		\! G_t\mathbb 1\{\!cG_t\le L_t\!\}.
	\end{equation}
	Taking expectations, summing over $t\in\mathcal T$, and using the definitions of the time-varying real-time revenues, we obtain
	\begin{equation}
		\frac{\Pi^\mathrm{s}_{\mathrm{srt}}(c)}{c}
		\le
		\frac{\Pi^\mathrm{s}_{\mathrm{prt}}(c)}{c},
		\qquad c>0.
		\nonumber
	\end{equation}
	Evaluating this inequality at $c=c_{\mathrm{prt}}^{\mathrm{ne}}$ and using~\eqref{cprt:e:eq} gives
	\begin{equation}
		\frac{\Pi^\mathrm{s}_{\mathrm{srt}}(c_{\mathrm{prt}}^{\mathrm{ne}})}
		{c_{\mathrm{prt}}^{\mathrm{ne}}}
		\le
		\frac{\Pi^\mathrm{s}_{\mathrm{prt}}(c_{\mathrm{prt}}^{\mathrm{ne}})}
		{c_{\mathrm{prt}}^{\mathrm{ne}}}
		=
		\pi_0.
		\nonumber
	\end{equation}
	On the other hand, by~\eqref{eq:srt:eq:tv},
	\[
	\frac{\Pi^\mathrm{s}_{\mathrm{srt}}(c_{\mathrm{srt}}^{\mathrm{ne}})}
	{c_{\mathrm{srt}}^{\mathrm{ne}}}
	=
	\pi_0.
	\]
	Therefore,
	\begin{equation}\label{eq:srt:compare:tv}
		\frac{\Pi^\mathrm{s}_{\mathrm{srt}}(c_{\mathrm{prt}}^{\mathrm{ne}})}
		{c_{\mathrm{prt}}^{\mathrm{ne}}}
		\le
		\frac{\Pi^\mathrm{s}_{\mathrm{srt}}(c_{\mathrm{srt}}^{\mathrm{ne}})}
		{c_{\mathrm{srt}}^{\mathrm{ne}}}.
	\end{equation}
	Moreover, the per-unit single-product real-time revenue is non-increasing in $c$. Indeed, for any $0<c_1<c_2$ and each $t\in\mathcal T$,
	\[
	\pi_{\mathrm {u}_t}G_t\mathbb 1\{c_2G_t\le L_t\}
	\le
	\pi_{\mathrm {u}_t}G_t\mathbb 1\{c_1G_t\le L_t\}.
	\]
	Taking expectations, summing over $t\in\mathcal T$, and using the definition of $\Pi^\mathrm{s}_{\mathrm{srt}}(\cdot)$ gives
	\begin{equation}\label{eq:srt:monotone:tv}
		\frac{\Pi^\mathrm{s}_{\mathrm{srt}}(c_2)}{c_2}
		\le
		\frac{\Pi^\mathrm{s}_{\mathrm{srt}}(c_1)}{c_1},
		\qquad 0<c_1<c_2.
	\end{equation}
	Combining~\eqref{eq:srt:compare:tv} with the monotonicity property in~\eqref{eq:srt:monotone:tv}, and using the uniqueness argument for the single-product equilibrium condition, we obtain
	\[
	c_{\mathrm{srt}}^{\mathrm{ne}}\le c_{\mathrm{prt}}^{\mathrm{ne}}.
	\]
	Combining this with $c_{\mathrm{prt}}^{\mathrm{ne}}=c_{\mathrm{opt}}$ gives
	\[
	c_{\mathrm{srt}}^{\mathrm{ne}}
	\le
	c_{\mathrm{prt}}^{\mathrm{ne}}
	=
	c_{\mathrm{opt}}.
	\]
	
It remains to show that $c_{\mathrm{srt}}^{\mathrm{ne}}\le c_{\mathrm{cb}}^{\mathrm{ne}}$. 
Using the definition of $\mu_{G_{t,\mathrm{tr}}}(\cdot)$ in~\eqref{mug:time_varying}, the equilibrium demand of buyer $i$ in the contract-based market is characterized by
\begin{equation}\label{cb:heter:cond:mu}
	\sum_{t\in\mathcal T}
	\left(\pi_{\mathrm {u}_t}+\epsilon\tilde v_i\right)
	\mu_{G_{t,\mathrm{tr}}}
	\left(
	d_i^\star
	\right)
	=
	\pi_0\frac{T}{\widetilde T}.
\end{equation}
On the other hand, the single-product real-time equilibrium condition gives
\begin{equation}\label{srt:heter:cond:mu}
	\sum_{t\in\mathcal T}
	\pi_{\mathrm {u}_t}
	\mu_{G_{t,\mathrm{tr}}}
	\left(
	c_{\mathrm{srt}}^{\mathrm{ne}}
	\right)
	=
	\pi_0\frac{T}{\widetilde T}.
\end{equation}
Evaluating the left-hand side of the contract-based buyer condition at $c_{\mathrm{srt}}^{\mathrm{ne}}$ gives
\[
\sum_{t\in\mathcal T}
\left(\pi_{\mathrm {u}_t}+\epsilon\tilde v_i\right)
\mu_{G_{t,\mathrm{tr}}}
\left(
c_{\mathrm{srt}}^{\mathrm{ne}}
\right)
\ge
\sum_{t\in\mathcal T}
\pi_{\mathrm {u}_t}
\mu_{G_{t,\mathrm{tr}}}
\left(
c_{\mathrm{srt}}^{\mathrm{ne}}
\right)
=
\pi_0\frac{T}{\widetilde T}.
\]
Since each $\mu_{G_{t,\mathrm{tr}}}(\cdot)$ is non-increasing, the left-hand side of the contract-based buyer condition is non-increasing in demand. Therefore, the demand level that solves the contract-based buyer condition cannot be smaller than $c_{\mathrm{srt}}^{\mathrm{ne}}$, i.e.,
\[
d_i^\star\ge c_{\mathrm{srt}}^{\mathrm{ne}},
\qquad i\in\mathcal I_{\mathrm b}.
\]
Integrating over all buyers gives
\[
c_{\mathrm{cb}}^{\mathrm{ne}}
=
\int_{\mathcal I_{\mathrm b}} d_i^\star\,\mathrm di
\ge
c_{\mathrm{srt}}^{\mathrm{ne}}.
\]
Thus,
\[
c_{\mathrm{srt}}^{\mathrm{ne}}
\le
c_{\mathrm{prt}}^{\mathrm{ne}}
=
c_{\mathrm{opt}},
\qquad
c_{\mathrm{srt}}^{\mathrm{ne}}
\le
c_{\mathrm{cb}}^{\mathrm{ne}}.
\]
\end{IEEEproof}

\begin{IEEEproof}[No solar premium]
	If $\epsilon = 0$, then the two sides of~\eqref{term:integrand:time} coincide for every operation period $t\in\mathcal T$ and every $c>0$. Therefore, taking expectations and summing over $t\in\mathcal T$ gives
	\[
	\frac{\Pi^\mathrm{s}_{\mathrm{prt}}(c)}{c}
	=
	\frac{\Pi^\mathrm{s}_{\mathrm{srt}}(c)}{c},
	\qquad c>0.
	\]
	Hence, the equilibrium conditions~\eqref{cprt:e:eq} and~\eqref{eq:srt:eq:tv} become identical. Using the uniqueness of the corresponding equilibrium condition, we obtain
	\[
	c_\mathrm{opt}
	=
	c^\mathrm{ne}_\mathrm{prt}
	=
	c^\mathrm{ne}_\mathrm{srt}.
	\]
	
	It remains to show that $c^\mathrm{ne}_\mathrm{cb}=c^\mathrm{ne}_\mathrm{srt}$. When $\epsilon=0$, the contract-based equilibrium condition~\eqref{cb:heter:cond:mu} reduces to
	\[
	\sum_{t\in\mathcal T}
	\pi_{\mathrm {u}_t}
	\mu_{G_{t,\mathrm{tr}}}
	\left(
	d_i^\star
	\right)
	=
	\pi_0\frac{T}{\widetilde T}.
	\]
	Comparing this equation with the single-product equilibrium condition~\eqref{srt:heter:cond:mu}, and using the uniqueness of the single-product equilibrium condition, we obtain
	\[
	d_i^\star
	=
	c_{\mathrm{srt}}^{\mathrm{ne}},
	\qquad i\in\mathcal I_{\mathrm b}.
	\]
	Finally, applying~\eqref{d:star:tv} gives
	\[
	c^\mathrm{ne}_\mathrm{cb}
	=
	\int_{\mathcal I_{\mathrm b}}
	d_i^\star\,\mathrm di
	=
	c^\mathrm{ne}_\mathrm{srt}.
	\]
	Thus,
	\[
	c^\mathrm{ne}_\mathrm{srt}
	=
	c^\mathrm{ne}_\mathrm{prt}
	=
	c_\mathrm{opt}
	=
	c^\mathrm{ne}_\mathrm{cb}.
	\]
\end{IEEEproof}

\begin{IEEEproof}[Small solar premium]
	The steps follow the same structure as in the proof of Theorem~\ref{thm:main}. Let $c_0:=c_{\mathrm{srt}}^{\mathrm{ne}}$. We begin by introducing two lemmas analogous to Lemmas~\ref{lemma4} and~\ref{lemma5}.
	
	\begin{lemma}[First-order approximation of $c_{\mathrm{prt}}^{\mathrm{ne}}$ around $\epsilon = 0$ with heterogeneous periods]\label{lemma11}
		For a sufficiently small $\epsilon>0$, we have
		\begin{align}
			c_\mathrm{prt}^{\mathrm{ne}}
			=&c_0-\frac{\epsilon \,\sum_{t\in\mathcal{T}} \mathbb E \left [\overline{F}_{\widetilde V}^{-1}\left(\frac{c_0G_{t}}{L_{t}}\right)\,G_{t}\,\mathbb{1} \bigl\{c_0\,G_{t} \leq L_{t} \bigl\}\right]}{\sum_{t\in\mathcal{T}}\pi_{\mathrm {u}_t}\mu_{G_{t,\mathrm{tr}}}'(c_0)}\nonumber \\
			&+O\left(\epsilon^2\right).\label{c_0:epsilon:lemma:tv:prt}
		\end{align}
	\end{lemma}
	
	\begin{lemma}[First-order approximation of $c_{\mathrm{cb}}^{\mathrm{ne}}$ around $\epsilon = 0$ with heterogeneous periods]\label{lemma12}
		For a sufficiently small $\epsilon>0$, we have
		\begin{equation}\label{di:c_0:epsilon:lemma:tv}
			d_{i}^\star\left(\frac{\pi_0 T}{\widetilde T}\right)
			=
			c_0
			-
			\frac{
				\epsilon \tilde v_{i}\sum_{t\in\mathcal{T}}\mu_{G_{t,\mathrm{tr}}}(c_0)
			}
			{
				\sum_{t\in\mathcal{T}}\pi_{\mathrm {u}_t} \mu_{G_{t,\mathrm{tr}}}'(c_0)
			}
			+O\left(\epsilon^2\right),
		\end{equation}
		and therefore,
		\begin{equation}\label{c_0:epsilon:lemma:tv}
			c_\mathrm{cb}^{\mathrm{ne}}
			=
			c_0
			-
			\frac{
				\epsilon \,\mathbb E[\widetilde V]\sum_{t\in\mathcal{T}}\mu_{G_{t,\mathrm{tr}}}(c_0)
			}
			{
				\sum_{t\in\mathcal{T}}\pi_{\mathrm {u}_t} \mu_{G_{t,\mathrm{tr}}}'(c_0)
			}
			+O\left(\epsilon^2\right).
		\end{equation}
	\end{lemma}
	
	First, note that Lemmas~\ref{lemma11} and~\ref{lemma12} have the same denominator. Define
	\begin{equation}
		D_{\mathrm{tv}}
		:=
		\sum_{t\in\mathcal{T}}
		\pi_{\mathrm {u}_t}
		\mu_{G_{t,\mathrm{tr}}}'(c_0).\nonumber
	\end{equation}
	We show that $D_{\mathrm{tv}}<0$. For each $t\in\mathcal{T}$,
	\begin{equation}
		\mu_{G_{t,\mathrm{tr}}}'(c_0)
		=
		-\frac{L_t^2}{c_0^3}
		f_{G_t}\left(\frac{L_t}{c_0}\right)
		\mathbb{1}\left\{\frac{L_t}{c_0}<\overline g_t\right\}
		\le 0.
		\nonumber
	\end{equation}
	Moreover, condition~\eqref{fg:constant:tv} evaluated at $p=1$ implies that
	\begin{equation}
		\sum_{t\in\mathcal{T}}
		\left(\frac{L_t}{c_0}\right)^2
		f_{G_t}\left(\frac{L_t}{c_0}\right)
		\mathbb{1}\left\{\frac{L_t}{c_0}<\overline g_t\right\}
		>0.
		\nonumber
	\end{equation}
	Hence, at least one term $\mu_{G_{t,\mathrm{tr}}}'(c_0)$ is strictly negative. Since $\pi_{\mathrm {u}_t}>0$ for all $t$, it follows that
	\[
	D_{\mathrm{tv}}<0.
	\]
	
	Thus, since the common denominator in Lemmas~\ref{lemma11} and~\ref{lemma12} is strictly negative, it remains to compare the corresponding numerators. Specifically, we prove
	\begin{equation}\label{agg:ineq:tv}
		\underbrace{
			\sum_{t\in\mathcal{T}} \mathbb E  \left [\overline{F}_{\widetilde V}^{-1}\left(\frac{c_0G_{t}}{L_{t}}\right)G_{t}\mathbb{1} \bigl\{c_0G_{t} \leq L_{t} \bigl\}\right]
		}_{=:A_{\mathrm{tv}}}
		\!\!<\!
		\underbrace{
			\mathbb E[\widetilde V]\sum_{t\in\mathcal{T}}\mu_{G_{t,\mathrm{tr}}}(c_0)
		}_{=:B_{\mathrm{tv}}}\!.
	\end{equation}
	
	As in the homogeneous case, we use the local abbreviation
	\begin{equation}
		\tilde v(p):=\overline F_{\widetilde V}^{-1}(p), \qquad p\in[0,1],
		\nonumber
	\end{equation}
	for the decreasing inverse associated with $\overline F_{\widetilde V}$. Since $\tilde v(p)$ is non-increasing, we have $-\tilde v'(p)\ge0$. Next, define
	\begin{equation}
		S(p)
		:=
		\sum_{t\in\mathcal{T}}
		\mathbb E\left[
		G_t
		\mathbb 1\left\{
		\frac{c_0G_t}{L_t}\le p
		\right\}
		\right],
		\qquad p\in[0,1].
		\nonumber
	\end{equation}
	Using the definition of $\mu_{G_{t,\mathrm{tr}}}(\cdot)$ in~\eqref{mug:time_varying}, we have
	\begin{equation}
		S(1)=\sum_{t\in\mathcal{T}}\mu_{G_{t,\mathrm{tr}}}(c_0).
		\nonumber
	\end{equation}
	Using integration by parts, together with $S(0)=0$ and $\tilde v(1)=0$, we obtain
	\begin{align}
		A_{\mathrm{tv}}
		&=
		\int_0^1
		\left(-\tilde v'(p)\right)S(p)\,\mathrm{d}p,
		\label{eq:A:tv:S}
	\end{align}
	and
	\begin{equation}\label{eq:B:tv:S}
		B_{\mathrm{tv}}
		=
		\int_0^1
		\left(-\tilde v'(p)\right)pS(1)\,\mathrm{d}p.
	\end{equation}
	
	We next relate $S(p)$ to the aggregate density condition. Under the bounded-support assumption,
	\begin{equation}
		S(p)
		=
		\sum_{t\in\mathcal{T}}
		\int_0^{\min\{L_t p/c_0,\overline g_t\}}
		g_t f_{G_t}(g_t)\,\mathrm{d}g_t.
		\nonumber
	\end{equation}
	Therefore,
	\begin{equation}\label{eq:Sprime:tv}
		S'(p)
		=
		p
		\sum_{t\in\mathcal{T}}
		\left(\frac{L_t}{c_0}\right)^2
		f_{G_t}\left(\frac{L_t p}{c_0}\right)
		\mathbb{1}\left\{\frac{L_t p}{c_0}< \overline g_t\right\}.
	\end{equation}
	By condition~\eqref{fg:constant:tv}, there exist $R>0$ and a sufficiently small $\delta>0$ such that, for all $p\in[0,1]$,
	\begin{equation}\label{eq:aggregate:density:bound:tv}
		(1-\delta)R
		\le
		\sum_{t\in\mathcal{T}}
		\left(\frac{L_t}{c_0}\right)^2
		\! f_{G_t}\left(\frac{L_t p}{c_0}\right)
		\mathbb{1}\left\{\frac{L_t p}{c_0}< \overline g_t\right\}
		\!\le
		(1+\delta)R.
	\end{equation}
	
	Since $S(0)=0$, we have
	\begin{equation}
		S(p)=\int_0^p S'(x)\,\mathrm{d}x.
		\nonumber
	\end{equation}
	Using~\eqref{eq:Sprime:tv} and the upper bound in~\eqref{eq:aggregate:density:bound:tv}, we get
	\begin{align}
		S(p)
		&=
		\int_0^p
		x
		\sum_{t\in\mathcal{T}}
		\left(\frac{L_t}{c_0}\right)^2
		f_{G_t}\left(\frac{L_t x}{c_0}\right)
		\mathbb{1}\left\{\frac{L_t x}{c_0}< \overline g_t\right\}
		\mathrm{d}x \nonumber\\
		&\le
		\int_0^p x(1+\delta)R\,\mathrm{d}x
		=
		(1+\delta)R\frac{p^2}{2}.
		\label{eq:Supper:tv}
	\end{align}
	Similarly, using the lower bound in~\eqref{eq:aggregate:density:bound:tv},
	\begin{align}
		S(1)
		&=
		\int_0^1
		x
		\sum_{t\in\mathcal{T}}
		\left(\frac{L_t}{c_0}\right)^2
		f_{G_t}\left(\frac{L_t x}{c_0}\right)
		\mathbb{1}\left\{\frac{L_t x}{c_0}< \overline g_t\right\}
		\mathrm{d}x \nonumber\\
		&\ge
		\int_0^1 x(1-\delta)R\,\mathrm{d}x
		=
		(1-\delta)R\frac{1}{2}.
		\label{eq:Sone:lower:tv}
	\end{align}
	Combining~\eqref{eq:Supper:tv} and~\eqref{eq:Sone:lower:tv} yields
	\begin{equation}\label{eq:Sratio:tv}
		S(p)
		\le
		\frac{1+\delta}{1-\delta}p^2S(1).
	\end{equation}
	
	Substituting~\eqref{eq:Sratio:tv} into~\eqref{eq:A:tv:S} gives
	\begin{equation}\label{eq:Aupper:tv}
		A_{\mathrm{tv}}
		\le
		\frac{1+\delta}{1-\delta}
		S(1)
		\int_0^1
		\left(-\tilde v'(p)\right)p^2\,\mathrm{d}p.
	\end{equation}
	For notational compactness within this proof, define
	\begin{equation}
		J_2:=
		\int_0^1
		\left(-\tilde v'(p)\right)p^2\,\mathrm{d}p,
		\qquad
		J_1:=
		\int_0^1
		\left(-\tilde v'(p)\right)p\,\mathrm{d}p.
		\nonumber
	\end{equation}
	Because $0\le p^2\le p$ on $[0,1]$, and the inequality is strict for $p\in(0,1)$, we have $J_2<J_1$. Hence, choosing $\delta>0$ sufficiently small, there exists $\theta_{\mathrm{tv}}\in(0,1)$ such that
	\begin{equation}\label{eq:theta:tv}
		\frac{1+\delta}{1-\delta}J_2
		\le
		\frac{1+\theta_{\mathrm{tv}}}{2}J_1.
	\end{equation}
	Using~\eqref{eq:Aupper:tv},~\eqref{eq:theta:tv}, and~\eqref{eq:B:tv:S}, we obtain
	\begin{equation}
		A_{\mathrm{tv}}
		\le
		\frac{1+\theta_{\mathrm{tv}}}{2}B_{\mathrm{tv}}.
		\nonumber
	\end{equation}
	Equivalently,
	\begin{equation}\label{eq:agg:ineq:tv:strong}
		B_{\mathrm{tv}}
		\ge
		\frac{2}{1+\theta_{\mathrm{tv}}}A_{\mathrm{tv}}.
	\end{equation}
	Since $\theta_{\mathrm{tv}}\in(0,1)$,~\eqref{eq:agg:ineq:tv:strong} implies~\eqref{agg:ineq:tv}.
	
	Finally, using Lemmas~\ref{lemma11} and~\ref{lemma12}, together with the definitions of $A_{\mathrm{tv}}$, $B_{\mathrm{tv}}$, and $D_{\mathrm{tv}}$, we obtain
	\begin{equation}
		c_\mathrm{cb}^{\mathrm{ne}}-c_\mathrm{prt}^{\mathrm{ne}}
		=
		-\epsilon\frac{B_{\mathrm{tv}}-A_{\mathrm{tv}}}{D_{\mathrm{tv}}}
		+O(\epsilon^2).
		\nonumber
	\end{equation}
	From~\eqref{eq:agg:ineq:tv:strong}, we have
	\[
	B_{\mathrm{tv}}-A_{\mathrm{tv}}
	\ge
	\frac{1-\theta_{\mathrm{tv}}}{1+\theta_{\mathrm{tv}}}A_{\mathrm{tv}}.
	\]
	Since $D_{\mathrm{tv}}<0$, define
	\begin{equation*}
		\beta_{\mathrm{tv}}
		:=
		\frac{
			\left(1-\theta_{\mathrm{tv}}\right)
			A_{\mathrm{tv}}
		}
		{
			-\left(1+\theta_{\mathrm{tv}}\right)
			D_{\mathrm{tv}}
		}
		>0.
	\end{equation*}
	Therefore,
	\begin{equation}\label{thm:main:case2:detail:tv}
		c_\mathrm{cb}^{\mathrm{ne}}-c_\mathrm{prt}^{\mathrm{ne}}
		\ge
		\beta_{\mathrm{tv}}\epsilon
		-
		O\left(\epsilon^2\right).
	\end{equation}
	Combining this with the general-case result $c^\mathrm{ne}_\mathrm{srt}\le c^\mathrm{ne}_\mathrm{prt}=c_\mathrm{opt}$ gives
	\begin{equation}
		c^\mathrm{ne}_\mathrm{srt}
		\le
		c^\mathrm{ne}_\mathrm{prt}
		=
		c_\mathrm{opt}
		\lesssim
		c^\mathrm{ne}_\mathrm{cb},\nonumber
	\end{equation}
	where the last inequality holds in the sense of~\eqref{thm:main:case2:detail:tv}. This completes the proof.
\end{IEEEproof}

\section{Proof of Lemma~\ref{lemma11} and Lemma~\ref{lemma12}}
\subsection{Proof of Lemma~\ref{lemma11}}

Using~\eqref{cprt:e:eq}, define the time-varying analogue of the product-differentiated equilibrium function as
\begin{equation}\label{prt:eps:k:g:tv}
	\phi_{\mathrm{prt}}^{\mathrm{tv}}(\epsilon,c)
	:=
	\pi_0 \frac{T}{\widetilde T}
	-
	\sum_{t\in\mathcal{T}}\pi_{\mathrm {u}_t}\,\mu_{G_{t,\mathrm{tr}}}(c)
	-
	\epsilon \sum_{t\in\mathcal{T}} k_t(c),
\end{equation}
where, for any $c\ge0$,
\begin{equation}\label{def;k_{t}(c)}
	k_{t}(c)
	:=
	\mathbb{E} \left[
	\overline{F}_{\widetilde V}^{-1}\left(\frac{c G_{t}}{L_{t}}\right)
	G_{t}
	\mathbb{1} \left\{c G_{t} \leq L_{t} \right\}
	\right].
\end{equation}
Then $c_{\mathrm{prt}}^{\mathrm{ne}}$ is characterized by
\[
\phi_{\mathrm{prt}}^{\mathrm{tv}}(\epsilon,c_{\mathrm{prt}}^{\mathrm{ne}})=0.
\]

Let $\Psi_{\mathrm{tv}}(\epsilon)$ denote the solution to~\eqref{prt:eps:k:g:tv}. Thus,
\[
c_{\mathrm{prt}}^{\mathrm{ne}}=\Psi_{\mathrm{tv}}(\epsilon),
\qquad
\Psi_{\mathrm{tv}}(0)=c_0.
\]
We first note that $\Psi_{\mathrm{tv}}(\epsilon)$ is Lipschitz continuous in a neighborhood of $\epsilon=0$. Indeed, at $(0,c_0)$,
\[
\frac{\partial \phi_{\mathrm{prt}}^{\mathrm{tv}}}{\partial c}(0,c_0)
=
-\sum_{t\in\mathcal T}
\pi_{\mathrm {u}_t}
\mu_{G_{t,\mathrm{tr}}}'(c_0).
\]
Moreover, under the bounded-support assumption,
\[
\mu_{G_{t,\mathrm{tr}}}'(c_0)
=
-\frac{L_t^2}{c_0^3}
f_{G_t}\left(\frac{L_t}{c_0}\right)
\mathbb{1}\left\{\frac{L_t}{c_0}<\overline g_t\right\}
\le 0.
\]
Condition~\eqref{fg:constant:tv} evaluated at $p=1$ implies that at least one of these derivative terms is strictly negative. Since $\pi_{\mathrm {u}_t}>0$ for all $t$, it follows that
\[
\frac{\partial \phi_{\mathrm{prt}}^{\mathrm{tv}}}{\partial c}(0,c_0)>0.
\]
Therefore, by continuity, this derivative is bounded away from zero in a neighborhood of $(0,c_0)$. Also, $\partial \phi_{\mathrm{prt}}^{\mathrm{tv}}/\partial \epsilon=-\sum_{t\in\mathcal T}k_t(c)$ is bounded in that neighborhood because $\widetilde V$ and $G_t$ have bounded support. Hence, by the Mean Value Theorem, there exists a constant $L_{\Psi,\mathrm{tv}}>0$ such that
\begin{equation}\label{prt:tv:error}
	|c_{\mathrm{prt}}^{\mathrm{ne}}-c_0|
	=
	|\Psi_{\mathrm{tv}}(\epsilon)-\Psi_{\mathrm{tv}}(0)|
	\le
	L_{\Psi,\mathrm{tv}}\epsilon.
\end{equation}

We now follow the same expansion argument as in the proof of Lemma~\ref{lemma4}. Taylor expansion of $\phi_{\mathrm{prt}}^{\mathrm{tv}}(\epsilon,c_{\mathrm{prt}}^{\mathrm{ne}})$ around $(0,c_0)$ gives
\begin{align}
	0
	&=
	\phi_{\mathrm{prt}}^{\mathrm{tv}}(\epsilon,c_{\mathrm{prt}}^{\mathrm{ne}})
	\nonumber\\
	&=
	-\epsilon\sum_{t\in\mathcal T}k_t(c_0)
	-
	(c_{\mathrm{prt}}^{\mathrm{ne}}-c_0)
	\sum_{t\in\mathcal T}
	\pi_{\mathrm {u}_t}\mu_{G_{t,\mathrm{tr}}}'(c_0)
	+
	O(\epsilon^2),
	\nonumber
\end{align}
where~\eqref{prt:tv:error} is used to write the remainder as $O(\epsilon^2)$. Isolating $c_{\mathrm{prt}}^{\mathrm{ne}}$ yields
\[
c_{\mathrm{prt}}^{\mathrm{ne}}
=
c_0
-
\frac{
	\epsilon\sum_{t\in\mathcal T}k_t(c_0)
}
{
	\sum_{t\in\mathcal T}
	\pi_{\mathrm {u}_t}\mu_{G_{t,\mathrm{tr}}}'(c_0)
}
+
O(\epsilon^2).
\]
Substituting the definition of $k_t(\cdot)$ from~\eqref{def;k_{t}(c)} evaluated at $c=c_0$ yields~\eqref{c_0:epsilon:lemma:tv:prt}, completing the proof.
$\qquad \qquad \qquad \, \quad \, \blacksquare$

\subsection{Proof of Lemma~\ref{lemma12}}

Using the contract-based equilibrium condition~\eqref{cb:heter:cond:mu}, define
\begin{equation}\label{cb:h:tv}
	\phi_{\mathrm{cb},i}^{\mathrm{tv}}(\epsilon,d)
	:=
	\pi_0 \frac{T}{\widetilde T}
	-
	\sum_{t\in\mathcal{T}}
	\left(\pi_{\mathrm {u}_t}+\epsilon\tilde v_i\right)
	\mu_{G_{t,\mathrm{tr}}}(d).
\end{equation}
Then $d_i^\star\left(\frac{\pi_0 T}{\widetilde T}\right)$ is characterized by
\[
\phi_{\mathrm{cb},i}^{\mathrm{tv}}
\left(
\epsilon,
d_i^\star\left(\frac{\pi_0 T}{\widetilde T}\right)
\right)
=0.
\]
When $\epsilon=0$, this condition coincides with the single-product equilibrium condition~\eqref{srt:heter:cond:mu}. Hence,
\[
\phi_{\mathrm{cb},i}^{\mathrm{tv}}(0,c_0)=0.
\]

To emphasize the dependence of $d_i^\star\left(\frac{\pi_0 T}{\widetilde T}\right)$ on $\epsilon$, let $\Psi_{i,\mathrm{tv}}(\epsilon)$ denote the solution to~\eqref{cb:h:tv}. Thus,
\[
d_i^\star\left(\frac{\pi_0 T}{\widetilde T}\right)
=
\Psi_{i,\mathrm{tv}}(\epsilon),
\qquad
c_0=\Psi_{i,\mathrm{tv}}(0).
\]
We next show that $\Psi_{i,\mathrm{tv}}(\epsilon)$ is Lipschitz continuous in a neighborhood of $\epsilon=0$, uniformly over $i\in\mathcal I_{\mathrm b}$. The derivative of $\phi_{\mathrm{cb},i}^{\mathrm{tv}}(\epsilon,d)$ with respect to $d$ is
\[
\frac{\partial \phi_{\mathrm{cb},i}^{\mathrm{tv}}}{\partial d}(\epsilon,d)
=
-\sum_{t\in\mathcal T}
\left(\pi_{\mathrm {u}_t}+\epsilon\tilde v_i\right)
\mu_{G_{t,\mathrm{tr}}}'(d).
\]
At $(0,c_0)$, this becomes
\[
\frac{\partial \phi_{\mathrm{cb},i}^{\mathrm{tv}}}{\partial d}(0,c_0)
=
-\sum_{t\in\mathcal T}
\pi_{\mathrm {u}_t}
\mu_{G_{t,\mathrm{tr}}}'(c_0).
\]
As shown in the proof of the small-solar-premium case, the denominator
\[
\sum_{t\in\mathcal T}
\pi_{\mathrm {u}_t}
\mu_{G_{t,\mathrm{tr}}}'(c_0)
\]
is strictly negative. Hence,
\[
\frac{\partial \phi_{\mathrm{cb},i}^{\mathrm{tv}}}{\partial d}(0,c_0)>0.
\]
By continuity, this derivative remains bounded away from zero in a neighborhood of $(0,c_0)$. Moreover,
\[
\frac{\partial \phi_{\mathrm{cb},i}^{\mathrm{tv}}}{\partial \epsilon}(\epsilon,d)
=
-\tilde v_i
\sum_{t\in\mathcal T}
\mu_{G_{t,\mathrm{tr}}}(d),
\]
which is uniformly bounded over $i\in\mathcal I_{\mathrm b}$ because $\tilde v_i\le \overline v$ and $\mu_{G_{t,\mathrm{tr}}}(d)$ is bounded under the bounded-support assumption. Therefore, by the Mean Value Theorem, there exists a constant $L_{\Psi,\mathrm{tv}}>0$, independent of $i$, such that
\begin{equation}\label{cb:tv:error}
	\left|
	d_i^\star\left(\frac{\pi_0 T}{\widetilde T}\right)
	-c_0
	\right|
	=
	\left|
	\Psi_{i,\mathrm{tv}}(\epsilon)-\Psi_{i,\mathrm{tv}}(0)
	\right|
	\le
	L_{\Psi,\mathrm{tv}}\epsilon.
\end{equation}

We now expand
\[
\phi_{\mathrm{cb},i}^{\mathrm{tv}}
\left(
\epsilon,
d_i^\star\left(\frac{\pi_0 T}{\widetilde T}\right)
\right)
\]
around $(0,c_0)$. Using~\eqref{cb:tv:error} to write the remainder as $O(\epsilon^2)$ uniformly over $i\in\mathcal I_{\mathrm b}$, we obtain
\begin{align}
	0
	=&\
	\phi_{\mathrm{cb},i}^{\mathrm{tv}}
	\left(
	\epsilon,
	d_i^\star\left(\frac{\pi_0 T}{\widetilde T}\right)
	\right)
	\nonumber\\
	=&
	-\epsilon\tilde v_i
	\!\sum_{t\in\mathcal T}\mu_{G_{t,\mathrm{tr}}}(c_0)
	-\!
	\left(\!
	d_i^\star\left(\frac{\pi_0 T}{\widetilde T}\right)\!-\!c_0
	\!\right)
	\sum_{t\in\mathcal T}
	\!\pi_{\mathrm {u}_t}\mu_{G_{t,\mathrm{tr}}}'\!(c_0)\nonumber\\
	&+
	O(\epsilon^2).
	\nonumber
\end{align}
Isolating $d_i^\star\left(\frac{\pi_0 T}{\widetilde T}\right)$ gives
\[
d_i^\star\left(\frac{\pi_0 T}{\widetilde T}\right)
=
c_0
-
\frac{
	\epsilon\tilde v_i
	\sum_{t\in\mathcal T}\mu_{G_{t,\mathrm{tr}}}(c_0)
}
{
	\sum_{t\in\mathcal T}
	\pi_{\mathrm {u}_t}\mu_{G_{t,\mathrm{tr}}}'(c_0)
}
+
O(\epsilon^2),
\]
which yields~\eqref{di:c_0:epsilon:lemma:tv}. Finally, applying~\eqref{d:star:tv} and using the uniform bound in~\eqref{cb:tv:error}, which implies that the $O(\epsilon^2)$ remainder above is uniform over $i\in\mathcal I_{\mathrm b}$, we obtain
\[
c_\mathrm{cb}^{\mathrm{ne}}
=
c_0
-
\frac{
	\epsilon\,\mathbb E[\widetilde V]
	\sum_{t\in\mathcal T}\mu_{G_{t,\mathrm{tr}}}(c_0)
}
{
	\sum_{t\in\mathcal T}
	\pi_{\mathrm {u}_t}\mu_{G_{t,\mathrm{tr}}}'(c_0)
}
+
O(\epsilon^2),
\]
which gives~\eqref{c_0:epsilon:lemma:tv} and completes the proof of Lemma~\ref{lemma12}.
$\,\,\blacksquare$

\section{Proof for Subsection~\ref{risk:analysis:subsection}}

\subsection{Proof of Proposition~\ref{prop:cvar_continuous}}

Let $\xi_\alpha:=\xi_\alpha(X)$. Since $X$ has a continuous distribution, we have
\[
\mathbb P(X\ge \xi_\alpha)=1-\alpha
\]
and
\[
\mathbb P(X=\xi_\alpha)=0.
\]
Using the definition of CVaR in~\eqref{eq:cvar_general}, we can write
\begin{align}
	\mathrm{CVaR}_{\alpha}(X)
	&=
	\xi_\alpha
	+
	\frac{1}{1-\alpha}
	\mathbb E\left[
	\left(X-\xi_\alpha\right)_+
	\right] \nonumber\\
	&=
	\xi_\alpha
	+
	\frac{1}{1-\alpha}
	\mathbb E\left[
	\left(X-\xi_\alpha\right)
	\mathbb 1\{X\ge \xi_\alpha\}
	\right] \nonumber\\
	&=
	\xi_\alpha
	+
	\frac{
		\mathbb E\left[X\mathbb 1\{X\ge \xi_\alpha\}\right]
		-
		\xi_\alpha\mathbb P(X\ge \xi_\alpha)
	}
	{\mathbb P(X\ge \xi_\alpha)} \nonumber\\
	&=
	\frac{
		\mathbb E\left[X\mathbb 1\{X\ge \xi_\alpha\}\right]
	}
	{\mathbb P(X\ge \xi_\alpha)} \nonumber\\
	&=
	\mathbb E
	\left[
	X
	\mid
	X\ge \xi_\alpha
	\right].
\end{align}
Substituting back $\xi_\alpha=\xi_\alpha(X)$ gives the desired result.
$\quad \,\,\,\,\,\blacksquare$

\subsection{Proof of Lemma~\ref{lem:risk_aversion_capacity}}

We first prove the claim for the single-product real-time market, i.e.,
$
c_{\mathrm{srt},\eta}^{\mathrm{ne}}
\le
c_{\mathrm{srt}}^{\mathrm{ne}}.
$
Recall from~\eqref{eq:srt:expect} that
\begin{equation}\label{eq:proof:risk:srt:neutral}
	\pi_{\mathrm u}
	\mu_{G_{\mathrm{tr}}}
	\left(c_{\mathrm{srt}}^{\mathrm{ne}}\right)
	=
	\frac{\pi_0}{\widetilde T}.
\end{equation}
Moreover, as established earlier, $\mu_{G_{\mathrm{tr}}}(\cdot)$ is non-increasing on $\mathbb R_+$ and is strictly decreasing on $(L/\overline g,\infty)$; in particular, its derivative in the relevant interior region is given by~\eqref{mug_prime_homo}. Also, by the uniqueness argument for~\eqref{eq:srt:expect} under~\eqref{eq:pi0cond}, we have
\[
c_{\mathrm{srt}}^{\mathrm{ne}}
>
\frac{L}{\overline g}.
\]

Now consider the risk-adjusted equilibrium capacity
$c_{\mathrm{srt},\eta}^{\mathrm{ne}}$. If
$c_{\mathrm{srt},\eta}^{\mathrm{ne}}=0$, then the desired inequality holds immediately. Thus, suppose that
$c_{\mathrm{srt},\eta}^{\mathrm{ne}}>0$. By the risk-adjusted equilibrium condition,
\[
\Pi_{\mathrm{srt},\eta}^{\mathrm{s}}
\left(c_{\mathrm{srt},\eta}^{\mathrm{ne}}\right)
=
\pi_0 c_{\mathrm{srt},\eta}^{\mathrm{ne}}.
\]
Using the definition of the risk-adjusted revenue under the single-product real-time market, we obtain
\begin{align}
	\frac{\pi_0}{\widetilde T}
	&=
	\pi_{\mathrm u}
	\mu_{G_{\mathrm{tr}}}
	\left(c_{\mathrm{srt},\eta}^{\mathrm{ne}}\right)
	-
	\frac{\eta}
	{c_{\mathrm{srt},\eta}^{\mathrm{ne}}}
	\mathrm{CVaR}_{\alpha}
	\left(
	X_{\mathrm{srt}}
	\left(c_{\mathrm{srt},\eta}^{\mathrm{ne}},G\right)
	\right) \nonumber\\
	&\le
	\pi_{\mathrm u}
	\mu_{G_{\mathrm{tr}}}
	\left(c_{\mathrm{srt},\eta}^{\mathrm{ne}}\right),
	\label{eq:proof:risk:srt:ineq}
\end{align}
where the inequality follows from $\eta\ge0$ and
$\mathrm{CVaR}_{\alpha}\left(X_{\mathrm{srt}}(c,G)\right)\ge0$. Combining~\eqref{eq:proof:risk:srt:neutral} and~\eqref{eq:proof:risk:srt:ineq} gives
\[
\pi_{\mathrm u}
\mu_{G_{\mathrm{tr}}}
\left(c_{\mathrm{srt}}^{\mathrm{ne}}\right)
\le
\pi_{\mathrm u}
\mu_{G_{\mathrm{tr}}}
\left(c_{\mathrm{srt},\eta}^{\mathrm{ne}}\right).
\]
If $c_{\mathrm{srt},\eta}^{\mathrm{ne}}\le L/\overline g$, then the desired inequality follows immediately because
$c_{\mathrm{srt}}^{\mathrm{ne}}>L/\overline g$. Otherwise,
$c_{\mathrm{srt},\eta}^{\mathrm{ne}}>L/\overline g$, and both capacities lie in the region where
$\mu_{G_{\mathrm{tr}}}(\cdot)$ is strictly decreasing. Hence,
\[
c_{\mathrm{srt},\eta}^{\mathrm{ne}}
\le
c_{\mathrm{srt}}^{\mathrm{ne}}.
\]
This proves the first claim.

We next prove the claim for the product-differentiated real-time market, i.e.,
$
c_{\mathrm{prt},\eta}^{\mathrm{ne}}
\le
c_{\mathrm{prt}}^{\mathrm{ne}}.
$
Recall from~\eqref{eq:eqc:prt} that the risk-neutral equilibrium capacity $c_{\mathrm{prt}}^{\mathrm{ne}}$ satisfies
\begin{equation}\label{eq:proof:risk:prt:neutral}
	\frac{\Pi_{\mathrm{prt}}^{\mathrm{s}}
		\left(c_{\mathrm{prt}}^{\mathrm{ne}}\right)}
	{\widetilde T c_{\mathrm{prt}}^{\mathrm{ne}}}
	=
	\frac{\pi_0}{\widetilde T}.
\end{equation}
Also, by the same pointwise monotonicity argument used earlier for the product-differentiated real-time market, the per-unit expected revenue
\[
c\mapsto
\frac{\Pi_{\mathrm{prt}}^{\mathrm{s}}(c)}
{\widetilde T c}
\]
is non-increasing in $c$ and is strictly decreasing on $(L/\overline g,\infty)$. Moreover, by~\eqref{eq:pi0cond} and the uniqueness argument for the product-differentiated equilibrium condition, we have
\[
c_{\mathrm{prt}}^{\mathrm{ne}}
>
\frac{L}{\overline g}.
\]

Now consider the risk-adjusted equilibrium capacity
$c_{\mathrm{prt},\eta}^{\mathrm{ne}}$. If
$c_{\mathrm{prt},\eta}^{\mathrm{ne}}=0$, then the desired inequality holds immediately. Thus, suppose that
$c_{\mathrm{prt},\eta}^{\mathrm{ne}}>0$. By the risk-adjusted equilibrium condition,
\[
\Pi_{\mathrm{prt},\eta}^{\mathrm{s}}
\left(c_{\mathrm{prt},\eta}^{\mathrm{ne}}\right)
=
\pi_0 c_{\mathrm{prt},\eta}^{\mathrm{ne}}.
\]
Using the definition of the risk-adjusted revenue under the product-differentiated real-time market, we obtain
\begin{align}
	\frac{\pi_0}{\widetilde T}
	&=
	\frac{\Pi_{\mathrm{prt}}^{\mathrm{s}}
		\left(c_{\mathrm{prt},\eta}^{\mathrm{ne}}\right)}
	{\widetilde T c_{\mathrm{prt},\eta}^{\mathrm{ne}}}
	-
	\frac{\eta}
	{c_{\mathrm{prt},\eta}^{\mathrm{ne}}}
	\mathrm{CVaR}_{\alpha}
	\left(
	X_{\mathrm{prt}}
	\left(c_{\mathrm{prt},\eta}^{\mathrm{ne}},G\right)
	\right) \nonumber\\
	&\le
	\frac{\Pi_{\mathrm{prt}}^{\mathrm{s}}
		\left(c_{\mathrm{prt},\eta}^{\mathrm{ne}}\right)}
	{\widetilde T c_{\mathrm{prt},\eta}^{\mathrm{ne}}},
	\label{eq:proof:risk:prt:ineq}
\end{align}
where the inequality follows from $\eta\ge0$ and
$\mathrm{CVaR}_{\alpha}\left(X_{\mathrm{prt}}(c,G)\right)\ge0$. Combining~\eqref{eq:proof:risk:prt:neutral} and~\eqref{eq:proof:risk:prt:ineq} gives
\[
\frac{\Pi_{\mathrm{prt}}^{\mathrm{s}}
	\left(c_{\mathrm{prt}}^{\mathrm{ne}}\right)}
{\widetilde T c_{\mathrm{prt}}^{\mathrm{ne}}}
\le
\frac{\Pi_{\mathrm{prt}}^{\mathrm{s}}
	\left(c_{\mathrm{prt},\eta}^{\mathrm{ne}}\right)}
{\widetilde T c_{\mathrm{prt},\eta}^{\mathrm{ne}}}.
\]
If $c_{\mathrm{prt},\eta}^{\mathrm{ne}}\le L/\overline g$, then the desired inequality follows immediately because
$c_{\mathrm{prt}}^{\mathrm{ne}}>L/\overline g$. Otherwise,
$c_{\mathrm{prt},\eta}^{\mathrm{ne}}>L/\overline g$, and both capacities lie in the region where the per-unit product-differentiated expected revenue is strictly decreasing. Hence,
\[
c_{\mathrm{prt},\eta}^{\mathrm{ne}}
\le
c_{\mathrm{prt}}^{\mathrm{ne}}.
\]
This proves the second claim.

We finally prove the claim for the contract-based market, i.e.,
$
c_{\mathrm{cb},\eta}^{\mathrm{ne}}
=
c_{\mathrm{cb}}^{\mathrm{ne}}.
$
Let $\pi_{\mathrm{cb}}(c)$ denote the contract-based market equilibrium price when the aggregate panel capacity is $c$. From Lemma~\ref{lemma:ne:cb}, this price satisfies
\begin{equation}\label{eq:proof:risk:cb:price}
	c
	=
	\widehat d^\star
	\left(
	\pi_{\mathrm{cb}}(c)
	\right).
\end{equation}
Moreover, the risk-neutral equilibrium capacity under the contract-based market is characterized by
\begin{equation}\label{eq:proof:risk:cb:neutral}
	c_{\mathrm{cb}}^{\mathrm{ne}}
	=
	\widehat d^\star
	\left(
	\frac{\pi_0}{\widetilde T}
	\right).
\end{equation}

Now consider the risk-adjusted equilibrium capacity
$c_{\mathrm{cb},\eta}^{\mathrm{ne}}$. We focus on the nonzero equilibrium capacity characterized by the zero-profit condition, consistent with the equilibrium characterization in Lemma~\ref{lemma:nec}. Under the contract-based market, sellers receive the contract payment before the realization of $G$. Hence, conditional on $c$, the aggregate seller revenue is deterministic and equal to
\[
R_{\mathrm{cb}}^{\mathrm{s}}(c,G)
=
c\pi_{\mathrm{cb}}(c).
\]
Therefore,
\[
X_{\mathrm{cb}}(c,G)
=
c
\left(
\frac{\pi_0}{\widetilde T}
-
\pi_{\mathrm{cb}}(c)
\right)_+,
\]
and since the CVaR of a deterministic random variable is equal to the value of that random variable,
\[
\mathrm{CVaR}_{\alpha}
\left(
X_{\mathrm{cb}}(c,G)
\right)
=
c
\left(
\frac{\pi_0}{\widetilde T}
-
\pi_{\mathrm{cb}}(c)
\right)_+.
\]

By the risk-adjusted equilibrium condition,
\[
\Pi_{\mathrm{cb},\eta}^{\mathrm{s}}
\left(c_{\mathrm{cb},\eta}^{\mathrm{ne}}\right)
=
\pi_0 c_{\mathrm{cb},\eta}^{\mathrm{ne}}.
\]
Using the definition of the risk-adjusted revenue and dividing by
$c_{\mathrm{cb},\eta}^{\mathrm{ne}}>0$ gives
\begin{equation}\label{eq:proof:risk:cb:price_eq}
	\frac{\pi_0}{\widetilde T}
	=
	\pi_{\mathrm{cb}}
	\left(c_{\mathrm{cb},\eta}^{\mathrm{ne}}\right)
	-
	\eta
	\left(
	\frac{\pi_0}{\widetilde T}
	-
	\pi_{\mathrm{cb}}
	\left(c_{\mathrm{cb},\eta}^{\mathrm{ne}}\right)
	\right)_+.
\end{equation}
We now show that~\eqref{eq:proof:risk:cb:price_eq} implies
\[
\pi_{\mathrm{cb}}
\left(c_{\mathrm{cb},\eta}^{\mathrm{ne}}\right)
=
\frac{\pi_0}{\widetilde T}.
\]
If
\[
\pi_{\mathrm{cb}}
\left(c_{\mathrm{cb},\eta}^{\mathrm{ne}}\right)
\ge
\frac{\pi_0}{\widetilde T},
\]
then the positive-part term in~\eqref{eq:proof:risk:cb:price_eq} is zero, and the claim follows. If instead
\[
\pi_{\mathrm{cb}}
\left(c_{\mathrm{cb},\eta}^{\mathrm{ne}}\right)
<
\frac{\pi_0}{\widetilde T},
\]
then~\eqref{eq:proof:risk:cb:price_eq} becomes
\[
\frac{\pi_0}{\widetilde T}
=
\pi_{\mathrm{cb}}
\left(c_{\mathrm{cb},\eta}^{\mathrm{ne}}\right)
-
\eta
\left(
\frac{\pi_0}{\widetilde T}
-
\pi_{\mathrm{cb}}
\left(c_{\mathrm{cb},\eta}^{\mathrm{ne}}\right)
\right),
\]
or equivalently,
\[
(1+\eta)
\pi_{\mathrm{cb}}
\left(c_{\mathrm{cb},\eta}^{\mathrm{ne}}\right)
=
(1+\eta)
\frac{\pi_0}{\widetilde T}.
\]
Since $\eta\ge0$, this implies
\[
\pi_{\mathrm{cb}}
\left(c_{\mathrm{cb},\eta}^{\mathrm{ne}}\right)
=
\frac{\pi_0}{\widetilde T},
\]
which contradicts the strict inequality assumed above. Therefore,
\begin{equation}\label{eq:proof:risk:cb:price_final}
	\pi_{\mathrm{cb}}
	\left(c_{\mathrm{cb},\eta}^{\mathrm{ne}}\right)
	=
	\frac{\pi_0}{\widetilde T}.
\end{equation}

Substituting~\eqref{eq:proof:risk:cb:price_final} into~\eqref{eq:proof:risk:cb:price} gives
\[
c_{\mathrm{cb},\eta}^{\mathrm{ne}}
=
\widehat d^\star
\left(
\frac{\pi_0}{\widetilde T}
\right).
\]
Comparing this expression with~\eqref{eq:proof:risk:cb:neutral} yields
\[
c_{\mathrm{cb},\eta}^{\mathrm{ne}}
=
c_{\mathrm{cb}}^{\mathrm{ne}}.
\]
This proves the third claim and completes the proof of Lemma~\ref{lem:risk_aversion_capacity}.
$\qquad \qquad \qquad \qquad \qquad\qquad\qquad \qquad\qquad \quad \quad\blacksquare$

\subsection{Proof of Theorem~\ref{thm:risk_sensitive_comparison}}

We first prove the general case, i.e., for any $\epsilon\ge0$,
\begin{equation}
	c_{\mathrm{srt},\eta}^{\mathrm{ne}}
	\le
	c_{\mathrm{prt},\eta}^{\mathrm{ne}}
	\le
	c_{\mathrm{opt}},
	\qquad
	c_{\mathrm{srt},\eta}^{\mathrm{ne}}
	\le
	c_{\mathrm{cb},\eta}^{\mathrm{ne}}.
	\nonumber
\end{equation}
The bound $c_{\mathrm{prt},\eta}^{\mathrm{ne}}\le c_{\mathrm{opt}}$ follows directly from Lemma~\ref{lem:risk_aversion_capacity} and Theorem~\ref{thm:main}. In particular,
\begin{equation}
	c_{\mathrm{prt},\eta}^{\mathrm{ne}}
	\le
	c_{\mathrm{prt}}^{\mathrm{ne}}
	=
	c_{\mathrm{opt}}.
	\nonumber
\end{equation}
It remains to prove
\[
c_{\mathrm{srt},\eta}^{\mathrm{ne}}
\le
c_{\mathrm{prt},\eta}^{\mathrm{ne}}.
\]
If $c_{\mathrm{srt},\eta}^{\mathrm{ne}}=0$, the result is immediate. Thus, suppose
$c_{\mathrm{srt},\eta}^{\mathrm{ne}}>0$.

For $m\in\{\mathrm{srt},\mathrm{prt}\}$ and $c>0$, let
\[
Y_m(c,G):=\frac{R_m^{\mathrm{s}}(c,G)}{c}
\]
denote the realized per-unit seller revenue in the representative operation period. Then, by the pointwise comparison in~\eqref{prt:srt:integrand:homo}, we have, for every $c>0$ and every realization of $G$,
\begin{equation}\label{eq:proof:risk:Y_order}
	Y_{\mathrm{srt}}(c,G)
	\le
	Y_{\mathrm{prt}}(c,G).
\end{equation}

Define the risk-adjusted per-unit revenue function
\begin{equation}\label{eq:proof:risk:Psi_def}
	\Psi_{m,\eta}(c)
	:=
	\mathbb{E}
	\left[
	Y_m(c,G)
	\right]
	-
	\eta
	\mathrm{CVaR}_{\alpha}
	\left(
	\left(
	\frac{\pi_0}{\widetilde T}
	-
	Y_m(c,G)
	\right)_+
	\right).
\end{equation}
Indeed, since
\[
X_m(c,G)
=
\left(
\frac{\pi_0 c}{\widetilde T}
-
R_m^{\mathrm{s}}(c,G)
\right)_+
=
c
\left(
\frac{\pi_0}{\widetilde T}
-
Y_m(c,G)
\right)_+,
\]
and $\mathrm{CVaR}_{\alpha}$ is positively homogeneous, the risk-adjusted equilibrium condition can be written in the per-unit form
\begin{equation}\label{eq:proof:risk:Psi_eq}
	\Psi_{m,\eta}(c_{m,\eta}^{\mathrm{ne}})
	=
	\frac{\pi_0}{\widetilde T}.
\end{equation}

From~\eqref{eq:proof:risk:Y_order}, we have
\[
\mathbb{E}
\left[
Y_{\mathrm{srt}}(c,G)
\right]
\le
\mathbb{E}
\left[
Y_{\mathrm{prt}}(c,G)
\right],
\]
and
\[
\left(
\frac{\pi_0}{\widetilde T}
-
Y_{\mathrm{prt}}(c,G)
\right)_+
\le
\left(
\frac{\pi_0}{\widetilde T}
-
Y_{\mathrm{srt}}(c,G)
\right)_+.
\]
By the monotonicity of $\mathrm{CVaR}_{\alpha}$ and since $\eta\ge0$, it follows that
\begin{equation}\label{eq:proof:risk:Psi_order}
	\Psi_{\mathrm{srt},\eta}(c)
	\le
	\Psi_{\mathrm{prt},\eta}(c),
	\qquad c>0.
\end{equation}

Next, note that $\Psi_{\mathrm{prt},\eta}(c)$ is non-increasing in $c$. Indeed, as established in the proof of Lemma~\ref{lem:risk_aversion_capacity}, the realized per-unit seller revenue under the product-differentiated real-time market is pointwise non-increasing in $c$. Hence, for any $0<c_1<c_2$,
\[
Y_{\mathrm{prt}}(c_2,G)
\le
Y_{\mathrm{prt}}(c_1,G)
\]
for every realization of $G$. Therefore,
\[
\mathbb{E}
\left[
Y_{\mathrm{prt}}(c_2,G)
\right]
\le
\mathbb{E}
\left[
Y_{\mathrm{prt}}(c_1,G)
\right],
\]
and
\[
\left(
\frac{\pi_0}{\widetilde T}
-
Y_{\mathrm{prt}}(c_2,G)
\right)_+
\ge
\left(
\frac{\pi_0}{\widetilde T}
-
Y_{\mathrm{prt}}(c_1,G)
\right)_+.
\]
Thus, the corresponding CVaR term is non-decreasing in $c$. Since this term is subtracted in~\eqref{eq:proof:risk:Psi_def}, $\Psi_{\mathrm{prt},\eta}(c)$ is non-increasing in $c$.

By~\eqref{eq:proof:risk:Psi_eq},
\[
\Psi_{\mathrm{srt},\eta}
\left(c_{\mathrm{srt},\eta}^{\mathrm{ne}}\right)
=
\frac{\pi_0}{\widetilde T}.
\]
Using~\eqref{eq:proof:risk:Psi_order}, we obtain
\[
\Psi_{\mathrm{prt},\eta}
\left(c_{\mathrm{srt},\eta}^{\mathrm{ne}}\right)
\ge
\frac{\pi_0}{\widetilde T}.
\]
Since $\Psi_{\mathrm{prt},\eta}(c)$ is non-increasing in $c$, the capacity at which the product-differentiated risk-adjusted per-unit revenue reaches the investment-cost level $\pi_0/\widetilde T$ cannot be smaller than $c_{\mathrm{srt},\eta}^{\mathrm{ne}}$. Therefore,
\[
c_{\mathrm{srt},\eta}^{\mathrm{ne}}
\le
c_{\mathrm{prt},\eta}^{\mathrm{ne}}.
\]
Combining this with $c_{\mathrm{prt},\eta}^{\mathrm{ne}}\le c_{\mathrm{opt}}$ gives
\[
c_{\mathrm{srt},\eta}^{\mathrm{ne}}
\le
c_{\mathrm{prt},\eta}^{\mathrm{ne}}
\le
c_{\mathrm{opt}}.
\]

It remains to establish the general-case comparison with the contract-based market. By Lemma~\ref{lem:risk_aversion_capacity},
\[
c_{\mathrm{srt},\eta}^{\mathrm{ne}}
\le
c_{\mathrm{srt}}^{\mathrm{ne}},
\qquad
c_{\mathrm{cb},\eta}^{\mathrm{ne}}
=
c_{\mathrm{cb}}^{\mathrm{ne}}.
\]
Moreover, by the general case of Theorem~\ref{thm:main},
\[
c_{\mathrm{srt}}^{\mathrm{ne}}
\le
c_{\mathrm{cb}}^{\mathrm{ne}}.
\]
Therefore,
\[
c_{\mathrm{srt},\eta}^{\mathrm{ne}}
\le
c_{\mathrm{cb},\eta}^{\mathrm{ne}}.
\]
This proves the first claim.

We next prove the small-solar-premium case. Suppose that $\epsilon>0$ is sufficiently small and the assumptions of the small-solar-premium case of Theorem~\ref{thm:main} hold. From the general case of this theorem, we already have
\[
c_{\mathrm{srt},\eta}^{\mathrm{ne}}
\le
c_{\mathrm{prt},\eta}^{\mathrm{ne}}
\le
c_{\mathrm{opt}}.
\]
By Lemma~\ref{lem:risk_aversion_capacity},
\[
c_{\mathrm{cb},\eta}^{\mathrm{ne}}
=
c_{\mathrm{cb}}^{\mathrm{ne}}.
\]
Moreover, by the small-solar-premium case of Theorem~\ref{thm:main},
\[
c_{\mathrm{opt}}
=
c_{\mathrm{prt}}^{\mathrm{ne}}
\lesssim
c_{\mathrm{cb}}^{\mathrm{ne}}.
\]
Therefore,
\[
c_{\mathrm{opt}}
\lesssim
c_{\mathrm{cb},\eta}^{\mathrm{ne}}.
\]
Combining the above relations gives
\begin{equation}
	c_{\mathrm{srt},\eta}^{\mathrm{ne}}
	\le
	c_{\mathrm{prt},\eta}^{\mathrm{ne}}
	\le
	c_{\mathrm{opt}}
	\lesssim
	c_{\mathrm{cb},\eta}^{\mathrm{ne}}.\nonumber
\end{equation}

It remains to state the gap bound corresponding to the last inequality. By Lemma~\ref{lem:risk_aversion_capacity},
\[
c_{\mathrm{prt},\eta}^{\mathrm{ne}}
\le
c_{\mathrm{prt}}^{\mathrm{ne}},
\qquad
c_{\mathrm{cb},\eta}^{\mathrm{ne}}
=
c_{\mathrm{cb}}^{\mathrm{ne}}.
\]
Thus,
\begin{align}
	c_{\mathrm{cb},\eta}^{\mathrm{ne}}
	-
	c_{\mathrm{prt},\eta}^{\mathrm{ne}}
	&=
	c_{\mathrm{cb}}^{\mathrm{ne}}
	-
	c_{\mathrm{prt},\eta}^{\mathrm{ne}}
	\nonumber\\
	&\ge
	c_{\mathrm{cb}}^{\mathrm{ne}}
	-
	c_{\mathrm{prt}}^{\mathrm{ne}}.
\end{align}
Using~\eqref{thm:main:case2:detail}, we obtain
\begin{equation}
	c_{\mathrm{cb},\eta}^{\mathrm{ne}}
	-
	c_{\mathrm{prt},\eta}^{\mathrm{ne}}
	\ge
	\beta\epsilon
	-
	O(\epsilon^2).\nonumber
\end{equation}
This proves the second claim.

We finally prove the no-solar-premium case. Suppose that $\epsilon=0$. Then the product-differentiated premium term vanishes. Equivalently, the pointwise comparison in~\eqref{prt:srt:integrand:homo} holds with equality, and therefore
\[
Y_{\mathrm{srt}}(c,G)
=
Y_{\mathrm{prt}}(c,G),
\qquad c>0.
\]
Hence,
\[
\Psi_{\mathrm{srt},\eta}(c)
=
\Psi_{\mathrm{prt},\eta}(c),
\qquad c>0.
\]
Thus, the risk-adjusted equilibrium equations for the single-product and product-differentiated real-time markets coincide, and under the same equilibrium selection,
\[
c_{\mathrm{srt},\eta}^{\mathrm{ne}}
=
c_{\mathrm{prt},\eta}^{\mathrm{ne}}.
\]
Moreover, by the general case of this theorem,
\[
c_{\mathrm{prt},\eta}^{\mathrm{ne}}
\le
c_{\mathrm{opt}}.
\]
It remains to identify the contract-based capacity. By Lemma~\ref{lem:risk_aversion_capacity},
\[
c_{\mathrm{cb},\eta}^{\mathrm{ne}}
=
c_{\mathrm{cb}}^{\mathrm{ne}}.
\]
On the other hand, by~\eqref{eq:thm:p3}, when $\epsilon=0$,
\[
c_{\mathrm{opt}}
=
c_{\mathrm{cb}}^{\mathrm{ne}}.
\]
Thus,
\[
c_{\mathrm{opt}}
=
c_{\mathrm{cb},\eta}^{\mathrm{ne}}.
\]
Combining the above relations gives
\[
c_{\mathrm{srt},\eta}^{\mathrm{ne}}
=
c_{\mathrm{prt},\eta}^{\mathrm{ne}}
\le
c_{\mathrm{opt}}
=
c_{\mathrm{cb},\eta}^{\mathrm{ne}}.
\]
This proves the third claim and completes the proof of Theorem~\ref{thm:risk_sensitive_comparison}.
$\qquad \qquad\qquad \qquad\qquad\qquad\qquad\qquad\qquad\,\,\,\quad \blacksquare$

\end{document}